\documentclass[%
amsmath,amssymb,
aps,
PRL,
reprint,
]{revtex4-2}
\usepackage{graphicx}% Include figure files
\usepackage{dcolumn}% Align table columns on decimal point
\usepackage{bm}% bold math
\usepackage{color}
\usepackage{xcolor}
\usepackage{subfigure}
\usepackage{xspace}
\usepackage{textpos}
\usepackage[english]{babel}
\usepackage{multirow}
\usepackage{overpic}
\usepackage{colortbl}
\usepackage{float}
\usepackage{array}
\usepackage{makecell}
\usepackage{enumerate}
\usepackage{soul}
\usepackage{adjustbox}
\usepackage{titlesec}
\usepackage{flushend}
\usepackage[colorlinks=true, linkcolor=blue, filecolor=blue, urlcolor=blue, citecolor=blue]{hyperref}
\definecolor{aa}{RGB}{0,0,139}
\newcommand{\comment}[1]{}

\newcommand{\BR}{{\cal B}}

\newcommand{\Jpsi}{$J/\psi$}
\newcommand{\psip}{$\psi(3686)$}
\newcommand{\pspp}{\psi(3770)}
\newcommand{\psipp}{$\pspp$}
\newcommand{\Ks}{$K_S^0$}
\newcommand{\KL}{$K_L^0$}

\newcommand{\EE}{e^+e^-}
\newcommand{\KsKL}{$K_S^0K_L^0$}
\newcommand{\pipi}{$\pi^+\pi^-$}

\newcommand{\eetoKsKL}{$e^+e^-\to K_S^0K_L^0$}
\newcommand{\psiptoKsKL}{$\psi(3686) \to K_S^0K_L^0$}
\newcommand{\psipptoKsKL}{$\psi(3770) \to K_S^0K_L^0$}
\newcommand{\ISRpsip}{$e^+e^-\to\gamma^{\rm ISR}\psi(3686)$}
\newcommand{\eetoKstarK}{$e^+e^-\to K^{\ast0}(892)\bar{K}^0+c.c.$}

\newcommand{\Kstopipi}{$K_S^0\to\pi^+\pi^-$}
\newcommand{\piotogg}{$\pi^0\to\gamma\gamma$}

\begin{document}

\title{
%(RM NOTE: TITLE)
\boldmath 
Observation of significant flavor-SU(3) breaking in the kaon wave function at \texorpdfstring{$12~{\rm GeV}^2<Q^2<25~{\rm GeV}^2$}{} and discovery of the charmless decay \texorpdfstring{$\psi(3770)\to K_S^0K_L^0$}{}
}

%%\author{BESIII Collaboration}
\author{
    %% Saved at => 2023-07-20
M.~Ablikim$^{1}$, M.~N.~Achasov$^{4,b}$, P.~Adlarson$^{75}$, X.~C.~Ai$^{81}$, R.~Aliberti$^{35}$, A.~Amoroso$^{74A,74C}$, M.~R.~An$^{39}$, Q.~An$^{71,58}$, Y.~Bai$^{57}$, O.~Bakina$^{36}$
% ...
\\
\vspace{0.2cm}
(BESIII Collaboration)\\
\vspace{0.2cm} {\it
$^{1}$ Institute of High Energy Physics, Beijing 100049, People's Republic of China\\
% ...
$^{4}$ Budker Institute of Nuclear Physics SB RAS (BINP), Novosibirsk 630090, Russia\\
% ...
$^{35}$ Johannes Gutenberg University of Mainz, Johann-Joachim-Becher-Weg 45, D-55099 Mainz, Germany\\
$^{36}$ Joint Institute for Nuclear Research, 141980 Dubna, Moscow region, Russia\\
% ...
$^{39}$ Liaoning Normal University, Dalian 116029, People's Republic of China\\
% ...
$^{57}$ Southeast University, Nanjing 211100, People's Republic of China\\
$^{58}$ State Key Laboratory of Particle Detection and Electronics, Beijing 100049, Hefei 230026, People's Republic of China\\
% ...
$^{71}$ University of Science and Technology of China, Hefei 230026, People's Republic of China\\
% ...
$^{74}$ University of Turin and INFN, (A)University of Turin, I-10125, Turin, Italy; (B)University of Eastern Piedmont, I-15121, Alessandria, Italy; (C)INFN, I-10125, Turin, Italy\\
$^{75}$ Uppsala University, Box 516, SE-75120 Uppsala, Sweden\\
% ...
$^{81}$ Zhengzhou University, Zhengzhou 450001, People's Republic of China\\

\vspace{0.2cm}
% ...
$^{b}$ Also at the Novosibirsk State University, Novosibirsk, 630090, Russia\\
% ...
}
%% ends here %%
}
%\date{\today}
\date{December 17, 2023}

%%%%%%%%%%%%%%%%%%%%%%%%%%%%%%%%%%%%%%%%%%%%%%%%%%%%%%%%%%%%%
\begin{abstract}

%(RM NOTE: ABSTRACT)
We present cross sections for the reaction $e^+e^-\to K_S^0K_L^0$ at center-of-mass energies ranging from 3.51 GeV to 4.95 GeV using data samples collected in the BESIII experiment, corresponding to a total integrated luminosity of 26.5 fb$^{-1}$.
The ratio of neutral-to-charged kaon form factors at large momentum transfers ($12~{\rm GeV}^2<Q^2<25~{\rm GeV}^2$) is determined to be $0.21\pm 0.01$, which indicates a small but significant effect of flavor-SU(3) breaking in the kaon wave function,
and consequently excludes the possibility that flavor-SU(3) breaking is the  primary reason for the strong experimental violation of the pQCD prediction $|F(\pi^{\pm})|/|F(K^{\pm})|=f^2_{\pi}/f^2_{K}$, where $F(\pi^{\pm})$ and $F(K^{\pm})$ are the form factors, and $f_{\pi}$ and $f_{K}$ are the decay constants of charged pions and kaons, respectively.
We also observe a significant signal for the charmless decay $\psi(3770)\to K_S^0K_L^0$ for the first time. 
Within a $1\sigma$ contour of the likelihood value, the the branching fraction for $\psi(3770)\to K_S^0K_L^0$ is determined to be ${\cal B}=(2.63_{-1.59}^{+1.40})\times 10^{-5}$, and the relative phase between the continuum and $\psi(3770)$ amplitudes is $\phi=(-0.39_{-0.10}^{+0.05})\pi$.
The branching fraction is in good agreement with the $\mathcal{S}$- and $\mathcal{D}$-wave charmonia mixing scheme proposed in the interpretation of the ``$\rho\pi$ puzzle'' between $J/\psi$ and $\psi(3686)$ decays.

\end{abstract}

%\pacs{Valid PACS appear here}

\maketitle

%%%%%%%%%%%%%%%%%%%%%%%%%%%%%%%%%%%%%%%%%%%%%%%%%%%%%%%%%%%%%
%\section{Introduction}

%(RM NOTE: BODY)
Understanding the internal quark-gluon structure of hadrons based on quantum chromodynamics (QCD) has been one of the fundamental aims of nuclear and particle physics. An important tool to achieve such a goal is to measure the electromagnetic form factors of hadrons at large momentum transfers ($Q^2$), 
where a photon acts as a probe that allows us to 
see the electric charges of the quarks and gluons inside of the hadron, rather than just the composite hadron~\cite{ff_pi,ff_pi2}.
Kaons are of great interest because their constituent quarks differ in mass by more than one order of magnitude, which leads to a broken flavor-SU(3) symmetry in the wave function and thus the form factor of neutral kaons develops an asymmetric component and deviates from zero~\cite{waveFun}. 
The measurements of the electromagnetic form factors of kaons at large momentum transfers provide a sensitive measure of flavor-SU(3) breaking effects, estimated by the ratio of the form factors $|F_{K_S^0K_L^0}(Q^2)|/|F_{K^+K^-}(Q^2)|$. A value close to $1$ indicates that the flavor-SU(3) breaking has a significant influence on the kaon's wave function, while a value close to $0$ indicates a minor effect~\cite{waveFun}.
Furthermore, this measurement also provides valuable insights into the hypothesis~\cite{waveFun} that flavor-SU(3) breaking may be the primary factor responsible for the large deviation between the perturbative QCD (pQCD) prediction $|F(\pi^{\pm})|/|F(K^{\pm})|=f^2_{\pi}/f^2_{K}$ \cite{piKConjecture,piKConjecture2} and the experimental results $|F(\pi^{\pm})|/|F(K^{\pm})|=1.09\pm 0.04$ at $|Q^2|=17.4$ GeV$^2$~\cite{eFormFactor2} and $f^2_{\pi}/f^2_{K}=0.84\pm0.01$ \cite{PDG}. Here $F(\pi^{\pm})$ and $F(K^{\pm})$ are the form factors, and $f_{\pi}$ and $f_{K}$ are the decay constants of charged pions and kaons, respectively.

The \KsKL~final state of vector charmonium decays is particularly interesting to understand the violation of the ``12\% rule''~\cite{12rule,12rule2,12rule3,12rule4}, where $Q_h = \frac{\BR(\psi(3686)\to h)}{\BR(J/\psi\to h)} = (13.3\pm0.3)\%$ according to pQCD \cite{PDG}. However, $Q_{K_S^0K_L^0}$ is found to be 
$(27.2\pm 3.6)\%$ 
%$(27.2\pm 3.55)\%$
which deviates from the ``12\% rule'' by more than $3\sigma$~\cite{BESresult2,BESresult22}. Here we focus on an explanation proposed by Rosner using $\mathcal{S}$- and $\mathcal{D}$-wave charmonia mixing~\cite{SDmixing}, where the \psip~and $\pspp$ are considered to be mixtures of the $\psi(2^3S_1)$ and $\psi(1^3D_1)$ states. In this scenario, the branching fraction of \psipptoKsKL~is predicted to be within 
%$0.07\pm0.05 \leq \BR(\psi^{\prime\prime} \to K_S^0K_L^0)\times10^5 \leq 3.7\pm1.6$~\cite{SDYuan}.
$[0.07\pm0.05,3.7\pm1.6]\times10^{-5}$~\cite{SDYuan}.

%Up to now, form factors at large momentum transfers are predicted~\cite{kaon_ff}, however 
Theoretical studies on the form factors at large momentum transfers based on different models have so far yielded conflicting results on the flavor-SU(3) breaking effects of kaons~\cite{kaon_ff,SU3Theo,SU3Theo2,SU3Theo3,SU3Theo4}, while experimental measurements are limited to small momentum transfers ($|Q^2|<9.49~\rm{GeV^2}$)~\cite{lowKaon,lowKaon2}. No significant signal was observed at the $\pspp$ peak before
%in BES and CLEO-c 
~\cite{singleKaon,singleKaon2}, and only 4 events are observed with an expectation of 0.3 background events at $\sqrt{s}=4.17$~GeV in the CLEO-c data~\cite{eFormFactor}, corresponding to a statistical significance of $3.6\sigma$.

In this Letter, we report the electromagnetic form factors of neutral kaons measured with 26.5~${\rm fb^{-1}}$ of data taken at center-of-mass (c.m.) energies ($\sqrt{s}$) from 3.51 GeV to 4.95~GeV (i.e. $12~{\rm GeV}^2<Q^2<25~{\rm GeV}^2$)~\cite{lumin,lumin2,lumin3,lumin4,lumin5}. We also report the first observation of the charmless decay $\pspp\to K_S^0K_L^0$ and discuss its impact on our understanding of charmonium decay dynamics. 
The data was recorded with the BESIII detector, which is described in detail in Ref.~\cite{BESIIIdetector}. 
Simulated data samples produced with a {\sc
geant4}-based~\cite{geant4} Monte Carlo (MC) package, which
includes the geometric description of the BESIII detector and the
detector response, are used to determine detection efficiencies
and to estimate backgrounds. The simulation models the beam
energy spread and initial state radiation (ISR) in the $e^+e^-$
annihilations with the generator {\sc
kkmc}~\cite{kkmc,kkmc2}.
Signal Monte Carlo samples of \eetoKsKL~are generated at each energy point with the Vector-Scalar-Scalar ({\sc vss}) model in which a vector particle decays into two scalars in {\sc evtgen}~\cite{evtgen,evtgen2}, and the subsequent decay \Kstopipi~is generated uniformly in phase space. 
To estimate the possible background, two kinds of exclusive MC samples are generated, which are \ISRpsip\ with \psiptoKsKL, and \eetoKstarK\ with $K^{\ast0}(892)\to \pi^0K_{S/L}^0$, \piotogg\ and $\bar{K}^0\to K_{L/S}^0$.
The inclusive MC samples generated at $\sqrt{s}=3.774$~GeV are used to study potential background. 
These samples include the production of $D\bar{D}$ pairs, the non-$D\bar{D}$ decays of the \psipp, the ISR production of the $J/\psi$ and \psip\ states, and the continuum processes incorporated in {\sc kkmc}. 

%%%%%%%%%%%%%%%%%%%%%%%%%%%%%%%%%%%%%%%%%%%%%%%%%%%%%%%%%%%%%
%\section{Event Selection}

Signal events contain a single \Ks\ reconstructed with \Kstopipi. 
%The \KL\ is not reconstructed and inferred from the momentum of \Ks\ with momentum conservation. 
The \KL\ is identified from the momentum difference between the \Ks\ and $e^+e^-$, assuming momentum conservation.
We retain signal events which have two charged tracks 
and either have no neutral clusters, in the 
case that the \KL\ passes through the detector without interaction,
or has special neutral clusters found in the expected \KL\ direction, in the case that the \KL\ interacts with the detector material. 
Neutral clusters reconstructed in the detector but not in the expected \KL\ direction are used to suppress background. 

For each good charged track detected in the multilayer drift chamber (MDC), the polar angle ($\theta$) is required to be within a range of $|\rm{cos\theta}|<0.93$, where $\theta$ is defined with respect to the $z$-axis, which is the symmetric axis of the MDC. 
%The distance of closest approach to the interaction point (IP) must be less than 10~cm along the $z$-axis, and less than 1~cm in the transverse plane. 
Exactly two good charged tracks with zero net charge are required. 
%The momentum of each track must be less than 1.7~GeV/$c$ and 
The deposited energy of each track in the electromagnetic calorimeter (EMC) is required to be less than 1.2~GeV to suppress Bhabha events. 
Particle identification is applied where the specific ionization energy loss ${\rm d}E/{\rm d}x$ measured by the MDC and the flight time measured by the time-of-flight (TOF) detector form likelihoods ${\cal L}(h)~(h=e,\pi)$ for each particle hypothesis. Both tracks are required to be identified as pions with ${\cal L}(\pi)>0.001$ and ${\cal L}(\pi)>{\cal L}(e)$.

If two charged tracks fulfill these criteria they are assigned as a \pipi~pair. They are constrained to originate from a secondary vertex of a \Ks~decay. The decay length of the \Ks~candidates starting from the interaction point (IP) is required to be greater than 2~cm and greater than twice the vertex resolution to suppress $\gamma_{\rm ISR}\mu\mu$ and $\gamma_{\rm ISR}\rho^0 (\rho^0\rightarrow\pi^+\pi^-)$. 
The $\chi^2$ of the secondary vertex fit is required to be less than 15, which is optimized using the figure-of-merit FOM=$\frac{S}{\sqrt{S+B}}$. Here, $S$ is the normalized number of events from the signal MC sample and $B$ is the normalized number of background events estimated from the inclusive MC sample, which only contains the background without $K^0_S$.
%normalized to the integrated luminosity of the data sample. 
The invariant mass of the \Ks~candidate is required to be within $0.478~{\rm GeV}/c^2 < m_{\pi^+\pi^-} < 0.518~{\rm GeV}/c^2$. To further improve the resolution, a one-constraint (1C) kinematic fit is performed on the \Ks\ mass, and  $\chi_{1C}^2<12$ is further required.
%The $\chi^2$ of the 1C fit is required to be less than 12, also from an optimization of FOM.
%An optimized $\chi^2$ value of the 1C fit of 12 was determined using the aforementioned FOM %method.

Neutral clusters are identified using showers in the EMC.  The deposited energy of each shower must be more than 25~MeV in the barrel region ($|\cos \theta|< 0.80$) and more than 50~MeV in the end cap region ($0.86 <|\cos \theta|< 0.92$). To suppress showers that originate from charged tracks, the angle subtended by the EMC shower and the position of the closest charged track at the EMC must be greater than 20 degrees as measured from the IP. To suppress electronic noise and showers unrelated to the event, the difference between the EMC time and the event start time is required to be within [0, 700]~ns.

Pure $e^+e^-\to K_S^0K_L^0$ ($K_S^0 \to \pi^+\pi^-$) events do not contain photons. However, \KL\ mesons~may interact with the detector material and produce neutral clusters in the EMC. 
In this case, we examine $K_L^0$ candidates within a cone with an opening angle of 20 degrees relative to the opposite direction of the \Ks~in the $e^+e^-$ c.m. frame. 
We then search for at least one neutral cluster satisfying a second moment~\cite{SM}
%divided by the shower energy deposition
$\frac{\Sigma_{i}(E_{i}r_{i}^2)}{\Sigma_{i}(E_{i})}$ greater than 20 $\rm cm^2$, where $E_{i}$ is the deposited energy in the $i$th crystal and $r_{i}$ refers to the radial distance of it from the cluster center. This requirement is useful to suppress $\gamma_{\rm ISR}\mu\mu$ background since the neutral clusters produced by \KL\ are wider than those produced by photons.

To suppress EMC noise and background events from $e^+e^-\to K^{\ast0}(892)\bar{K}^{0}+c.c.$, where $K^{\ast0}(892)\bar{K}^{0}\to \pi^{0}K_{S}^{0}K_{L}^{0}$, we require the total energy of the neutral clusters outside the cone to be less than 0.2~GeV. The invariant mass of every combination of two neutral clusters is calculated to search for $\pi^0$ candidates. Events where the invariant mass of any combination satisfies $M_{\gamma\gamma}\in [0.123,0.144]~{\rm GeV}/c^2$ are rejected.

%\section{SIGNAL EXTRACTION}

\begin{figure}[!htbp]
  \centering
  % Requires \usepackage{graphicx}
  \includegraphics[width=0.7\linewidth]{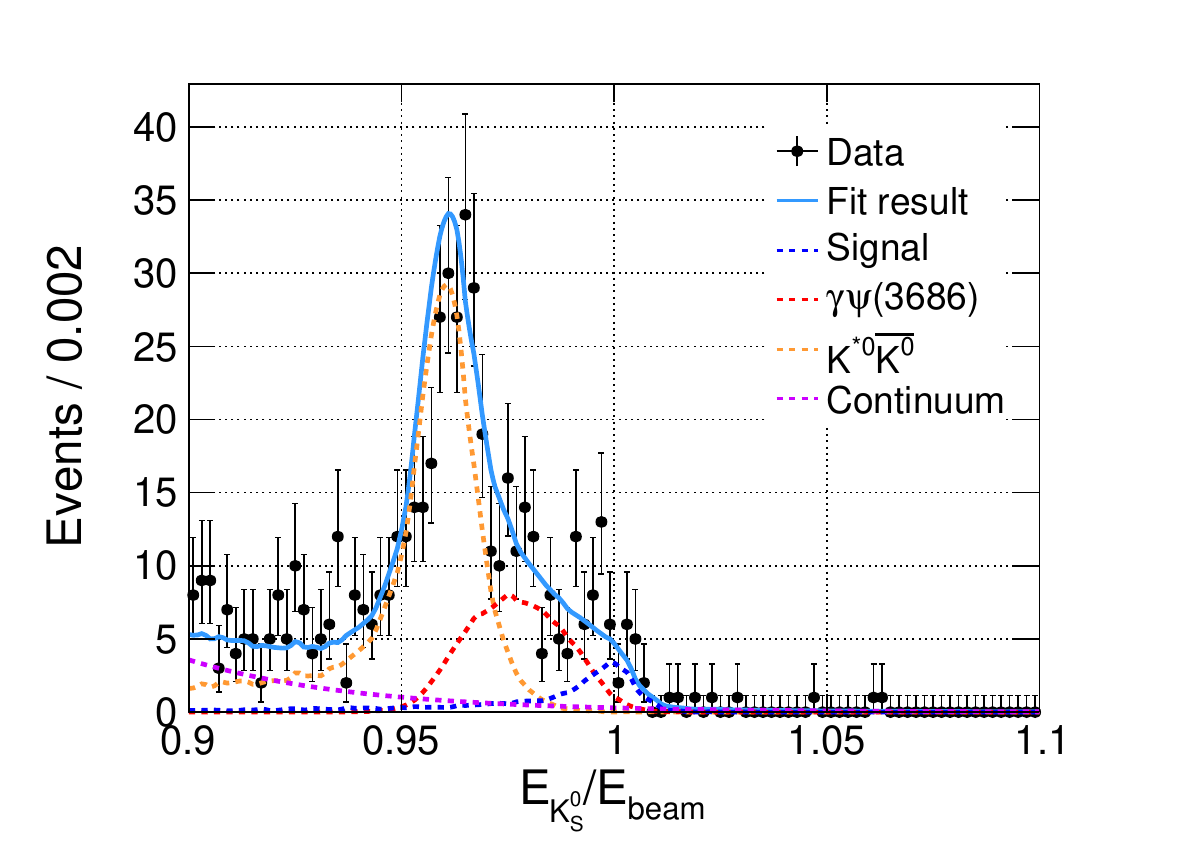}\\
  \includegraphics[width=0.7\linewidth]{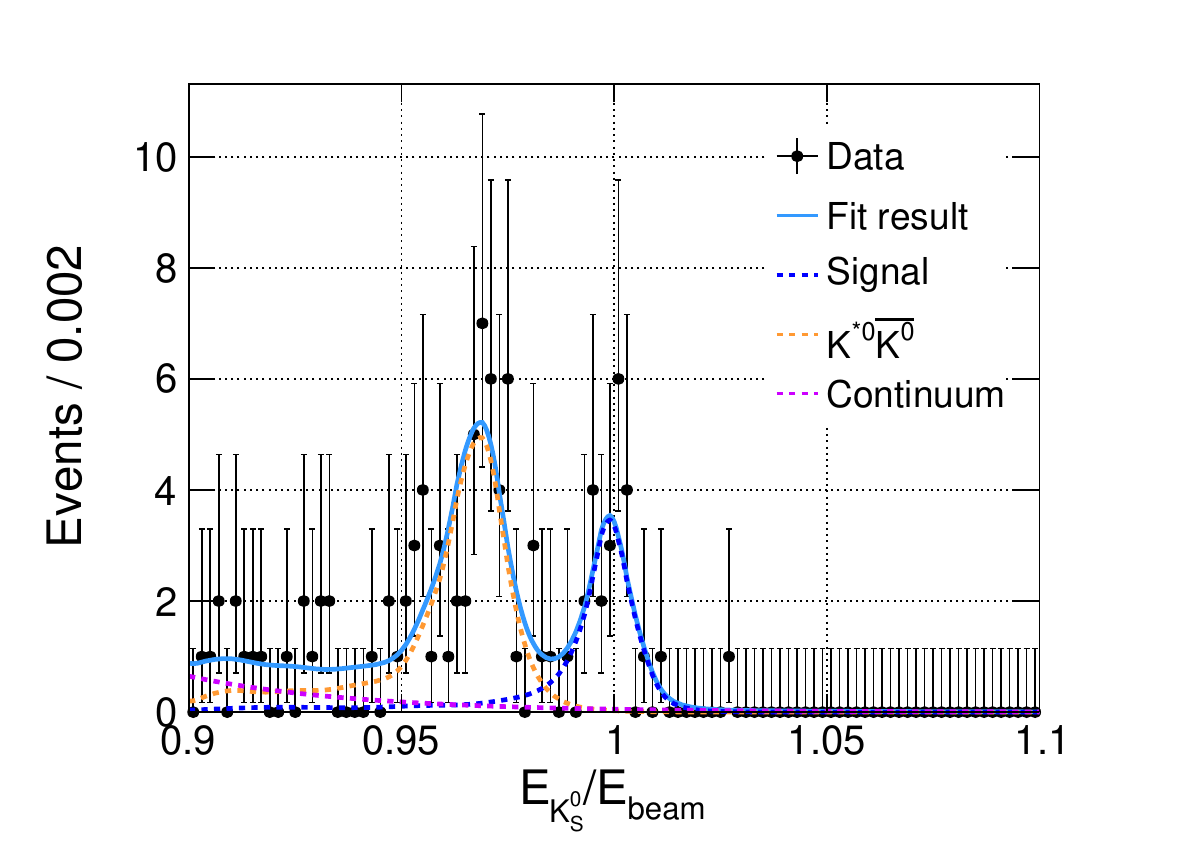}
\caption{Distributions of $X\equiv E_{K_S^0}/E_{\rm beam}$ at $\sqrt{s}=3.774$~GeV (top) and $4.226$~GeV (bottom), where the corresponding $\chi^2/n.d.f$ of the fits are 1.06 and 0.50, respectively. Dots with error bars are data. The blue solid lines are the total fit results, the blue dashed lines are the signal component, the red dashed lines represent the \ISRpsip\ component, the orange dashed lines represent the \eetoKstarK\ component, and the violet dashed lines represent exponential functions describing the remaining background. }
%{\color{red}{Each MC simulated shape is convolved with a Gaussian function with a fixed mean %and width, which are from the fitting to the $X$ distribution from $\psi^\prime\to %K_S^0K_L^0$.}}
%Events with $X>1.02$ come from misidentification.}
  \label{pic:eks}
\end{figure}

Figure~\ref{pic:eks} shows the $X\equiv E_{K_S^0}/E_{\rm beam}$ distributions at $\sqrt{s}=3.774$~GeV and $4.226$~GeV, where $E_{K_S^0}$ is the energy of the \Ks\ in the $\EE$ c.m. frame, and $E_{\rm beam}=\sqrt{s}/2$ is the beam energy. Clear \eetoKsKL\ signals are observed at $X=1$. The signal region defined as $X\in [0.98,1.02]$ is used for the optimization of the selection criteria. Similar distributions are observed at the other c.m. energies.

An unbinned maximum likelihood fit is performed using the $X$ distributions to extract the number of signal events. The definition of the likelihood can be found in the Supplemental Material~\cite{sp}. Three simulated shapes derived from MC samples are used to describe the dominant components of the data: the signal component, the \ISRpsip\ background, and the $e^+e^-\to K^{\ast0}(892)\bar{K}^{0}+c.c.$ background. 
Each is convolved with a Gaussian function to account for the  discrepancies in $X$ ($\Delta X$) and resolution ($\Delta \sigma$) between data and MC simulation. 
The $\Delta X=(3.8\pm 0.8)\times 10^{-4}$ and $\Delta \sigma=(7.5\pm 3.2)\times 10^{-4}$ are measured with the process \psiptoKsKL, which increases the goodness-of-fit of data.
%, which are $(3.8\pm 0.8)\times 10^{-4}$ and $(7.5\pm 3.2)\times 10^{-4}$, respectively. 
%The central values determined by the fit are used to fix the parameters for the three Gaussian %functions.
%In the fit, the same $\Delta X$ and $\Delta \sigma$ are fixed at the central values for all three Gaussian functions. 
The remaining background is described by an exponential function. 
The expected number of background events from the processes \ISRpsip\ and \eetoKstarK\ in the $\sqrt{s}=3.774$~GeV data sample are calculated to be $147\pm 15$ and $390\pm 22$ using the corresponding integrated luminosity and cross sections \cite{KstarKCS}. They are fixed at their central values in the fit at $\sqrt{s}=3.774$ GeV while floating at other energy points, the uncertainties are considered as one source of systematic uncertainty.
The simulated shape of \ISRpsip~is only used at 3.710 GeV and 3.774 GeV, its inclusion has negligible effects on the signal yield at other energy points.
The parameters of the exponential function are determined from the fit at energy points with high luminosity. For energy points with low luminosity, the parameters extracted at the nearest high luminosity point are used. 
%{\rr \st{The fit results at $\sqrt{s}=3.773$ and $4.230$~GeV are shown in Fig. {\ref{pic:eks}}, similar fits are obtained at other c.m. energies.}}

\begin{figure*}[!htbp]
  \centering
  \subfigure[]{
  \label{Fig.sub.1}
  \includegraphics[width=0.32\linewidth]{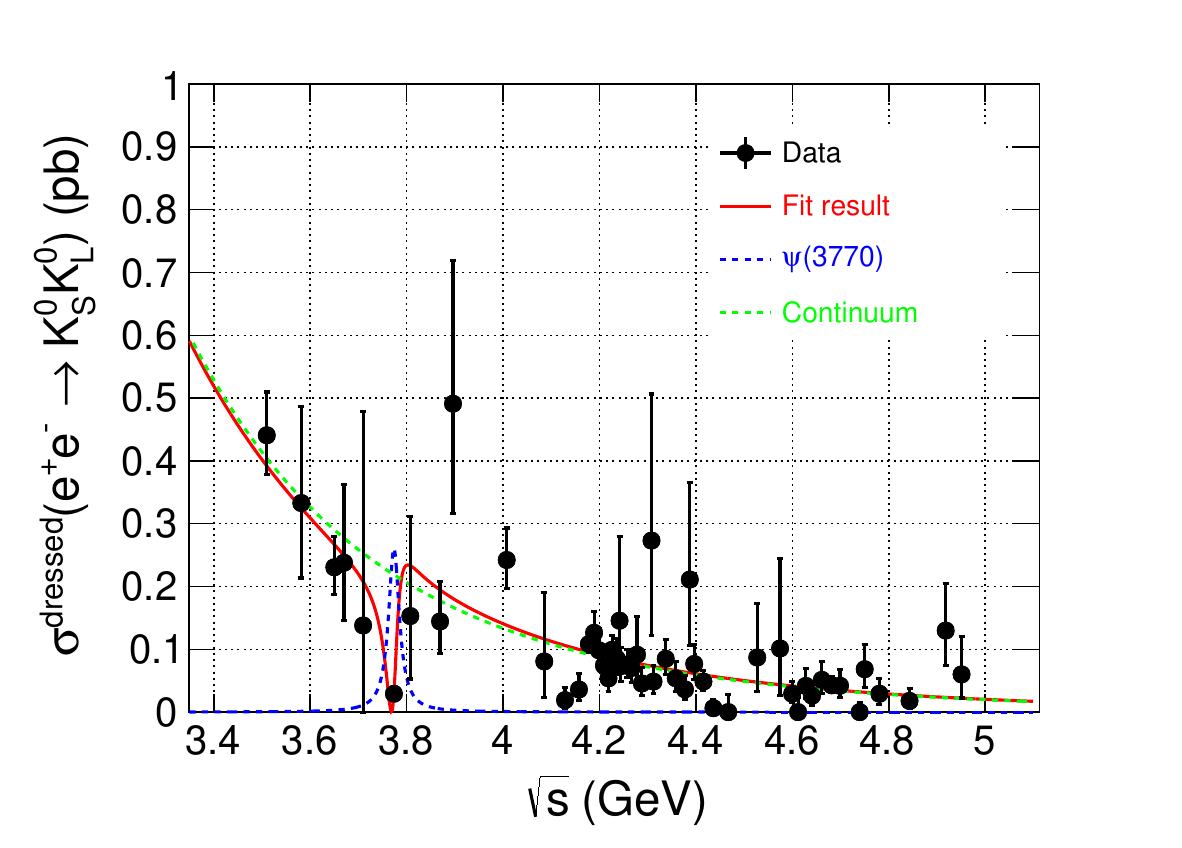}}
  \subfigure[]{
  \label{Fig.sub.2}
  \includegraphics[width=0.32\linewidth]{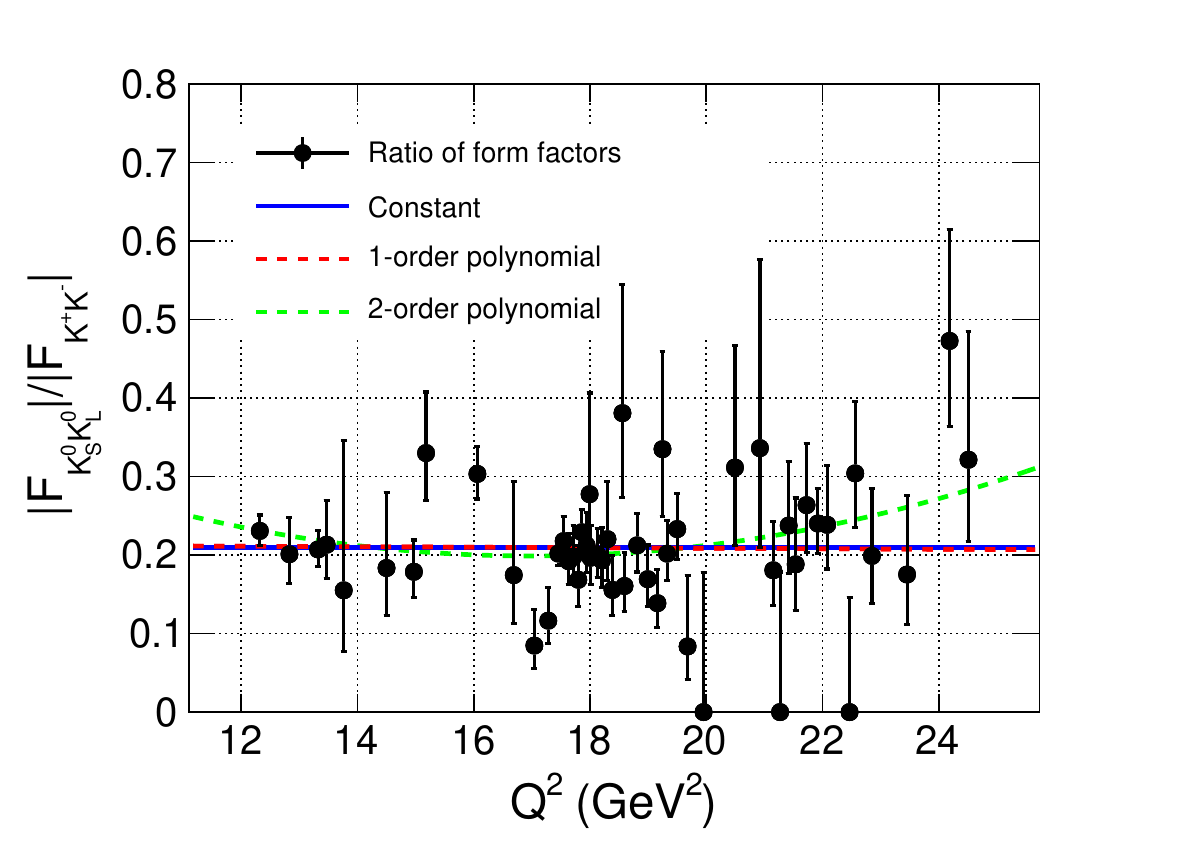}}
  \subfigure[]{
  \label{Fig.sub.3}
  \includegraphics[width=0.32\linewidth]{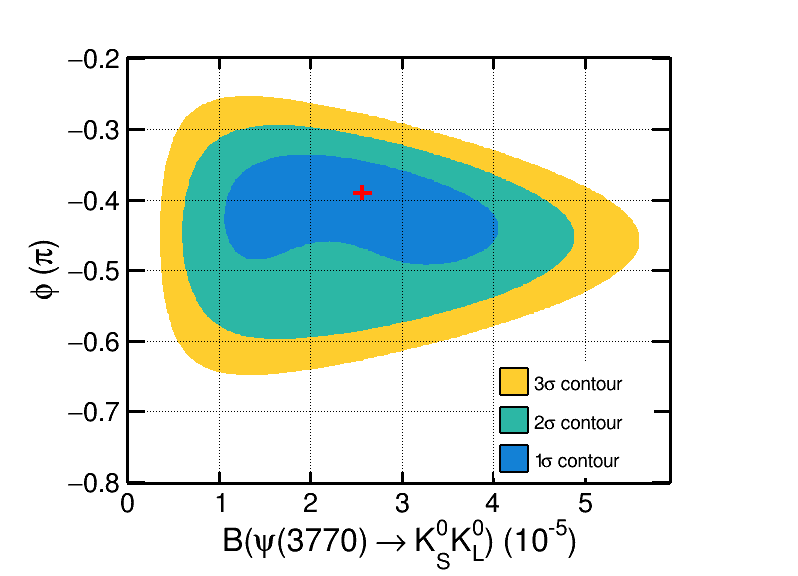}}\\
\caption{ (a) Dressed cross sections of \eetoKsKL\ and a fit with the coherent sum of a continuum and a $\pspp$ resonance amplitude, using the local minimum in (c) where $\BR=2.6\times10^{-5}$ and $\phi=-0.39\pi$. Dots with error bars are data. Red solid, green dashed, and blue dashed lines are the fit results, the continuum production, and the $\pspp$ production, respectively.
(b) The ratio of neutral-to-charged kaon form factors. The blue solid, red dashed, and green dashed lines result from fits with different order polynomials, the corresponding $\chi^2/{\rm n.d.f}$ are 1.26, 1.28, 1.19 respectively.
(c) The likelihood contours in the $\BR(\pspp\to K_S^0K_L^0)$ and the relative phase $\phi$ plane. The filled areas are up to $3\sigma$ likelihood contours. The red cross shows the local minimum.}
\label{pic:csFit}
\end{figure*}

The dressed cross section is determined as 
\begin{equation}\label{eq:cs_dressed}
 \sigma^{\rm dressed} = \frac{N^{\rm obs}}
      {\varepsilon\cdot {\cal L} \cdot (1+\delta) \cdot \BR(K_S^0 \to \pi^+\pi^-)},
\end{equation}
where $N^{\rm obs}$ is the number of signal events from the fit, ${\cal L}$ is the integrated luminosity, and $\BR(K_S^0 \to \pi^+\pi^-)=(69.20\pm0.05)\%$~\cite{PDG} is the branching fraction of $K_S^0\to \pi^+\pi^-$. The efficiency $\varepsilon$ and the ISR correction factor $(1+\delta)$~\cite{ISR} are obtained iteratively following the procedure used in Ref.~\cite{ISR_iteration}. Figure~\ref{pic:csFit}(a) shows the resulting cross sections. We find that the cross sections are at a sub picobarn level and decrease gradually with increasing c.m. energy. The cross section at $\sqrt{s}=3.774$~GeV, however, is significantly lower than the expected trend by more than 5$\sigma$, suggesting interference between the \eetoKsKL\ and \psipptoKsKL\ amplitudes. 

We use the cross sections (excluding $\sqrt{s}=3.774$~GeV data to avoid the \psipp\ resonance contribution) to calculate the electromagnetic form factors of the neutral kaon via
\begin{equation}
 |F_{K_S^0K_L^0}(s)|^2=\frac{\sigma^{\rm dressed}\cdot |1-\Pi|^2\cdot 3s}{\pi \alpha^2\beta^3}, \label{eq:ff_fun}
\end{equation}

where $\alpha$ is the fine-structure constant, $\beta=\sqrt{1-4m_{K^0}^2/s}$ is the velocity of the \Ks\ in the $\EE$ c.m. frame, and $\frac{1}{|1-\Pi|^2}$ is the vacuum polarization factor \cite{VP}. Figure \ref{pic:csFit}(b) shows the ratio of the neutral and the charged kaon form factors, where the data on charged kaons comes from Ref. \cite{ff_K+K-}. 
We fit the ratios with polynomials of different orders and find the shape
is close to constant at $|F_{K_S^0K_L^0}| / |F_{K^+K^-}| = 0.21\pm 0.01$. The coefficients of higher-order polynomial terms are not significant ($<2\sigma$) under hypothesis testing. %although the second-order polynomial fit result seems to increase at $Q^2>20~\rm{GeV^2}$. 
The electromagnetic form factors of the neutral kaon and the ratio of neutral-to-charged kaon form factors at each c.m. energy are listed in the Supplemental Material~\cite{sp}.

The contribution of the \psipptoKsKL\ amplitude is determined using a maximum likelihood fit of the dressed cross section, which we describe as a coherent sum of a Breit-Wigner function for the \psipp\ amplitude and a power function for the continuum amplitude:
\begin{equation}\label{eq:cs_dressed_fit}
  \sigma^{\rm dressed}=\left |BW\cdot e^{i\phi} + 
        \frac{a}{(\sqrt{s})^n}\cdot\sqrt{\Phi(\sqrt{s})}\right |^2,
\end{equation}
where $BW=\frac{\sqrt{12\pi\Gamma_{ee}\Gamma\BR}}{s-M^2+iM\Gamma}\sqrt{\frac{\Phi(\sqrt{s})}{\Phi(M)}}$ is the \psipp\ amplitude with $M$, $\Gamma$, $\Gamma_{ee}$, and $\BR$ being the mass, width, electronic width, and the branching fraction of \psipptoKsKL, respectively; $\Phi(\sqrt{s}) = \frac{q^{3}}{s}$ is a $\mathcal{P}$-wave phase space factor, in which $q$ is the $K^{0}_{S/L}$ momentum in the $e^{+}e^{-}$ c.m. frame;
$\phi$ is the relative phase between the continuum and \psipp\ amplitudes; 
The $a$ and $n$ are free parameters.
% as are $\BR(\psi^{\prime\prime} \to K_S^0K_L^0)$ and $\phi$.

Without using the data at $\sqrt{s}=3.774$~GeV, we fit the cross sections with the pure continuum amplitude and determine $a=(0.016\pm 0.007)~{\rm GeV^{n-0.5}pb^{0.5}}$ and $n=4.60\pm 0.31$. By including the data at $\sqrt{s}=3.774$~GeV and the $\pspp$ resonance amplitude, with the mass, width, and electronic width of the $\pspp$ fixed at their world average values~\cite{PDG}, we perform a likelihood scan in the $\BR$ versus $\phi$ plane.
%with $a$ and $n$ fixed to their central values obtained from the previous fit.
In this scan, the $a$ and $n$ are allowed to float until they reach the local minimum within their parameter space.
The results are shown in Fig.~\ref{pic:csFit}(c). 
Clear region of local minima was observed through the scan,
the $\BR$ and $\phi$ are determined to be $(2.63_{-1.59}^{+1.40})\times 10^{-5}$ and $(-0.39_{-0.10}^{+0.05})\pi$ within $1\sigma$ likelihood contour.
%Considering that the inclusive cross section of the continuum is expected to be much larger than that of the \psipp~\cite{non-physics}, we take those results corresponding to $\BR\sim 2.2\times 10^{-5}$ and larger as non-physical solutions since they indicate a resonance cross section that exceeds that of the continuum process.
%Two local minima were observed through the scan, one corresponding to $\BR\sim 1\times 10^{-5}$ and the other to $\BR\sim 4\times 10^{-5}$. The latter corresponds to a resonance cross section about twice as large as that of the continuum process so we take it as a non-physical solution since the inclusive cross section of the continuum is expected to be much larger than that of the \psipp~\cite{non-physics}. 
%The former suggests a \psipptoKsKL\ branching fraction of $10^{-5}$ or smaller.  
%The local minimum suggests
%a \psipptoKsKL\ branching fraction of 
%$\BR=1.0\times10^{-5}$, the corresponding fit result is shown in 
Figure~\ref{pic:csFit}(a) shows the fit result corresponding to the local minimum.
The significance of the $\pspp$ resonance contribution is determined to be $10\sigma$ in comparison to an alternative fit without including the resonance. This indicates that the charmless decay \psipptoKsKL\ is observed for the first time. 

%The Supplemental Material~\cite{sp1} lists various parameters used in Eq.~(\ref{eq:cs_dressed}) and the Born cross section using solutions shown in Fig.~\ref{pic:csFit}(a).
%Notice that there's no multiple-solution in this analysis following the method introduced in Ref.~\cite{multi-solution}, solution space in Fig. \ref{pic:csFit}(b) are due to limited statistics and insufficient data points near \psipp~peak.

%%%%%%%%%%%%%%%%%%%%%%%%%%%%%%%%%%%%%%%%%%%%%%%%%%%%%%%%%%%%%
%\section{SYSTEMATIC UNCERTAINTY}\label{sec:sys}

The systematic uncertainties of the cross section measurement are listed in Table~\ref{tab:sys}, where the total systematic uncertainty is the square root of the quadratic sum of all sources, assuming they are independent. The uncertainty of the integrated luminosity is  1.0\%~\cite{lumin,lumin2,lumin3,lumin4,lumin5}. The difference in the tracking efficiency between data and MC simulation is 1.0\% per track~\cite{tracking}. The uncertainties of the \Ks\ reconstruction, the \KL\ requirements, and the $\pi^0$ rejection are estimated with control samples and the efficiency differences between data and MC simulation are measured following the method in Ref.~\cite{maoqiang}. 
We use $\pspp\to K^{\ast0}(892)\bar{K}^0$, with $K^{\ast0}(892)\to K^{\pm}\pi^{\mp}$ and $\bar{K}^0\to K_S^0$ to determine an uncertainty of 1.3\% for the \Ks\ reconstruction after multiplying by a correction factor of 1.018 to the MC efficiency. Using the control sample \psiptoKsKL, we find that the \KL\ acceptance rate in MC simulation is $(5.0\pm 0.5)\%$ higher compared to data, therefore we correct the MC efficiency and quote an uncertainty of 0.5$\%$. Similarly for the $\pi^0$ rejection, the rate in MC simulation is $(3.0\pm 0.4)\%$ lower compared to data, therefore we correct the MC efficiency and quote an uncertainty of 0.4\%. 
%The uncertainty due to kinematic fit is estimated by removing the helix parameter corrections, the determined value is 0.7\% according to the difference in detection efficiency.

To study the uncertainties resulting from our choice of signal and background shapes when extracting the number of signal events, we select five data samples with large statistics to avoid statistical fluctuations, namely at c.m. energies of $\sqrt{s}=3.510$ GeV, 3.650 GeV, 3.774 GeV, 4.178 GeV, and 4.226~GeV.
The uncertainty resulting from the fixed parameters of the Gaussian function is estimated by varying the parameter values within $1\sigma$. The difference in the signal yields is negligible ($<0.5\%$) except for 3.774 GeV, which are 1.2\% for $\Delta X$ and 2.7\% for $\Delta \sigma$.
The numbers of \ISRpsip\ and \eetoKstarK\ events are fixed in the fit at $\sqrt{s}=3.774$ GeV, and differences are estimated by varying them by $1\sigma$ around their central values. The resulting uncertainty is 4.6\% for \ISRpsip\ and 0.3\% for \eetoKstarK, respectively. 
We perform alternative fits by replacing the exponential function with the background shape extracted from the inclusive MC sample. The relative differences in the signal yields are 2.6\%, 2.8\%, 9.4\%, 1.9\%, and 1.8\% for the
%{\rr \st{$\sqrt{s}=3.510$, 3.650, 3.773, 4.180, and 4.230~GeV}}
selected data samples. 
Therefore, after combining the uncertainties at 3.774 GeV, a luminosity-weighted uncertainty of 4.7\% is taken as the systematic uncertainty caused by the line shapes used in the fits for all energy points.
The uncertainty from the fitting range is examined by the ``Barlow test''~\cite{BarlowTest}. We choose 20 different fixed-length fitting ranges with a step size of 1~MeV to compare the deviation between different measurements. We find these deviations are due to statistical fluctuations and are thereby ignored in the systematic uncertainty.
The uncertainty from the branching fraction of $K^{0}_{S}\to \pi^+\pi^-$ is less than 0.1\%~according to the Particle Data Group, which is ignored.

\begin{table}[!htbp]
\caption{Relative systematic uncertainties of the cross section measurements.}\label{tab:sys}
\centering
\begin{tabular}{cc}
\hline\hline
Source &Systematic uncertainty (\%)\\
\hline
Luminosity &1.0\\
Tracking &2.0\\
\Ks~reconstruction &1.3\\
%Kinematic fit &0.7\\
\KL~requirements &0.5\\
$\pi^0$ rejection &0.4\\
Fitting line shape &4.7\\
\hline
Total &5.4\\
\hline\hline
\end{tabular}
\end{table}

\comment{
\begin{table}[!htbp]
\caption{List of several selection criteria ($v$), 
the correction factor with its uncertainty $f^{v}\pm\sigma_{f^{v}}$, 
the ultimate systematic uncertainty $\sigma_{f}$ of $v$, which is obtained from $\sigma_{f^{v}}$ or $|\Delta f^{v}|+|\sigma_{f^{v}}|$ according to the estimation method described in the Sec.~\ref{sec:sys},
and then the ultimate correction factor of MC efficiency $f^{v}_{\rm cor}$.}\label{tab:sys_estimation}
\begin{tabular}{cccccc}
    \hline\hline
    Selection criteria ($v$)  &  &$f^{v}\pm\sigma_{f^{v}}$   &$\sigma_{f}$  &$f^{v}_{\rm cor}$\\
    \hline
    \Ks~reconstruction  &   &$0.963\pm 0.010$   &0.01   &0.963  \\
    \KL~restriction     &   &$0.996\pm 0.006$   &0.01   &-      \\
    veto $\pi^0$      &   &$1.004\pm 0.003$   &0.003  &1.004  \\
    \hline\hline
\end{tabular}
\end{table}
}

%%%%%%%%%%%%%%%%%%%%%%%%%%%%%%%%%%%%%%%%%%%%%%%%%%%%%%%%%%%%%
%\section{SUMMARY and discussions}

In summary, we measure the electromagnetic form factors of the neutral kaons at large momentum transfers from 12~GeV$^2$ to 25~GeV$^2$ for the first time, which are consistent with previous measurement at 17.4~GeV$^2$~\cite{eFormFactor}.
The constant ratio ($0.21\pm 0.01$) of neutral-to-charged kaon form factors we obtained indicates a small but significant effect of flavor-SU(3) breaking on the kaon wave function, and consequently excludes the possibility that flavor-SU(3) breaking is the primary reason for the large observed deviation between the pQCD prediction and the experimental result.
The observed constant ratio of the kaon form factors is also in disagreement with the predicted trend using a single bound-state interaction kernel~\cite{kaon_ff}, providing more information for the investigation of the internal structure of neutral kaons.
 
We observe a significant signal of \psipptoKsKL\ for the first time. This is the first discovery of the charmless decay of the \psipp\ with a statistical significance exceeding $5\sigma$. 
Within the $1\sigma$ contour of the likelihood value, the branching fraction of $\pspp\to K_S^0K_L^0$ is 
determined to be $\BR=(2.63_{-1.59}^{+1.40})\times 10^{-5}$, and the relative phase between the continuum amplitude and the $\pspp$ decay amplitude is $\phi=(-0.39_{-0.10}^{+0.05})\pi$.
%restricted in a range of $[0.37,5.62]\times 10^{-5}$ and the relative phase between the continuum amplitude and the $\pspp$ decay amplitude $\phi\in[-0.64\pi,-0.26\pi]$. 
The uncertainties are mainly due to insufficient data points near the \psipp\ peak, thus a finer scan around the \psipp~will help to reveal the nature of the \psipp. The branching fraction is in good agreement with the prediction~\cite{SDYuan} of the $\mathcal{S}$- and $\mathcal{D}$-wave charmonia mixing model developed to interpret the ``$\rho\pi$ puzzle'' between \Jpsi\ and \psip\ decays~\cite{SDmixing}. 
Assuming a negligible contribution from the electromagnetic \psipp\ amplitude, a phase $\phi$, around $-\frac{\pi}{2}$, supports the proposition~\cite{Wang:2006zzg} that the relative phase between strong and electromagnetic charmonium decay amplitudes is universally $-\frac{\pi}{2}$ and agrees with experimental measurements in various final states~\cite{Zhu:2015bha}.

%%%%%%%%%%%%%%%%%%%%%%%%%%%%%%%%%%%%%%%%%%%%%%%%%%%%%%%%%%%%%
%\section{Acknowledgement}

%% Saved at => 2023-07-20
%\textbf{Acknowledgement}

The BESIII Collaboration thanks the staff of BEPCII and the IHEP computing center for their strong support. This work is supported in part by National Key R\&D Program of China under Contracts Nos. 2020YFA0406300, 2020YFA0406400; National Natural Science Foundation of China (NSFC) under Contracts Nos. 11635010, 11735014, 11835012, 11935015, 11935016, 11935018, 11961141012, 12025502, 12035009, 12035013, 12061131003, 12192260, 12192261, 12192262, 12192263, 12192264, 12192265, 12221005, 12225509, 12235017; the Chinese Academy of Sciences (CAS) Large-Scale Scientific Facility Program; the CAS Center for Excellence in Particle Physics (CCEPP); Joint Large-Scale Scientific Facility Funds of the NSFC and CAS under Contract No. U1832207; CAS Key Research Program of Frontier Sciences under Contracts Nos. QYZDJ-SSW-SLH003, QYZDJ-SSW-SLH040; 100 Talents Program of CAS; The Institute of Nuclear and Particle Physics (INPAC) and Shanghai Key Laboratory for Particle Physics and Cosmology; European Union's Horizon 2020 research and innovation programme under Marie Sklodowska-Curie grant agreement under Contract No. 894790; German Research Foundation DFG under Contracts Nos. 455635585, Collaborative Research Center CRC 1044, FOR5327, GRK 2149; Istituto Nazionale di Fisica Nucleare, Italy; Ministry of Development of Turkey under Contract No. DPT2006K-120470; National Research Foundation of Korea under Contract No. NRF-2022R1A2C1092335; National Science and Technology fund of Mongolia; National Science Research and Innovation Fund (NSRF) via the Program Management Unit for Human Resources \& Institutional Development, Research and Innovation of Thailand under Contract No. B16F640076; Polish National Science Centre under Contract No. 2019/35/O/ST2/02907; The Swedish Research Council; U. S. Department of Energy under Contract No. DE-FG02-05ER41374.

%\textbf{Other Fund Information}

%To be inserted with an additional sentence into papers that are relevant to the topic of special funding for specific topics. Authors can suggest which to Li Weiguo and/or the physics coordinator.
%        Example added sentence: This paper is also supported by the NSFC under Contract Nos. 10805053, 10979059, ....National Natural Science Foundation of China (NSFC), 10805053, PWANational Natural Science Foundation of China (NSFC), 10979059, Lund弦碎裂强子化模型及其通用强子产生器研究National Natural Science Foundation of China (NSFC), 10775075, National Natural Science Foundation of China (NSFC), 10979012, baryonsNational Natural Science Foundation of China (NSFC), 10979038, charmoniumNational Natural Science Foundation of China (NSFC), 10905034, psi(2S)->B BbarNational Natural Science Foundation of China (NSFC), 10975093, D 介子National Natural Science Foundation of China (NSFC), 10979033, psi(2S)->VPNational Natural Science Foundation of China (NSFC), 10979058, hcNational Natural Science Foundation of China (NSFC), 10975143, charmonium rare decays
%% ends here %%

%%%%%%%%%%%%%%%%%%%%%%%%%%%%%%%%%%%%%%%%%%%%%%%%%%%%%%%%%%%%%

\nocite{*}
%\bibliography{draft}

%\clearpage
%\begin{widetext}
%    \include{supplement_material}
%\end{widetext}

\end{document}

\title{
%\boldmath Observation of significant flavor-SU(3) breaking on the kaon wave function at $12~{\rm GeV}^2<Q^2<25~{\rm GeV}^2$ and discovery of the charmless decay $\psi(3770)\to K_S^0K_L^0$\\ 
Supplemental Material
}
%\date{\today}
%
\maketitle %%???

%\newpage
\section{Summary of various variables used to obtain cross sections}

\begin{table*}[!htbp]
\caption{List of c.m. energies ($\sqrt{s}$), integrated luminosities (${\cal L}_{int}$), signal yields ($N^{obs}$), event selection efficiencies ($\varepsilon$) and ISR correction factors ($(1+\delta)$) obtained from an iterative package \cite{ISR_iteration}, vacuum polarization factors ($\frac{1}{|1-\Pi|^{2}}$), dressed cross section ($\sigma^{\rm dressed}$), Born cross section ($\sigma^{\rm Born}$) and significance ($\cal S$) of signal yield. 
The integrated luminosities with uncertainty are obtained from offline measurements, while those without uncertainty are estimated online.
The first uncertainty of cross sections is statistical and the second is systematic.}
%shown in Fig. \ref{pic:csFit}. Both of the $\varepsilon$ and $(1+\delta)$ are extract using the line-shape shown in Figure \ref{pic:csFit}. The $\varepsilon$ are already corrected as discussed in Section \ref{sec:sys}.}\label{tab:cs}. Only statistical uncertainties are shown here.
\centering
\begin{tabular}{ccccccccc}
\hline\hline
$\sqrt{s}$ (GeV)     &${\cal L}_{int}$ ($\rm pb^{-1}$)    &$N^{obs}$   &$\varepsilon$   &$(1+\delta)$   &$\frac{1}{|1-\Pi|^{2}}$   &\makebox[0.17 \textwidth][c]{$\sigma^{\rm dressed}$ (pb)} &\makebox[0.17\textwidth][c]{$\sigma^{\rm Born}$ (pb)} &$\cal S$ \\
\hline
3.510  &$459  $  &$48.9_{-7.0}^{+7.7}$  &$0.402$  &$0.871$  &$1.045$  &$0.441_{-0.063}^{+0.069}\pm0.024$  &$0.422_{-0.060}^{+0.066}\pm0.023$  &$ 11.3\sigma$  \\  
3.582  &$85.7 $  &$7.2_{-2.6}^{+3.3}$  &$0.401$  &$0.915$  &$1.039$  &$0.333_{-0.119}^{+0.153}\pm0.018$  &$0.321_{-0.115}^{+0.148}\pm0.017$  &$ 4.9\sigma$  \\  
3.650  &$454  $  &$27.6_{-5.2}^{+5.9}$  &$0.401$  &$0.951$  &$1.021$  &$0.231_{-0.043}^{+0.049}\pm0.012$  &$0.226_{-0.042}^{+0.048}\pm0.012$  &$ 8.0\sigma$  \\  
3.670  &$84.7 $  &$5.3_{-2.1}^{+2.8}$  &$0.398$  &$0.964$  &$0.994$  &$0.238_{-0.092}^{+0.124}\pm0.013$  &$0.239_{-0.093}^{+0.125}\pm0.013$  &$ 4.8\sigma$  \\  
3.710  &$70.3 $  &$2.6_{-2.6}^{+6.5}$  &$0.391$  &$1.000$  &$1.090$  &$0.138_{-0.138}^{+0.340}\pm0.007$  &$0.127_{-0.127}^{+0.312}\pm0.007$  &$ 0.5\sigma$  \\  
3.774  &$2932\pm14 $  &$47.8_{-10.7}^{+11.7}$  &$0.367$  &$2.195$  &$1.056$  &$0.029_{-0.007}^{+0.007}\pm0.002$  &$0.028_{-0.006}^{+0.007}\pm0.001$  &$ 5.6\sigma$  \\  
3.808  &$50.5\pm0.5 $  &$2.0_{-1.3}^{+2.0}$  &$0.406$  &$0.908$  &$1.056$  &$0.153_{-0.100}^{+0.159}\pm0.008$  &$0.145_{-0.095}^{+0.151}\pm0.008$  &$ 2.0\sigma$  \\  
3.869  &$219.2$  &$8.1_{-2.8}^{+3.6}$  &$0.377$  &$0.975$  &$1.051$  &$0.144_{-0.051}^{+0.064}\pm0.008$  &$0.137_{-0.048}^{+0.061}\pm0.007$  &$ 4.5\sigma$  \\  
3.896  &$52.6\pm0.5 $  &$6.9_{-2.5}^{+3.2}$  &$0.391$  &$0.991$  &$1.049$  &$0.491_{-0.175}^{+0.228}\pm0.027$  &$0.468_{-0.166}^{+0.217}\pm0.025$  &$ 5.0\sigma$  \\  
4.008  &$482\pm5  $  &$30.8_{-5.7}^{+6.5}$  &$0.367$  &$1.039$  &$1.044$  &$0.242_{-0.045}^{+0.051}\pm0.013$  &$0.232_{-0.043}^{+0.049}\pm0.013$  &$ 10.6\sigma$  \\  
4.085  &$52.6\pm0.4 $  &$1.1_{-0.8}^{+1.6}$  &$0.366$  &$1.064$  &$1.051$  &$0.081_{-0.057}^{+0.110}\pm0.004$  &$0.077_{-0.054}^{+0.105}\pm0.004$  &$ 2.2\sigma$  \\  
4.128  &$400  $  &$1.9_{-1.3}^{+2.1}$  &$0.340$  &$1.080$  &$1.052$  &$0.019_{-0.013}^{+0.021}\pm0.001$  &$0.018_{-0.012}^{+0.020}\pm0.001$  &$ 1.9\sigma$  \\  
4.157  &$400  $  &$3.7_{-1.9}^{+2.7}$  &$0.337$  &$1.089$  &$1.053$  &$0.036_{-0.018}^{+0.026}\pm0.002$  &$0.034_{-0.017}^{+0.025}\pm0.002$  &$ 3.2\sigma$  \\  
4.178  &$3189\pm32 $  &$89.2_{-10.4}^{+11.2}$  &$0.341$  &$1.096$  &$1.054$  &$0.108_{-0.013}^{+0.014}\pm0.006$  &$0.103_{-0.012}^{+0.013}\pm0.006$  &$ 13.3\sigma$  \\  
4.189  &$570\pm2  $  &$18.0_{-4.2}^{+4.9}$  &$0.328$  &$1.100$  &$1.056$  &$0.126_{-0.029}^{+0.035}\pm0.007$  &$0.120_{-0.028}^{+0.033}\pm0.006$  &$ 8.3\sigma$  \\  
4.199  &$526\pm2  $  &$12.8_{-3.7}^{+4.5}$  &$0.325$  &$1.103$  &$1.056$  &$0.098_{-0.029}^{+0.035}\pm0.005$  &$0.093_{-0.027}^{+0.033}\pm0.005$  &$ 6.0\sigma$  \\  
4.209  &$572\pm2  $  &$10.4_{-3.2}^{+4.0}$  &$0.319$  &$1.106$  &$1.057$  &$0.074_{-0.023}^{+0.029}\pm0.004$  &$0.070_{-0.022}^{+0.027}\pm0.004$  &$ 5.5\sigma$  \\  
4.219  &$569\pm2  $  &$7.4_{-2.9}^{+3.7}$  &$0.316$  &$1.109$  &$1.056$  &$0.054_{-0.021}^{+0.027}\pm0.003$  &$0.051_{-0.020}^{+0.026}\pm0.003$  &$ 3.9\sigma$  \\  
4.226  &$1092\pm7 $  &$29.3_{-5.8}^{+6.7}$  &$0.348$  &$1.112$  &$1.056$  &$0.099_{-0.020}^{+0.023}\pm0.005$  &$0.094_{-0.019}^{+0.021}\pm0.005$  &$ 8.0\sigma$  \\  
4.236  &$530\pm3  $  &$11.2_{-3.4}^{+4.2}$  &$0.322$  &$1.114$  &$1.056$  &$0.085_{-0.026}^{+0.032}\pm0.005$  &$0.081_{-0.025}^{+0.030}\pm0.004$  &$ 5.9\sigma$  \\  
4.242  &$55.6\pm0.4 $  &$2.2_{-1.3}^{+2.0}$  &$0.347$  &$1.117$  &$1.055$  &$0.146_{-0.084}^{+0.134}\pm0.008$  &$0.138_{-0.080}^{+0.127}\pm0.007$  &$ 2.8\sigma$  \\  
4.244  &$538\pm3  $  &$9.8_{-3.2}^{+4.0}$  &$0.324$  &$1.118$  &$1.056$  &$0.073_{-0.024}^{+0.030}\pm0.004$  &$0.069_{-0.023}^{+0.028}\pm0.004$  &$ 5.0\sigma$  \\  
4.258  &$826\pm6  $  &$16.5_{-4.4}^{+5.2}$  &$0.338$  &$1.122$  &$1.054$  &$0.076_{-0.020}^{+0.024}\pm0.004$  &$0.072_{-0.019}^{+0.023}\pm0.004$  &$ 6.5\sigma$  \\  
4.267  &$531\pm3  $  &$9.4_{-3.1}^{+4.0}$  &$0.323$  &$1.124$  &$1.053$  &$0.070_{-0.024}^{+0.030}\pm0.004$  &$0.067_{-0.022}^{+0.028}\pm0.004$  &$ 5.3\sigma$  \\  
4.278  &$176\pm1  $  &$3.9_{-1.8}^{+2.6}$  &$0.310$  &$1.128$  &$1.053$  &$0.091_{-0.043}^{+0.061}\pm0.005$  &$0.087_{-0.040}^{+0.058}\pm0.005$  &$ 4.0\sigma$  \\  
4.288  &$500  $  &$5.6_{-2.3}^{+3.0}$  &$0.312$  &$1.130$  &$1.053$  &$0.046_{-0.019}^{+0.025}\pm0.002$  &$0.043_{-0.018}^{+0.024}\pm0.002$  &$ 4.1\sigma$  \\  
4.308  &$44.9\pm0.3 $  &$3.2_{-1.8}^{+2.8}$  &$0.333$  &$1.136$  &$1.052$  &$0.273_{-0.152}^{+0.233}\pm0.015$  &$0.260_{-0.144}^{+0.222}\pm0.014$  &$ 3.0\sigma$  \\  
4.312  &$500  $  &$5.9_{-2.3}^{+3.1}$  &$0.307$  &$1.137$  &$1.052$  &$0.049_{-0.019}^{+0.025}\pm0.003$  &$0.046_{-0.018}^{+0.024}\pm0.002$  &$ 4.8\sigma$  \\  
4.337  &$500  $  &$10.5_{-3.1}^{+3.9}$  &$0.309$  &$1.147$  &$1.051$  &$0.085_{-0.025}^{+0.031}\pm0.005$  &$0.081_{-0.024}^{+0.030}\pm0.004$  &$ 6.7\sigma$  \\  
4.358  &$540\pm4  $  &$7.6_{-3.0}^{+3.8}$  &$0.326$  &$1.153$  &$1.051$  &$0.054_{-0.021}^{+0.027}\pm0.003$  &$0.051_{-0.020}^{+0.026}\pm0.003$  &$ 4.0\sigma$  \\  
4.377  &$500  $  &$4.6_{-2.0}^{+2.8}$  &$0.302$  &$1.159$  &$1.051$  &$0.036_{-0.016}^{+0.022}\pm0.002$  &$0.034_{-0.015}^{+0.021}\pm0.002$  &$ 4.1\sigma$  \\  
4.387  &$55.2\pm0.4 $  &$3.1_{-1.5}^{+2.3}$  &$0.326$  &$1.162$  &$1.051$  &$0.211_{-0.105}^{+0.155}\pm0.011$  &$0.201_{-0.100}^{+0.147}\pm0.011$  &$ 3.5\sigma$  \\  
4.396  &$500  $  &$9.4_{-3.0}^{+3.8}$  &$0.301$  &$1.164$  &$1.051$  &$0.077_{-0.025}^{+0.031}\pm0.004$  &$0.073_{-0.023}^{+0.029}\pm0.004$  &$ 5.7\sigma$  \\  
4.416  &$1074\pm7 $  &$13.0_{-3.8}^{+4.7}$  &$0.303$  &$1.172$  &$1.052$  &$0.049_{-0.014}^{+0.017}\pm0.003$  &$0.046_{-0.014}^{+0.017}\pm0.002$  &$ 5.7\sigma$  \\  
4.436  &$570  $  &$0.9_{-0.9}^{+1.9}$  &$0.295$  &$1.178$  &$1.054$  &$0.006_{-0.006}^{+0.013}\pm0.000$  &$0.006_{-0.006}^{+0.013}\pm0.000$  &$ 0.8\sigma$  \\  
4.467  &$110\pm1  $  &$0.0_{-0.0}^{+0.8}$  &$0.292$  &$1.189$  &$1.055$  &$0.000_{-0.000}^{+0.028}\pm0.000$  &$0.000_{-0.000}^{+0.027}\pm0.000$  &$ 0.0\sigma$  \\  
4.527  &$110\pm1  $  &$2.3_{-1.4}^{+2.3}$  &$0.285$  &$1.211$  &$1.054$  &$0.087_{-0.054}^{+0.086}\pm0.005$  &$0.083_{-0.051}^{+0.081}\pm0.004$  &$ 2.2\sigma$  \\  
4.574  &$47.7\pm0 $  &$1.2_{-0.9}^{+1.7}$  &$0.278$  &$1.225$  &$1.054$  &$0.101_{-0.074}^{+0.144}\pm0.005$  &$0.096_{-0.071}^{+0.136}\pm0.005$  &$ 1.9\sigma$  \\  
4.599  &$567\pm4  $  &$4.0_{-1.9}^{+2.7}$  &$0.275$  &$1.236$  &$1.055$  &$0.029_{-0.014}^{+0.019}\pm0.002$  &$0.028_{-0.013}^{+0.018}\pm0.001$  &$ 3.1\sigma$  \\
4.612  &$104\pm1  $  &$0.0_{-0.0}^{+0.6}$  &$0.262$  &$1.242$  &$1.055$  &$0.000_{-0.000}^{+0.024}\pm0.000$  &$0.000_{-0.000}^{+0.022}\pm0.000$  &$ 0.0\sigma$  \\
4.628  &$522\pm3  $  &$4.8_{-2.4}^{+3.2}$  &$0.256$  &$1.246$  &$1.054$  &$0.042_{-0.021}^{+0.028}\pm0.002$  &$0.040_{-0.020}^{+0.027}\pm0.002$  &$ 2.8\sigma$  \\
4.641  &$552\pm3  $  &$3.2_{-1.9}^{+2.8}$  &$0.257$  &$1.252$  &$1.054$  &$0.026_{-0.016}^{+0.023}\pm0.001$  &$0.025_{-0.015}^{+0.022}\pm0.001$  &$ 2.2\sigma$  \\
4.661  &$529\pm3  $  &$6.0_{-2.6}^{+3.5}$  &$0.253$  &$1.259$  &$1.054$  &$0.051_{-0.022}^{+0.030}\pm0.003$  &$0.049_{-0.021}^{+0.028}\pm0.003$  &$ 3.5\sigma$  \\
4.682  &$1667\pm9 $  &$15.7_{-4.4}^{+5.2}$  &$0.251$  &$1.266$  &$1.054$  &$0.043_{-0.012}^{+0.014}\pm0.002$  &$0.040_{-0.011}^{+0.013}\pm0.002$  &$ 5.8\sigma$  \\
4.699  &$536\pm3  $  &$4.9_{-2.2}^{+3.0}$  &$0.248$  &$1.272$  &$1.055$  &$0.042_{-0.019}^{+0.026}\pm0.002$  &$0.040_{-0.018}^{+0.024}\pm0.002$  &$ 3.5\sigma$  \\
4.740  &$164\pm1  $  &$0.0_{-0.0}^{+0.6}$  &$0.255$  &$1.288$  &$1.055$  &$0.000_{-0.000}^{+0.016}\pm0.000$  &$0.000_{-0.000}^{+0.015}\pm0.000$  &$ 0.0\sigma$  \\
4.750  &$367\pm2  $  &$5.7_{-2.4}^{+3.3}$  &$0.255$  &$1.292$  &$1.055$  &$0.068_{-0.029}^{+0.040}\pm0.004$  &$0.065_{-0.028}^{+0.038}\pm0.003$  &$ 3.7\sigma$  \\
4.780  &$511\pm3  $  &$3.4_{-2.0}^{+2.9}$  &$0.250$  &$1.303$  &$1.055$  &$0.029_{-0.017}^{+0.025}\pm0.002$  &$0.028_{-0.016}^{+0.024}\pm0.001$  &$ 2.2\sigma$  \\
4.843  &$525\pm3  $  &$2.1_{-1.5}^{+2.3}$  &$0.240$  &$1.327$  &$1.056$  &$0.018_{-0.013}^{+0.020}\pm0.001$  &$0.017_{-0.012}^{+0.019}\pm0.001$  &$ 1.7\sigma$  \\
4.918  &$208\pm1  $  &$5.8_{-2.5}^{+3.4}$  &$0.230$  &$1.356$  &$1.056$  &$0.130_{-0.056}^{+0.075}\pm0.007$  &$0.123_{-0.053}^{+0.071}\pm0.007$  &$ 4.2\sigma$  \\
4.951  &$159\pm1  $  &$2.0_{-1.3}^{+2.0}$  &$0.223$  &$1.368$  &$1.056$  &$0.060_{-0.038}^{+0.060}\pm0.003$  &$0.057_{-0.036}^{+0.057}\pm0.003$  &$ 2.2\sigma$  \\
\hline\hline
\end{tabular}
\end{table*}

\clearpage
\section{Summary of Form factors}

\begin{table*}[!htbp]
\caption{List of the electromagnetic form factors of the neutral kaon and the ratio of neutral-to-charged kaon form factors at non-resonant energy points. %The charged kaon form factors are referred to Ref. \cite{ff_K+K-}.
The first uncertainty is statistical and the second is systematic. The form factors of charged kaon are taken from Ref. \cite{ff_K+K-}.
}\label{tab:ff}
\centering
\begin{tabular}{ccc}
\hline\hline
$\sqrt{s}$ (GeV)     &$|F_{K_{S}^{0}K_{L}^{0}}|\times 1000$     &$|F_{K_{S}^{0}K_{L}^{0}}(Q^{2})|/|F_{K^{+}K^{-}}(Q^{2})|$  \\
\hline
3.510  &$16.50_{-1.20}^{+1.30}\pm0.90$  &$0.23_{-0.02}^{+0.02}\pm0.01$  \\  
3.582  &$14.60_{-2.60}^{+3.40}\pm0.80$  &$0.20_{-0.04}^{+0.05}\pm0.01$  \\  
3.650  &$12.50_{-1.20}^{+1.30}\pm0.70$  &$0.21_{-0.02}^{+0.02}\pm0.01$  \\  
3.670  &$12.90_{-2.50}^{+3.40}\pm0.70$  &$0.21_{-0.04}^{+0.06}\pm0.01$  \\  
3.710  &$9.50_{-4.70}^{+11.70}\pm0.50$  &$0.16_{-0.08}^{+0.19}\pm0.00$  \\  
3.808  &$10.40_{-3.40}^{+5.40}\pm0.60$  &$0.18_{-0.06}^{+0.10}\pm0.01$  \\  
3.869  &$10.20_{-1.80}^{+2.30}\pm0.60$  &$0.18_{-0.03}^{+0.04}\pm0.01$  \\  
3.896  &$19.00_{-3.40}^{+4.40}\pm1.00$  &$0.33_{-0.06}^{+0.08}\pm0.01$  \\  
4.008  &$13.70_{-1.30}^{+1.40}\pm0.70$  &$0.30_{-0.03}^{+0.04}\pm0.01$  \\  
4.085  &$8.00_{-2.80}^{+5.50}\pm0.40$  &$0.17_{-0.06}^{+0.12}\pm0.01$  \\  
4.128  &$3.90_{-1.40}^{+2.20}\pm0.20$  &$0.08_{-0.03}^{+0.05}\pm0.00$  \\  
4.157  &$5.50_{-1.30}^{+2.00}\pm0.30$  &$0.12_{-0.03}^{+0.04}\pm0.00$  \\  
4.178  &$9.50_{-0.60}^{+0.60}\pm0.50$  &$0.20_{-0.02}^{+0.02}\pm0.01$  \\  
4.189  &$10.30_{-1.20}^{+1.40}\pm0.60$  &$0.22_{-0.03}^{+0.03}\pm0.01$  \\  
4.199  &$9.10_{-1.30}^{+1.60}\pm0.50$  &$0.19_{-0.03}^{+0.04}\pm0.01$  \\  
4.209  &$7.90_{-1.20}^{+1.50}\pm0.40$  &$0.20_{-0.03}^{+0.04}\pm0.01$  \\  
4.219  &$6.70_{-1.30}^{+1.70}\pm0.40$  &$0.17_{-0.03}^{+0.04}\pm0.01$  \\  
4.226  &$9.20_{-0.90}^{+1.00}\pm0.50$  &$0.23_{-0.03}^{+0.03}\pm0.01$  \\  
4.236  &$8.50_{-1.30}^{+1.60}\pm0.50$  &$0.21_{-0.03}^{+0.04}\pm0.01$  \\  
4.242  &$11.20_{-3.20}^{+5.10}\pm0.60$  &$0.28_{-0.08}^{+0.13}\pm0.01$  \\  
4.244  &$7.90_{-1.30}^{+1.60}\pm0.40$  &$0.20_{-0.03}^{+0.04}\pm0.01$  \\  
4.258  &$8.10_{-1.10}^{+1.30}\pm0.40$  &$0.20_{-0.03}^{+0.03}\pm0.01$  \\  
4.267  &$7.80_{-1.30}^{+1.60}\pm0.40$  &$0.19_{-0.03}^{+0.04}\pm0.01$  \\  
4.278  &$8.90_{-2.10}^{+3.00}\pm0.50$  &$0.22_{-0.05}^{+0.07}\pm0.01$  \\  
4.288  &$6.30_{-1.30}^{+1.70}\pm0.30$  &$0.16_{-0.03}^{+0.04}\pm0.00$  \\  
4.308  &$15.50_{-4.30}^{+6.60}\pm0.80$  &$0.38_{-0.11}^{+0.16}\pm0.01$  \\  
4.312  &$6.60_{-1.30}^{+1.70}\pm0.40$  &$0.16_{-0.03}^{+0.04}\pm0.01$  \\  
4.337  &$8.70_{-1.30}^{+1.60}\pm0.50$  &$0.21_{-0.03}^{+0.04}\pm0.01$  \\  
4.358  &$7.00_{-1.40}^{+1.80}\pm0.40$  &$0.17_{-0.03}^{+0.04}\pm0.01$  \\  
4.377  &$5.70_{-1.20}^{+1.70}\pm0.30$  &$0.14_{-0.03}^{+0.04}\pm0.00$  \\  
4.387  &$13.90_{-3.40}^{+5.10}\pm0.70$  &$0.33_{-0.09}^{+0.12}\pm0.01$  \\  
4.396  &$8.40_{-1.30}^{+1.70}\pm0.50$  &$0.20_{-0.03}^{+0.04}\pm0.01$  \\  
4.416  &$6.70_{-1.00}^{+1.20}\pm0.40$  &$0.23_{-0.04}^{+0.05}\pm0.01$  \\  
4.436  &$2.40_{-1.20}^{+2.60}\pm0.10$  &$0.08_{-0.04}^{+0.09}\pm0.00$  \\  
4.467  &$0.00_{-0.00}^{+5.20}\pm0.00$  &$0.00_{-0.00}^{+0.18}\pm0.00$  \\  
4.527  &$9.20_{-2.80}^{+4.50}\pm0.50$  &$0.31_{-0.10}^{+0.16}\pm0.01$  \\  
4.574  &$10.00_{-3.70}^{+7.10}\pm0.50$  &$0.34_{-0.13}^{+0.24}\pm0.01$  \\  
4.599  &$5.40_{-1.30}^{+1.80}\pm0.30$  &$0.18_{-0.05}^{+0.06}\pm0.01$  \\ 
4.612  &$0.00_{-0.00}^{+4.80}\pm0.00$  &$0.00_{-0.00}^{+0.18}\pm0.00$  \\
4.628  &$6.50_{-1.60}^{+2.20}\pm0.30$  &$0.24_{-0.06}^{+0.08}\pm0.01$  \\
4.641  &$5.10_{-1.50}^{+2.30}\pm0.30$  &$0.19_{-0.06}^{+0.08}\pm0.01$  \\
4.661  &$7.20_{-1.60}^{+2.10}\pm0.40$  &$0.26_{-0.06}^{+0.08}\pm0.01$  \\
4.682  &$6.60_{-0.90}^{+1.10}\pm0.40$  &$0.24_{-0.04}^{+0.04}\pm0.01$  \\
4.699  &$6.60_{-1.50}^{+2.00}\pm0.40$  &$0.24_{-0.06}^{+0.08}\pm0.01$  \\
4.740  &$0.00_{-0.00}^{+4.10}\pm0.00$  &$0.00_{-0.00}^{+0.15}\pm0.00$  \\
4.750  &$8.50_{-1.80}^{+2.50}\pm0.50$  &$0.30_{-0.07}^{+0.09}\pm0.01$  \\
4.780  &$5.60_{-1.60}^{+2.40}\pm0.30$  &$0.20_{-0.06}^{+0.09}\pm0.01$  \\
4.843  &$4.40_{-1.60}^{+2.50}\pm0.20$  &$0.18_{-0.06}^{+0.10}\pm0.01$  \\
4.918  &$12.10_{-2.60}^{+3.50}\pm0.70$  &$0.47_{-0.11}^{+0.14}\pm0.02$  \\
4.951  &$8.30_{-2.60}^{+4.10}\pm0.40$  &$0.32_{-0.10}^{+0.16}\pm0.01$  \\
\hline\hline
\end{tabular}
\end{table*}

\section{Definition of the likelihood in signal extraction}

The likelihood for signal extraction is defined as:

\begin{equation}
    {\cal L}(N_j,N_0,a)=\frac{\underset{j=0}{\overset{3}\prod} e^{-N_{j}}}{N_{tot}!}\underset{i=1}{\overset{N_{tot}}{\prod}}[\underset{j=1}{\overset{3}{\sum}} (c_{ij}\cdot{\rm PDF}_{j}\otimes Gauss(\mu,\sigma))+c_{i0}\cdot e^{ax}],
    %{\cal L}(N_j,N_r,a)=\frac{1}{N_{tot}!}\underset{i=1}{\overset{N_{tot}}{\prod}}[\underset{j=1}{\overset{3}{\sum}} ({\rm PDF}_{j}\otimes Gauss(\mu,\sigma))+e^{ax}],
\end{equation}
where 
\begin{itemize}

\item $N_j=\underset{i=1}{\overset{N_{tot}}{\sum}}c_{ij}$ is the number of events for the $j$th physics process, which are $e^+e^-\rightarrow K^0_SK^0_L,~e^+e^-\rightarrow \gamma^{ISR}\psi(3686),~e^+e^-\rightarrow K^{\ast0}(892)\bar{K}^0$. The probability density function (PDF) of $e^+e^-\rightarrow \gamma^{ISR}\psi(3686)$ is only used at 3.710 GeV and 3.774 GeV since it is negligible at other energy points. 

\item $N_0=\underset{i=1}{\overset{N_{tot}}{\sum}}c_{i0}$ is the number of events from the continuum background described by the exponential function, 

\item $N_{tot}$ is the total number of events, and $a$ is a parameter in the exponential function.

\item ${\rm PDF}_j$ is the probability density function extracted from the exclusive MC sample of the $j$th process, which are $e^+e^-\rightarrow K^0_SK^0_L,~e^+e^-\rightarrow \gamma^{ISR}\psi(3686),~e^+e^-\rightarrow K^{\ast0}(892)\bar{K}^0$.

\item $Gauss(\mu,\sigma)$ is a Gaussian function with a mean fixed to $\mu=0.38\times10^{-3}$ and a width fixed to $\sigma=0.75\times10^{-3}$, which are extracted from the fitting to the $X$ distribution from $\psi(3686)\to K_S^0K_L^0$.
\end{itemize}

\clearpage
\section{Fit results at each energy point}

\begin{figure}[!htbp]
  \centering
  % Requires \usepackage{graphicx}
  \includegraphics[width=0.3\textwidth]{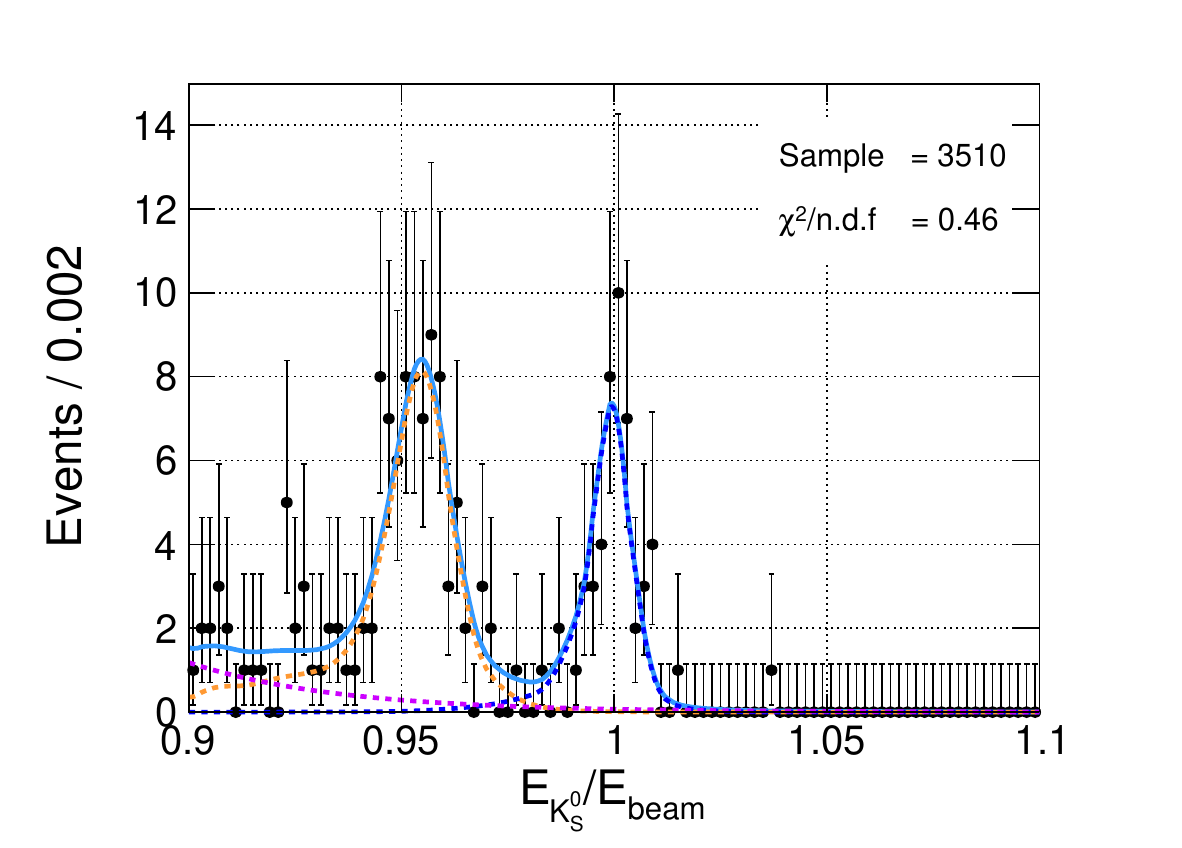}
  \includegraphics[width=0.3\textwidth]{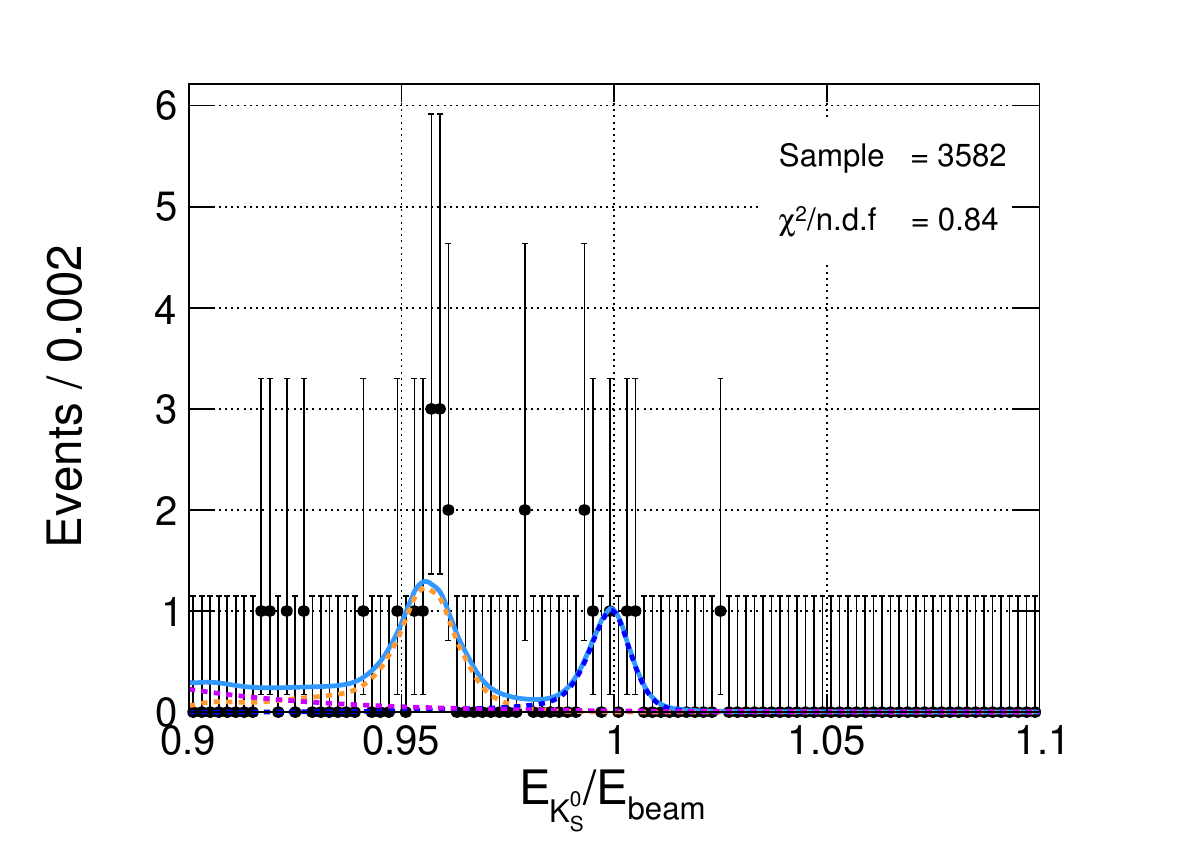}
  \includegraphics[width=0.3\textwidth]{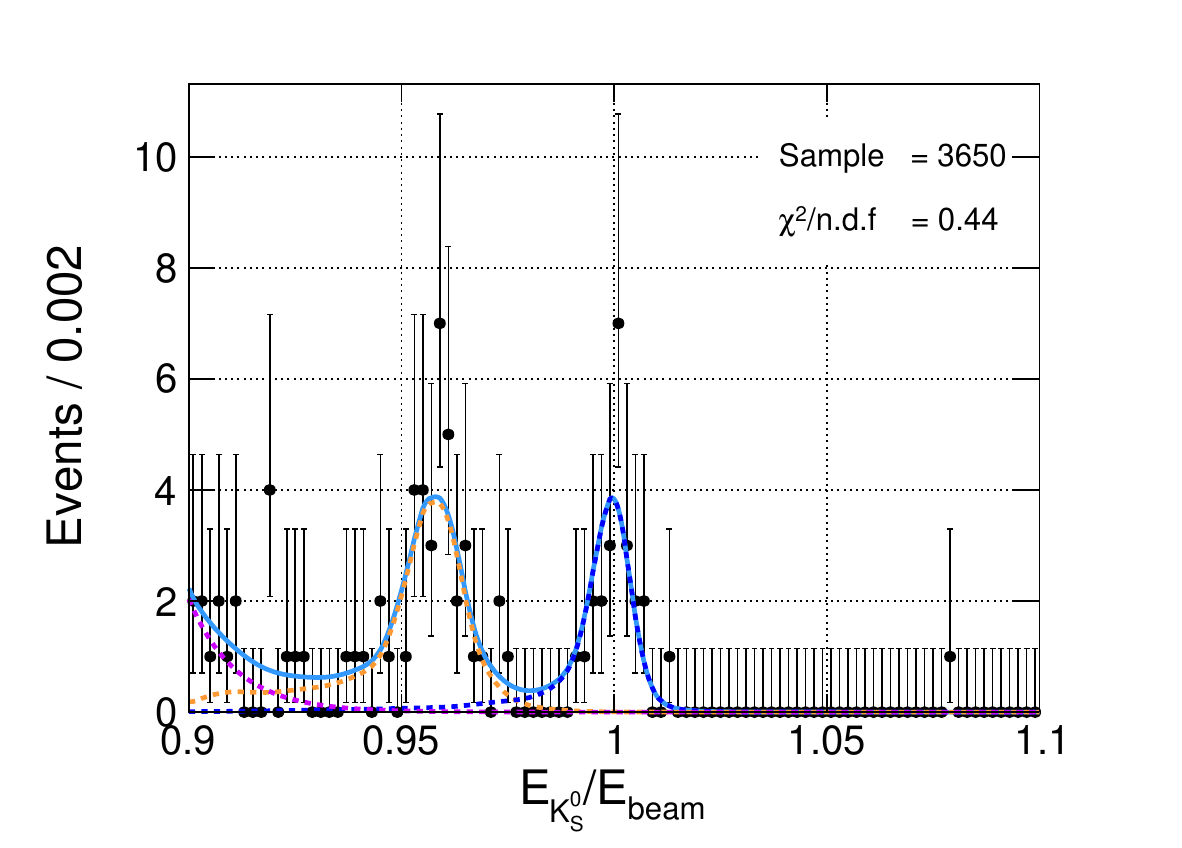}\\
  \includegraphics[width=0.3\textwidth]{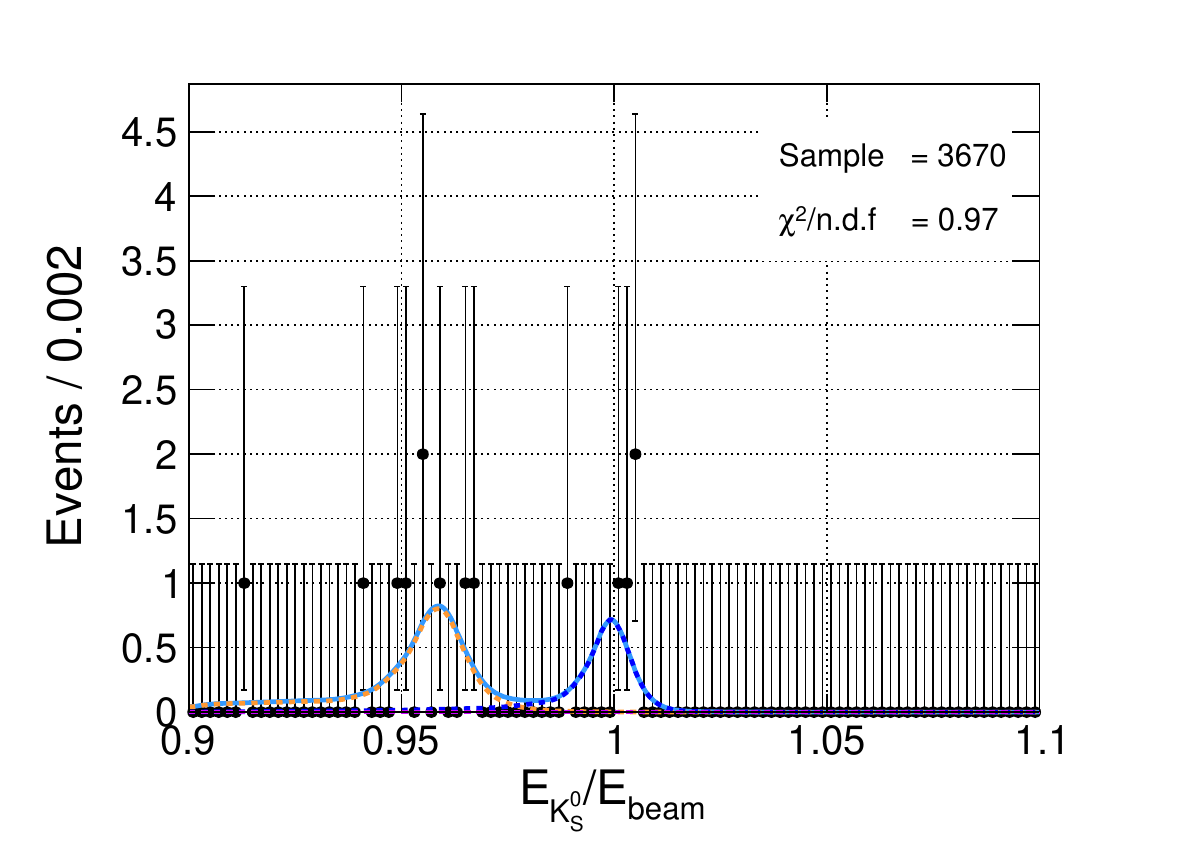}
  \includegraphics[width=0.3\textwidth]{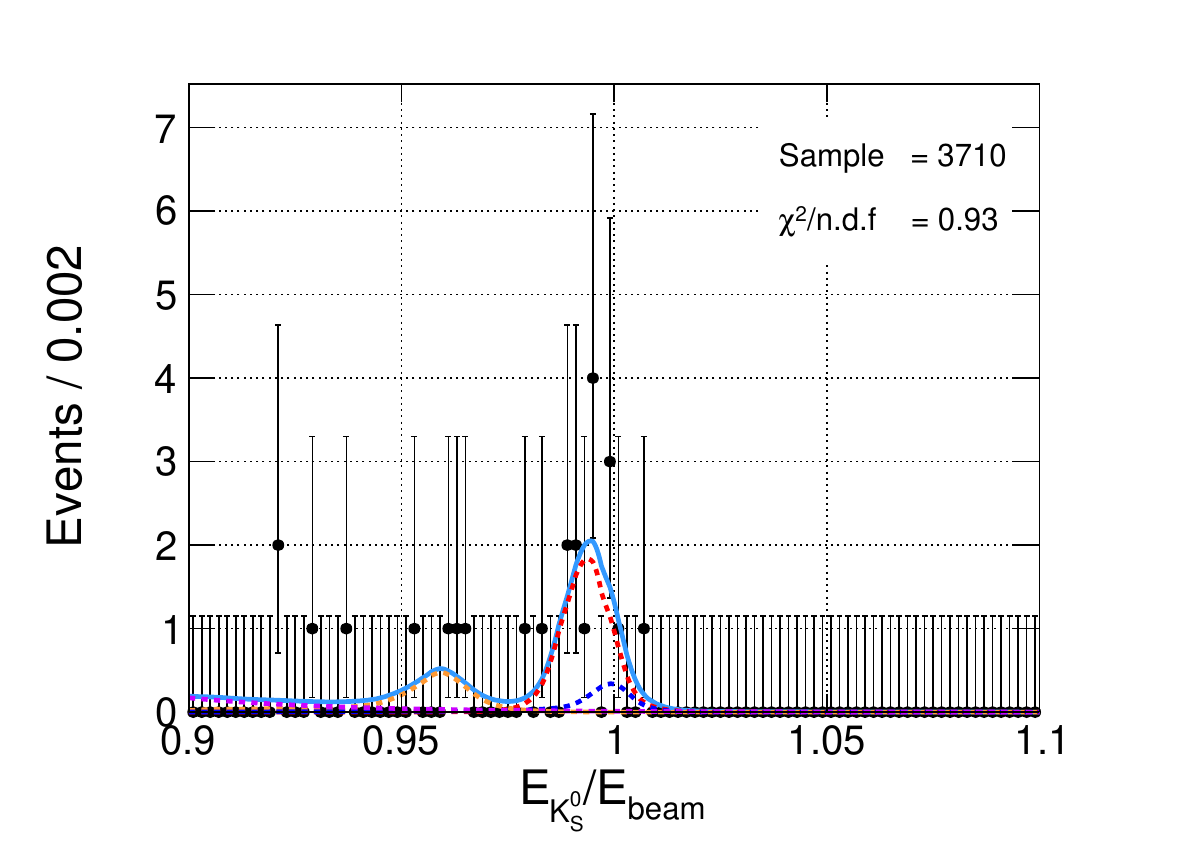}
  \includegraphics[width=0.3\textwidth]{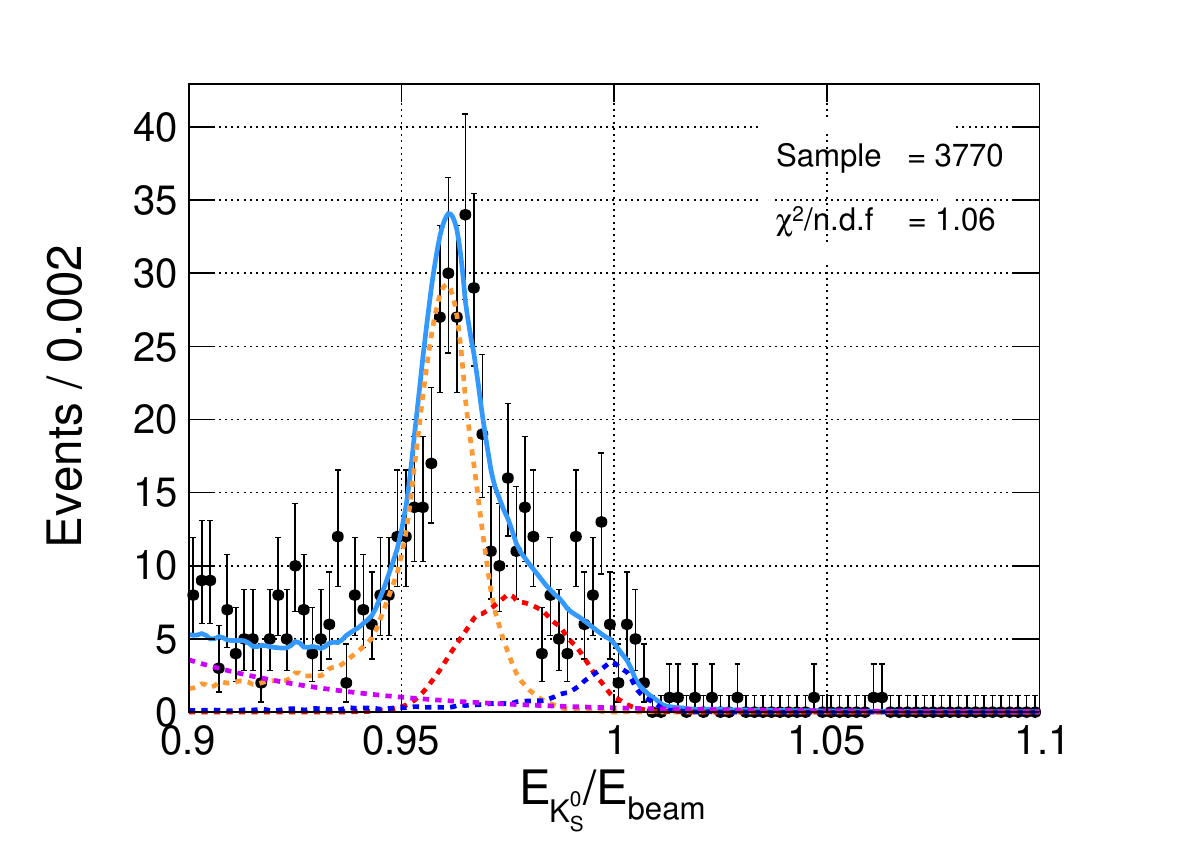}\\
  \includegraphics[width=0.3\textwidth]{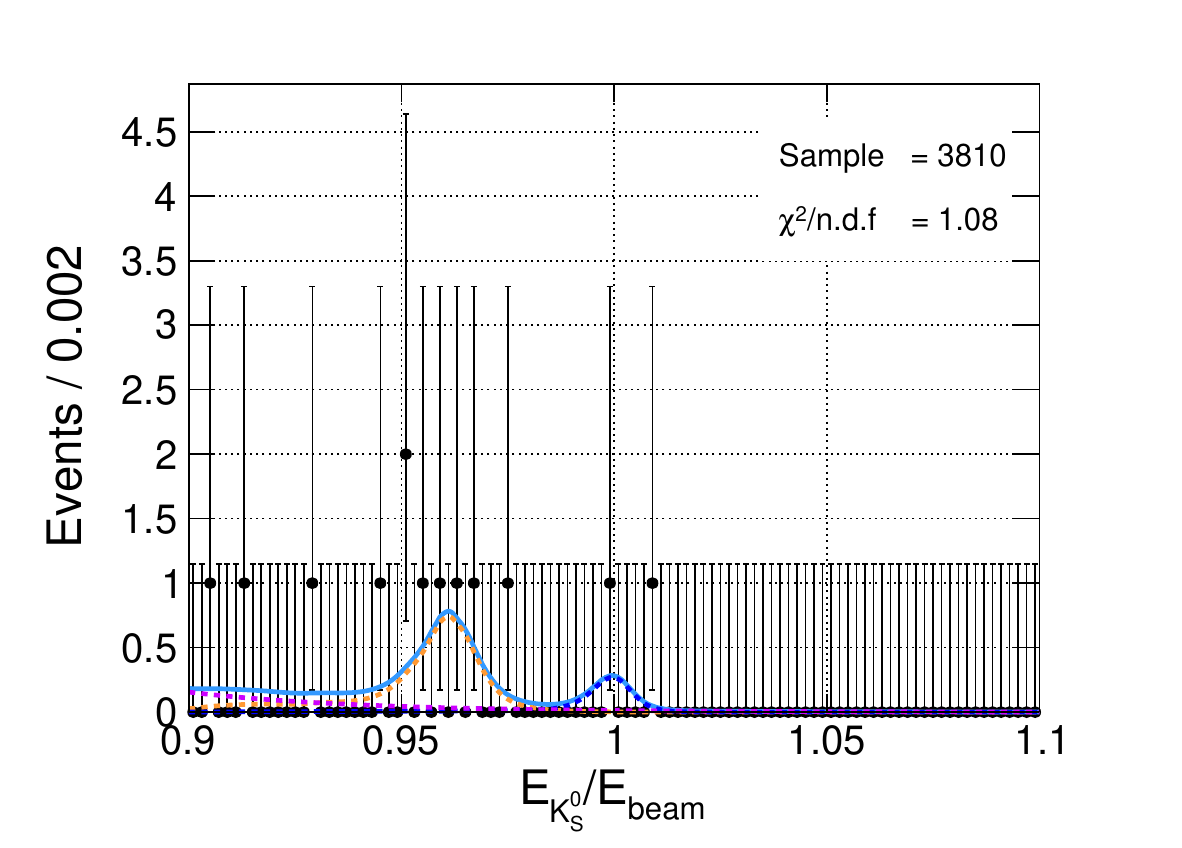}
  \includegraphics[width=0.3\textwidth]{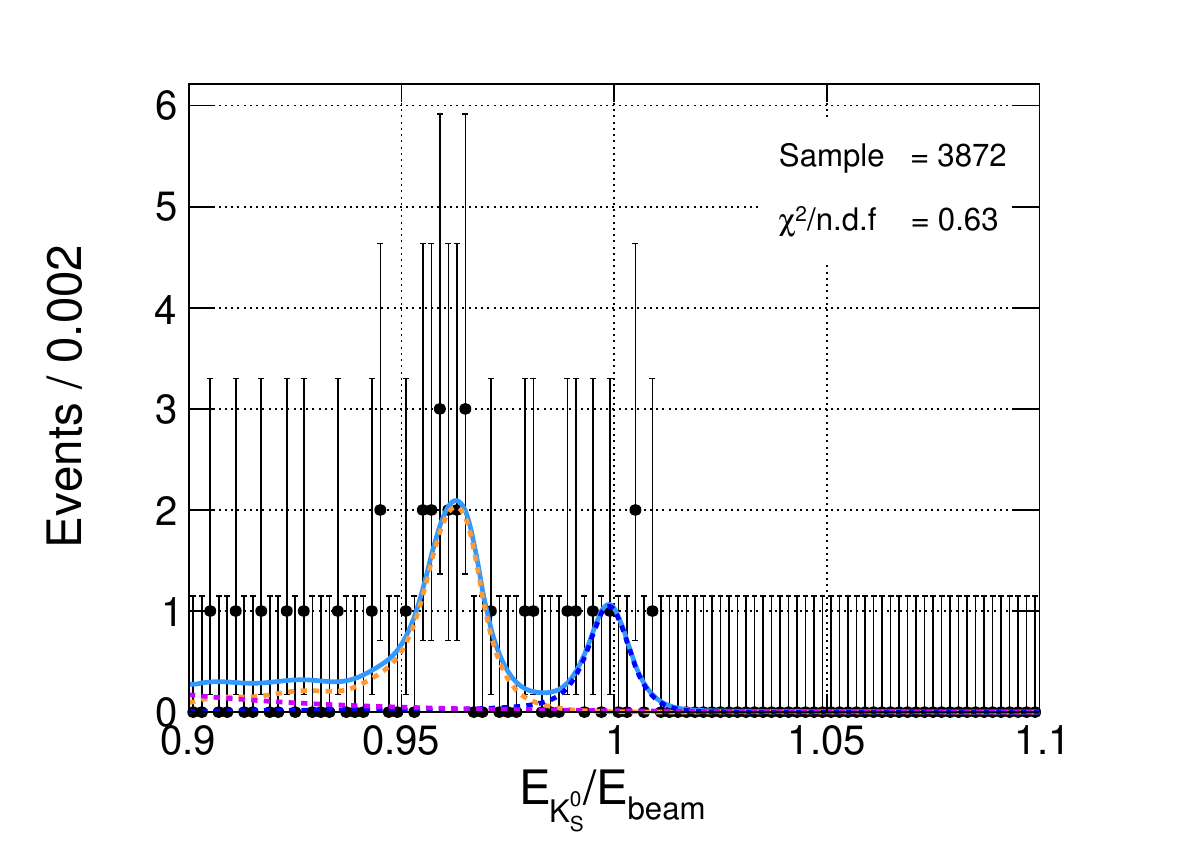}
  \includegraphics[width=0.3\textwidth]{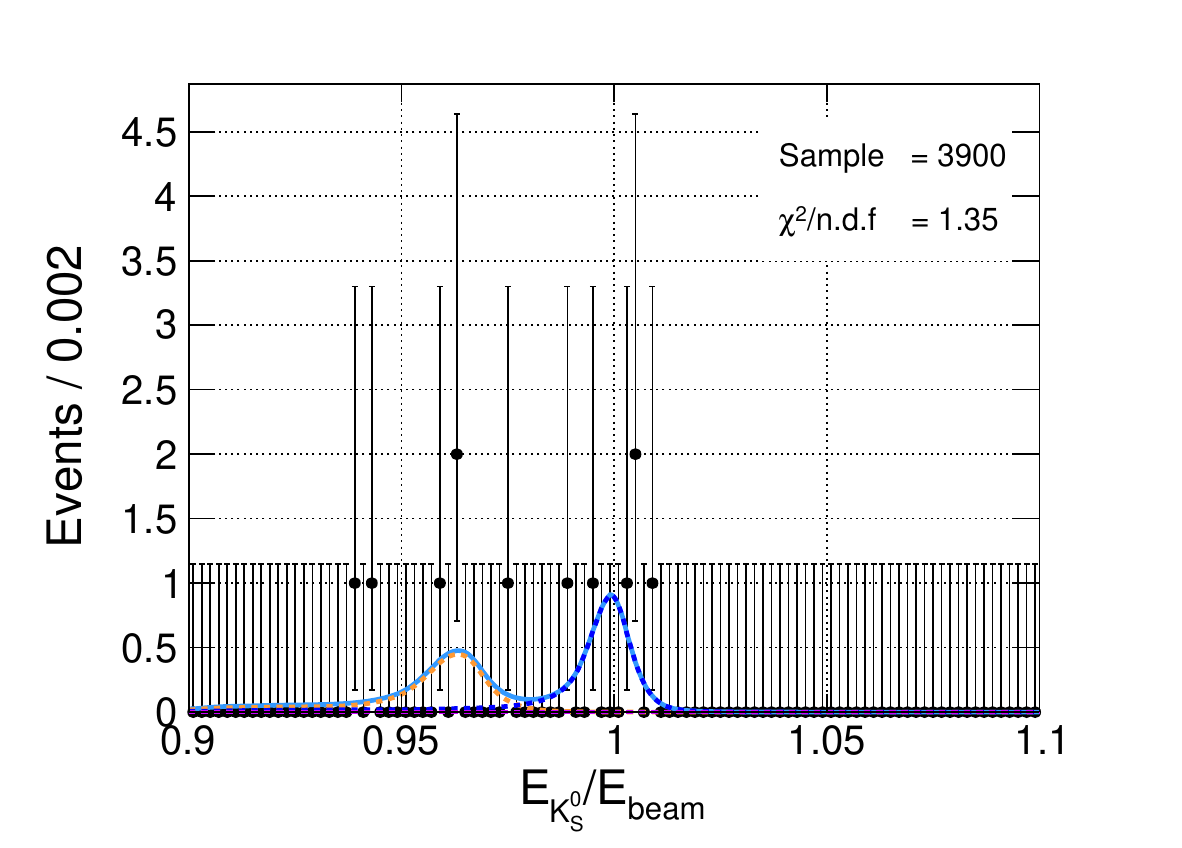}\\
  \includegraphics[width=0.3\textwidth]{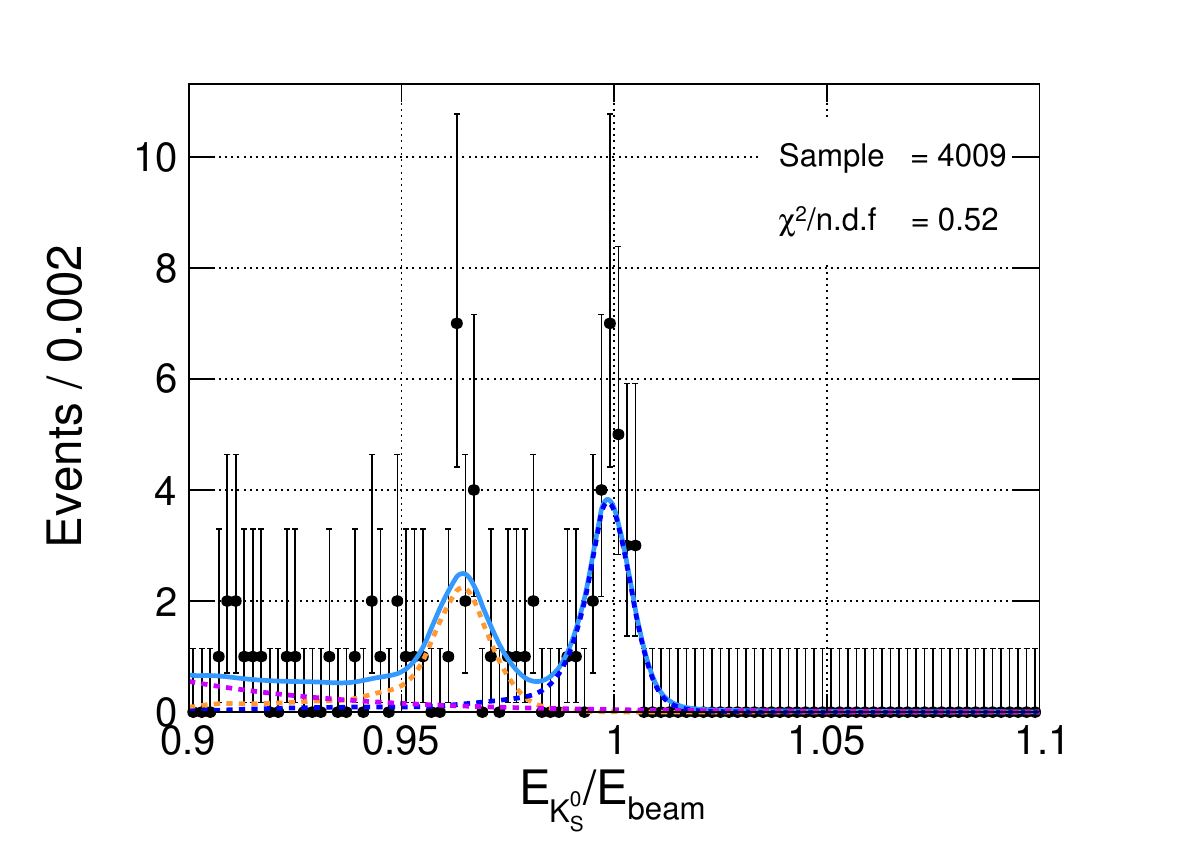}
  \includegraphics[width=0.3\textwidth]{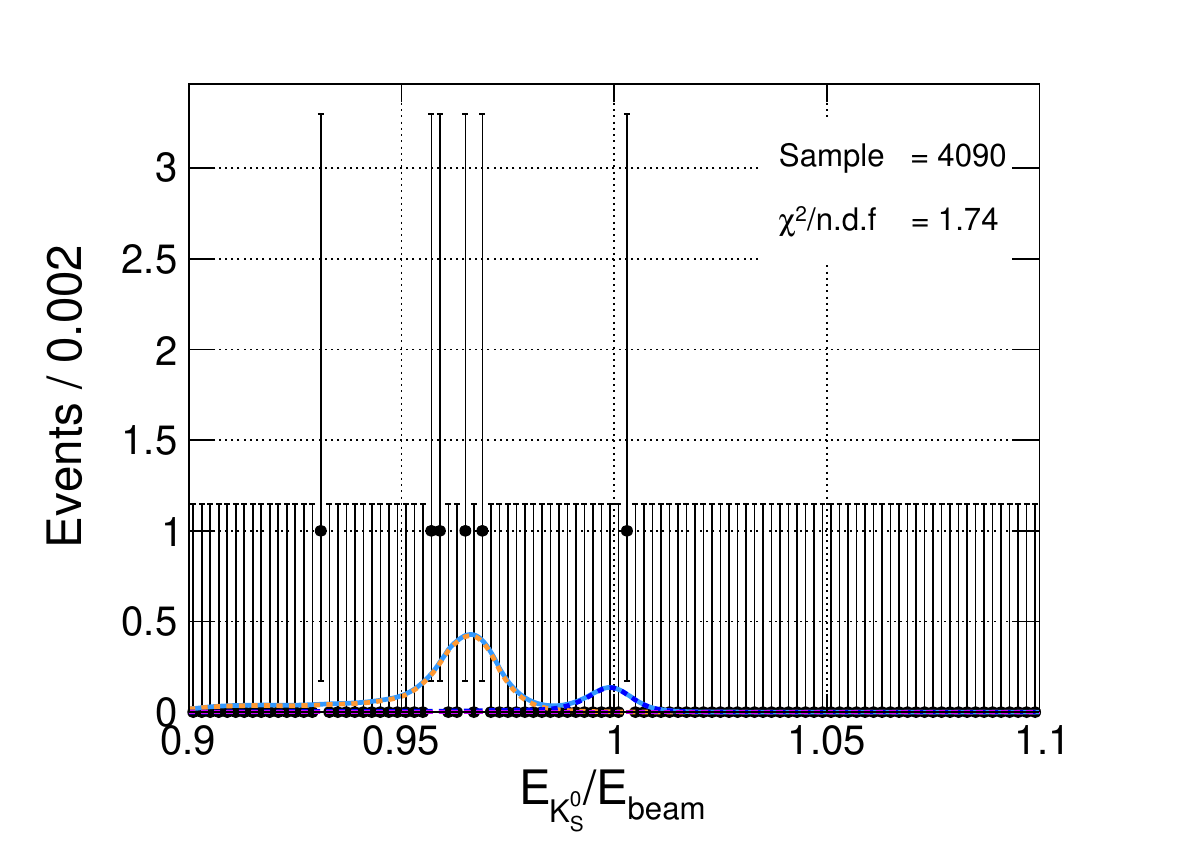}
  \includegraphics[width=0.3\textwidth]{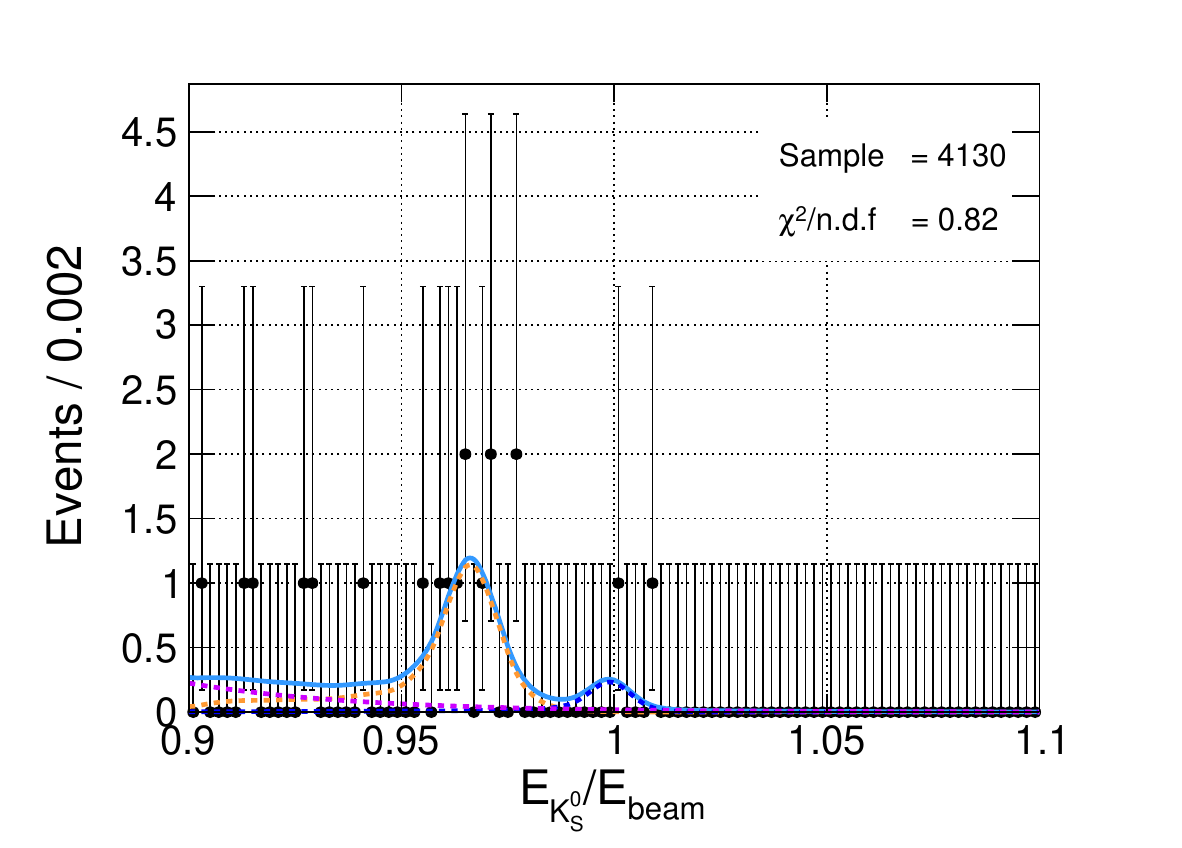}\\
  \includegraphics[width=0.3\textwidth]{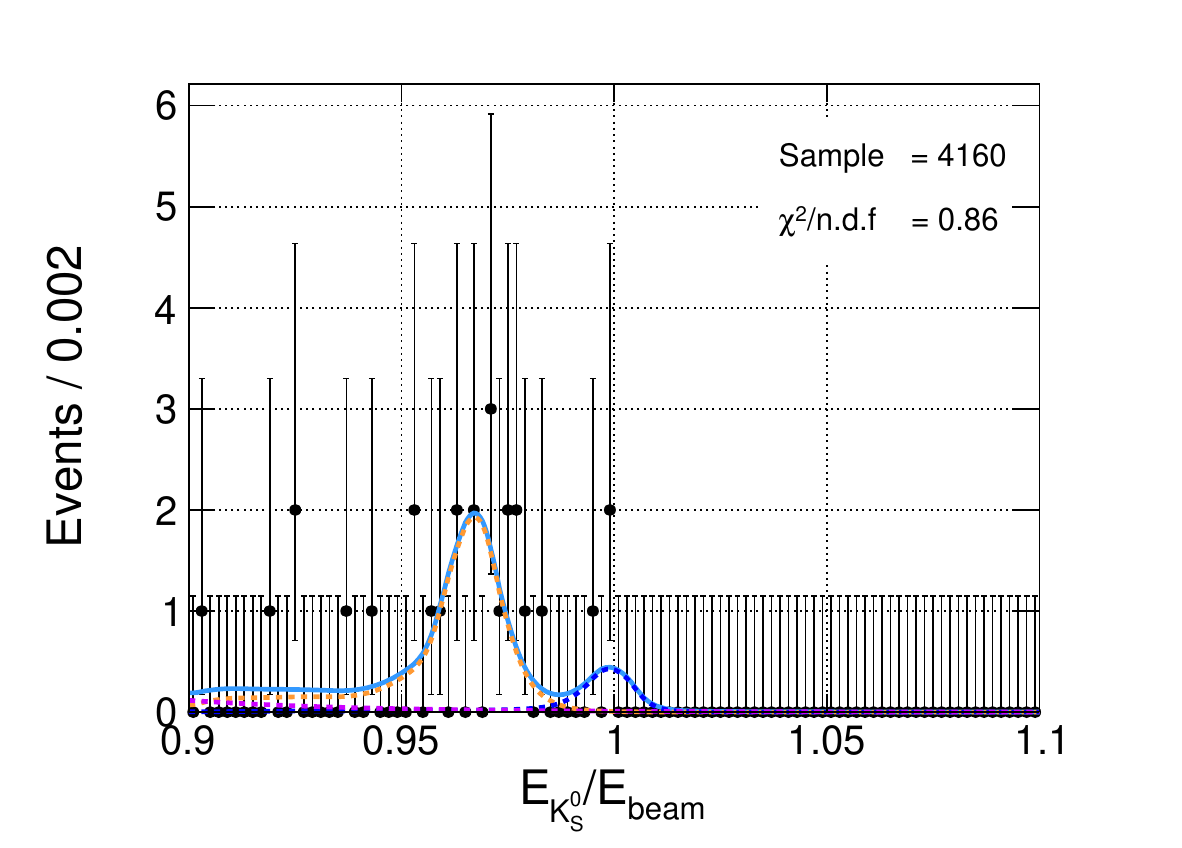}
  \includegraphics[width=0.3\textwidth]{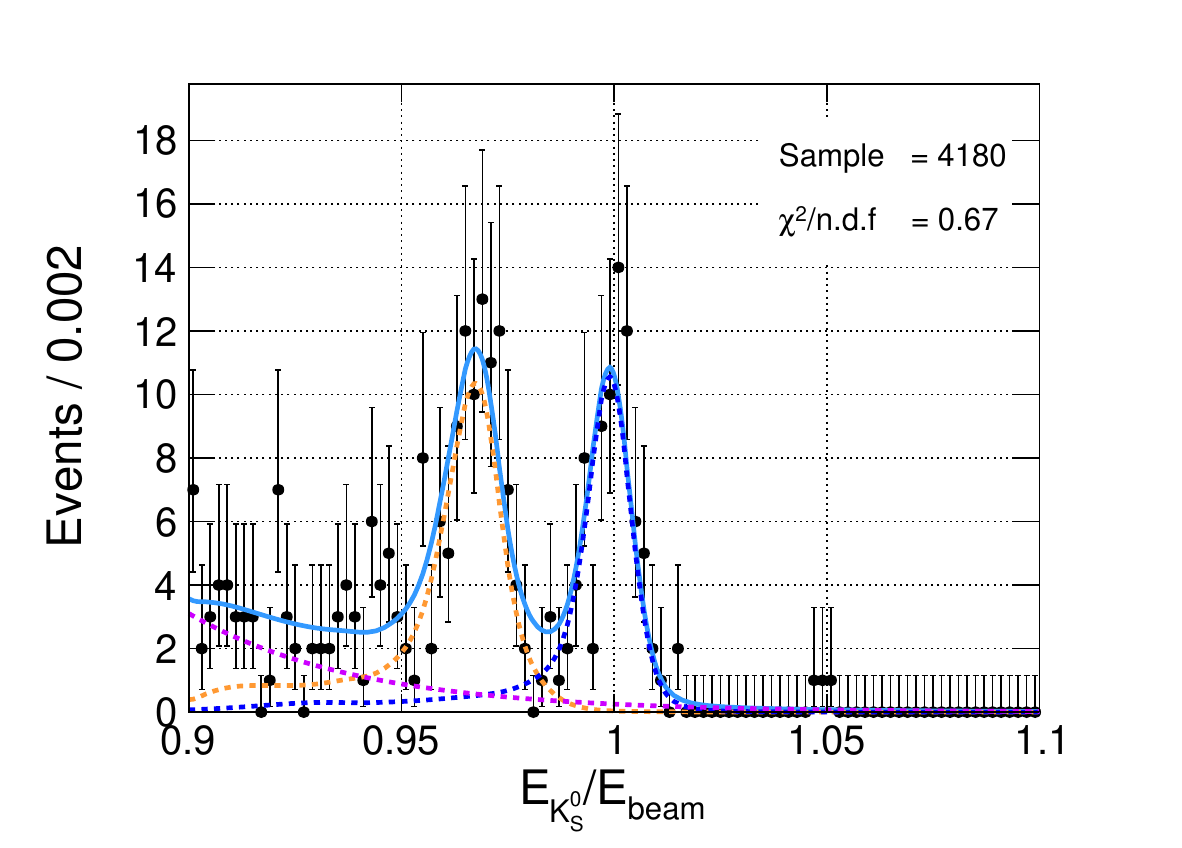}
  \includegraphics[width=0.3\textwidth]{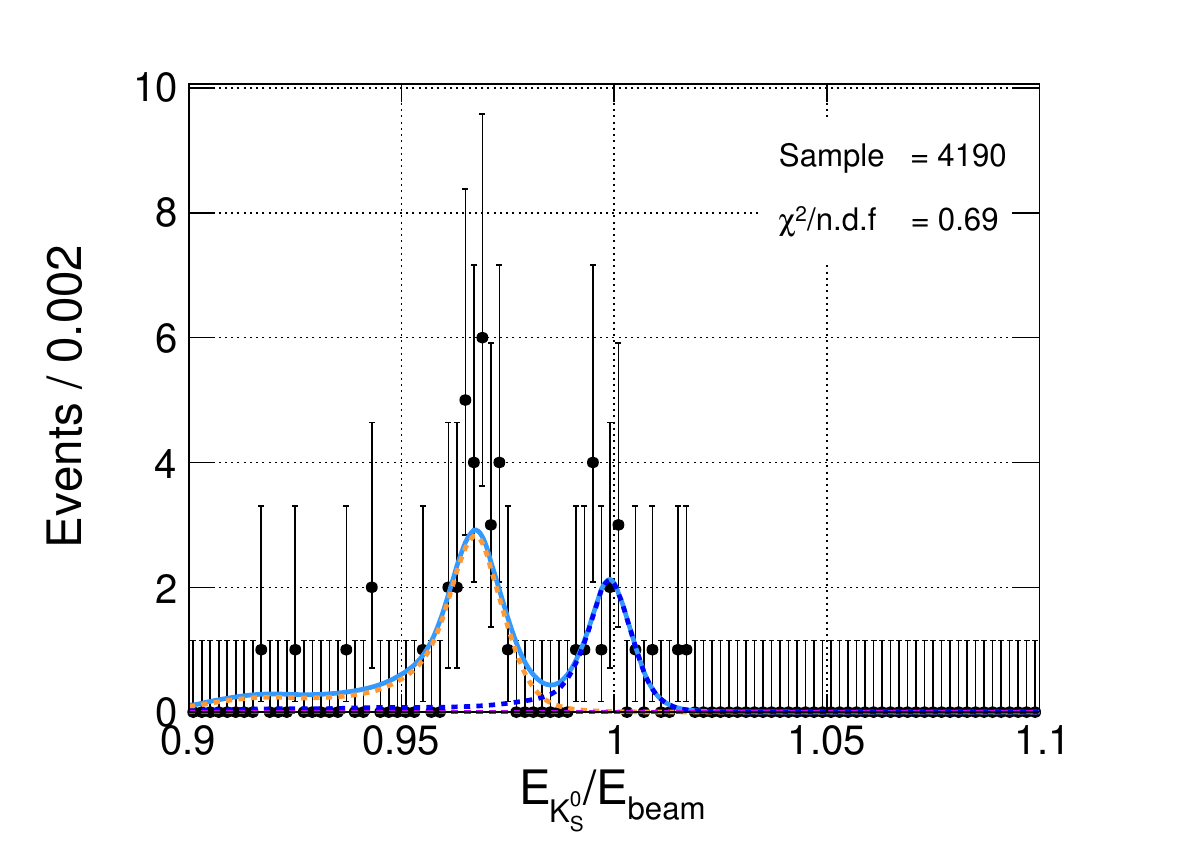}\\
  \caption{Distributions of $X\equiv E_{K_S^0}/E_{\rm beam}$. Dots with error bars are data. Blue solid, blue dashed, red dashed, orange dashed, and violet dashed lines are the total fit results, the signal MC samples, the MC samples of $e^+e^-\rightarrow \gamma^{ISR}\psi(3686)$, the MC samples of $e^+e^-\rightarrow K^{\ast0}(892)\bar{K^0}$, and the exponential functions describing the remaining background, respectively. Each MC simulated shape is convolved with a Gaussian function.}\label{data}
\end{figure}

\newpage
\begin{figure}[tp]
  \centering
  %\ContinuedFloat
  \includegraphics[width=0.3\textwidth]{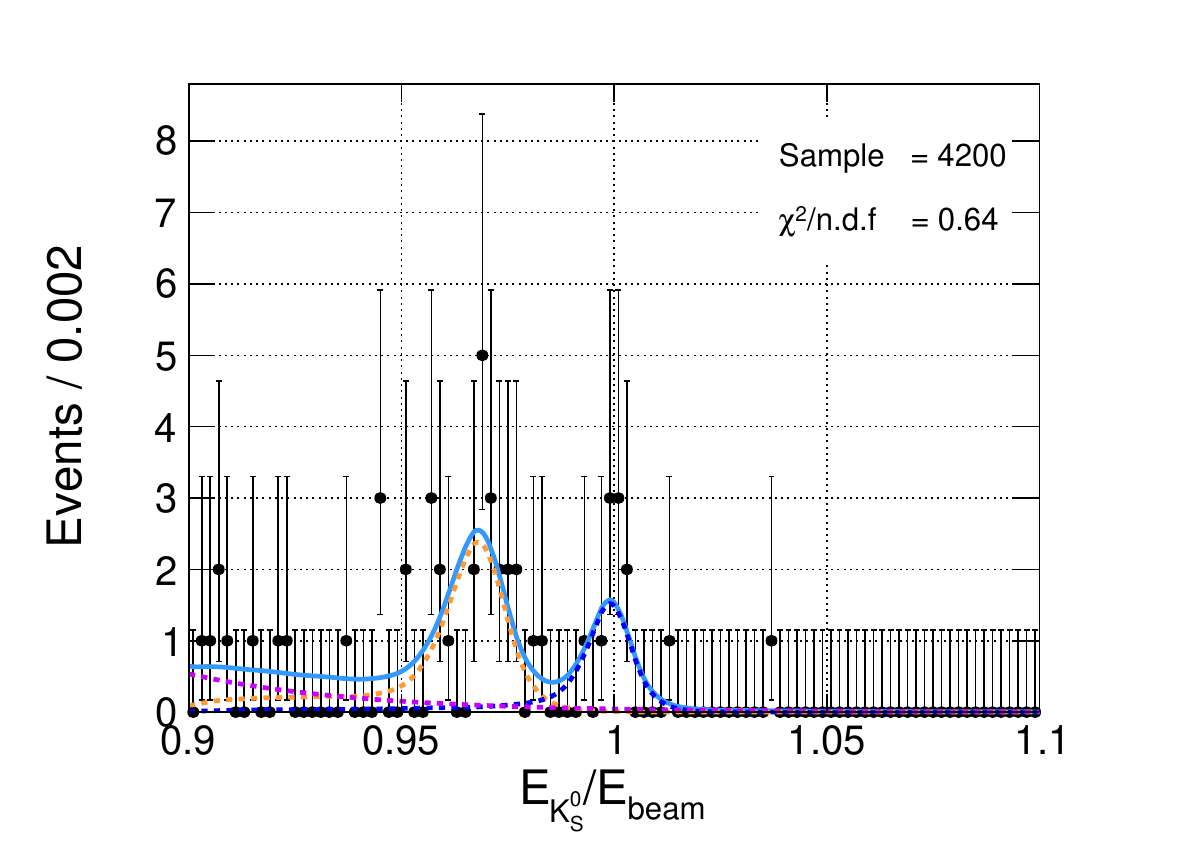}
  \includegraphics[width=0.3\textwidth]{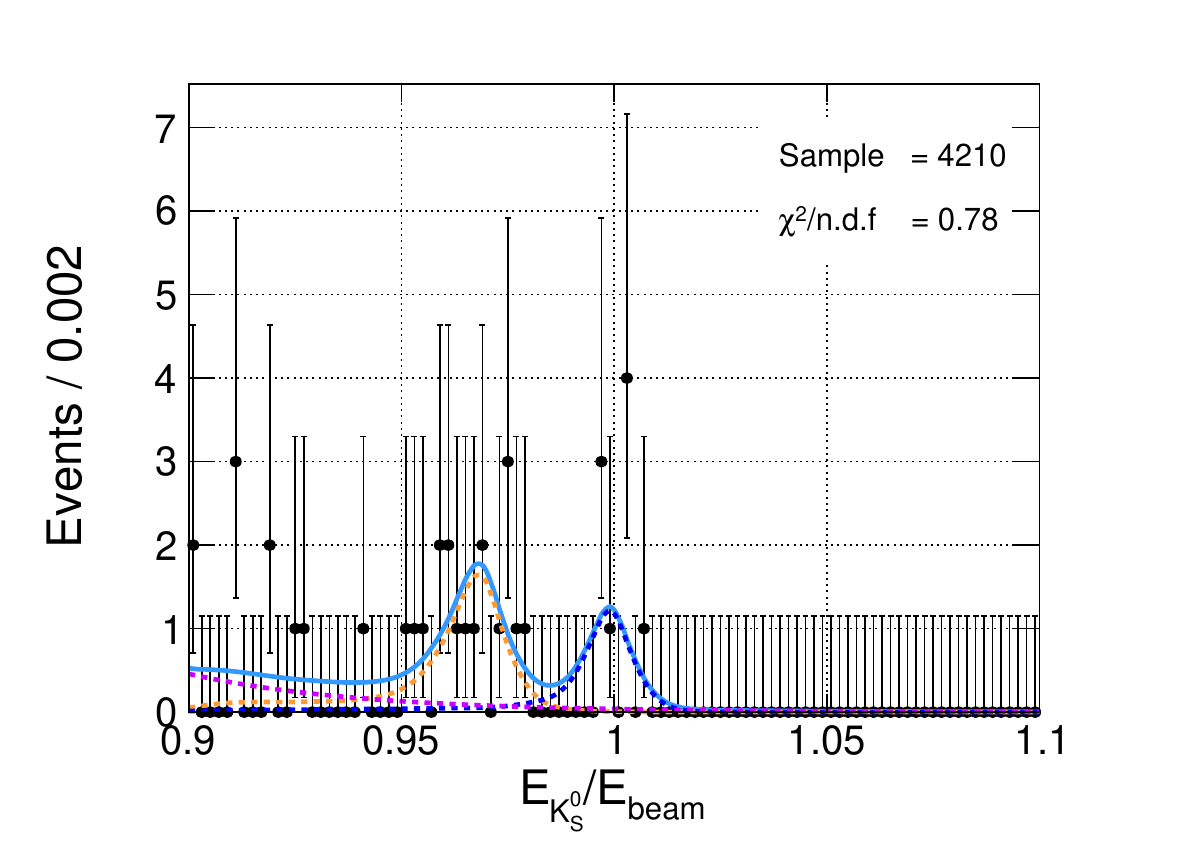}
  \includegraphics[width=0.3\textwidth]{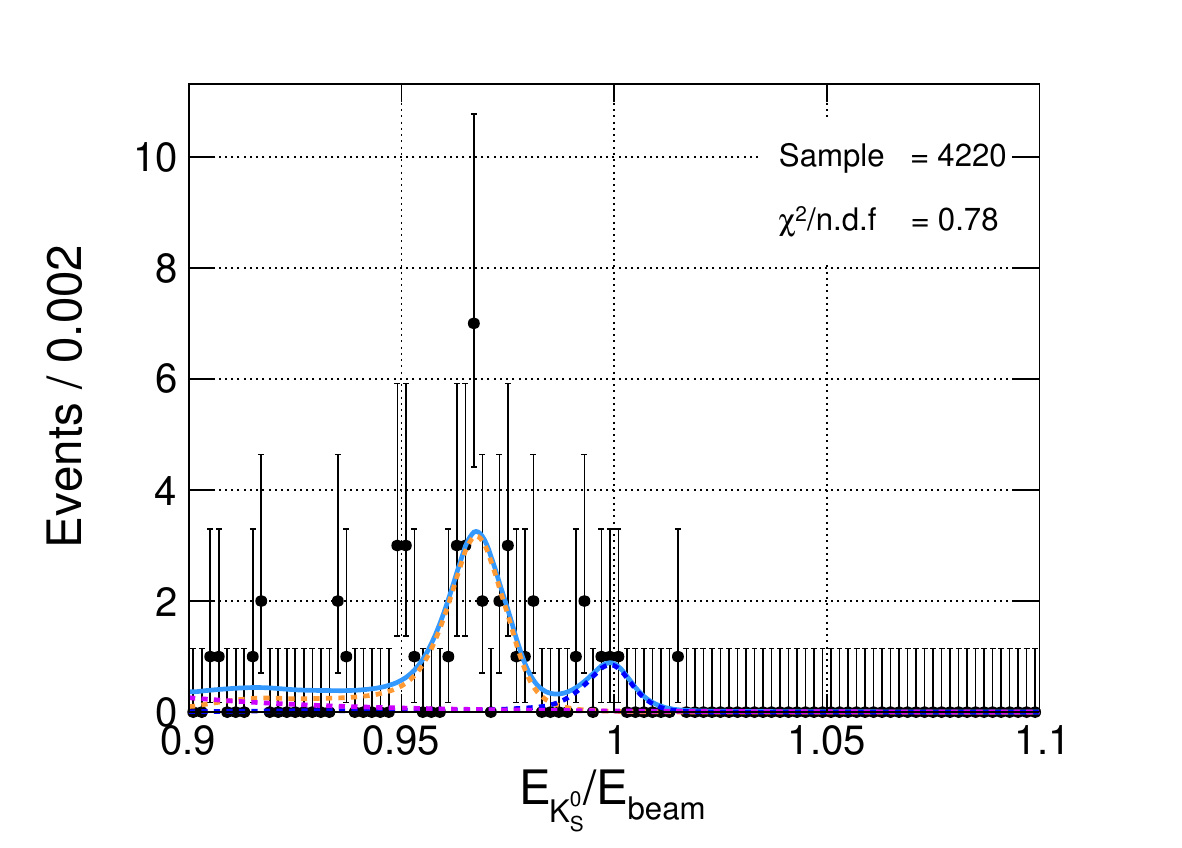}\\
  \includegraphics[width=0.3\textwidth]{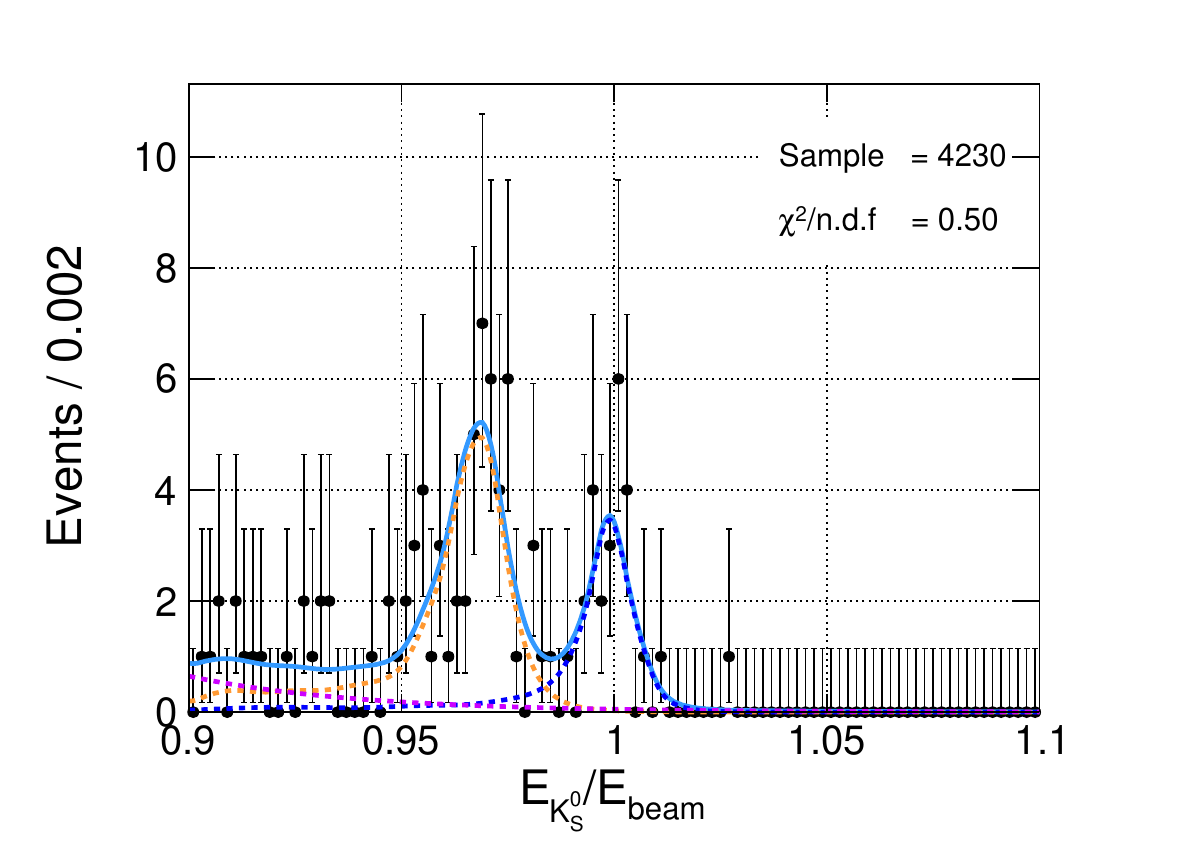}
  \includegraphics[width=0.3\textwidth]{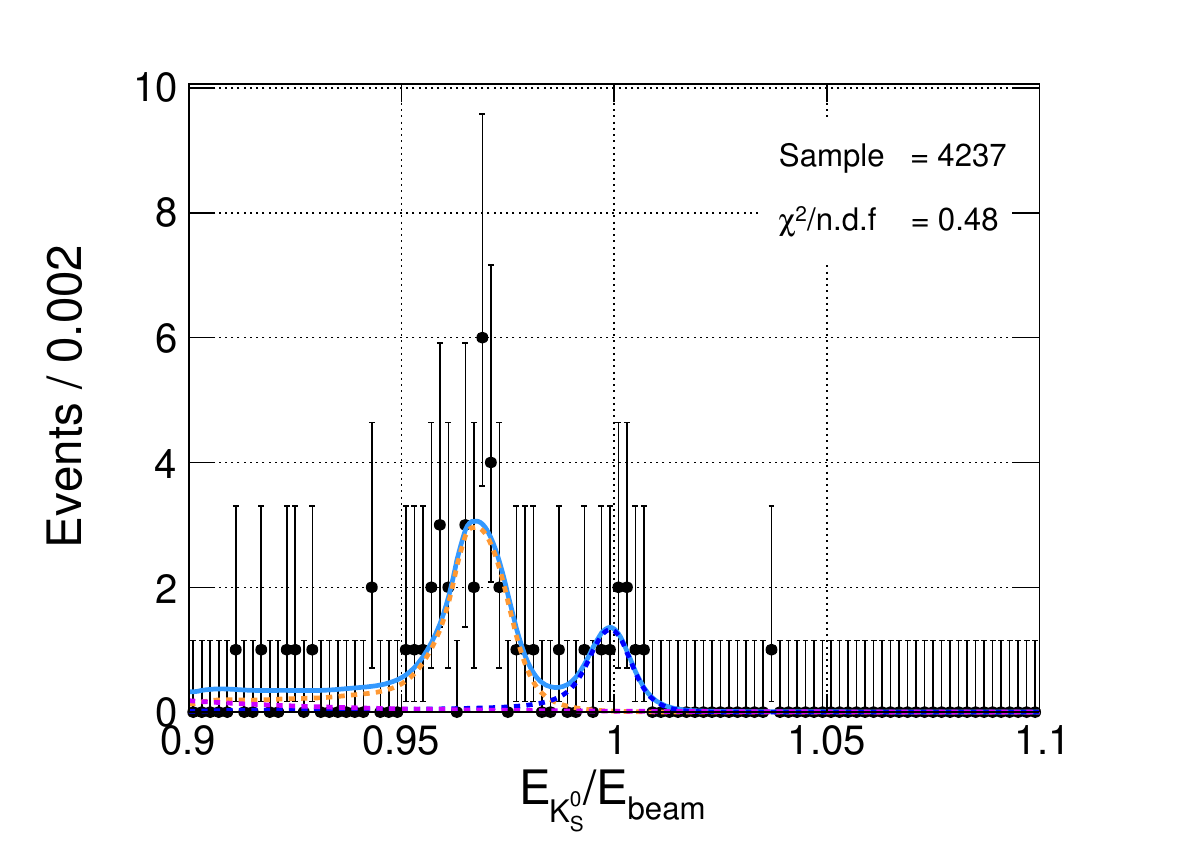}
  \includegraphics[width=0.3\textwidth]{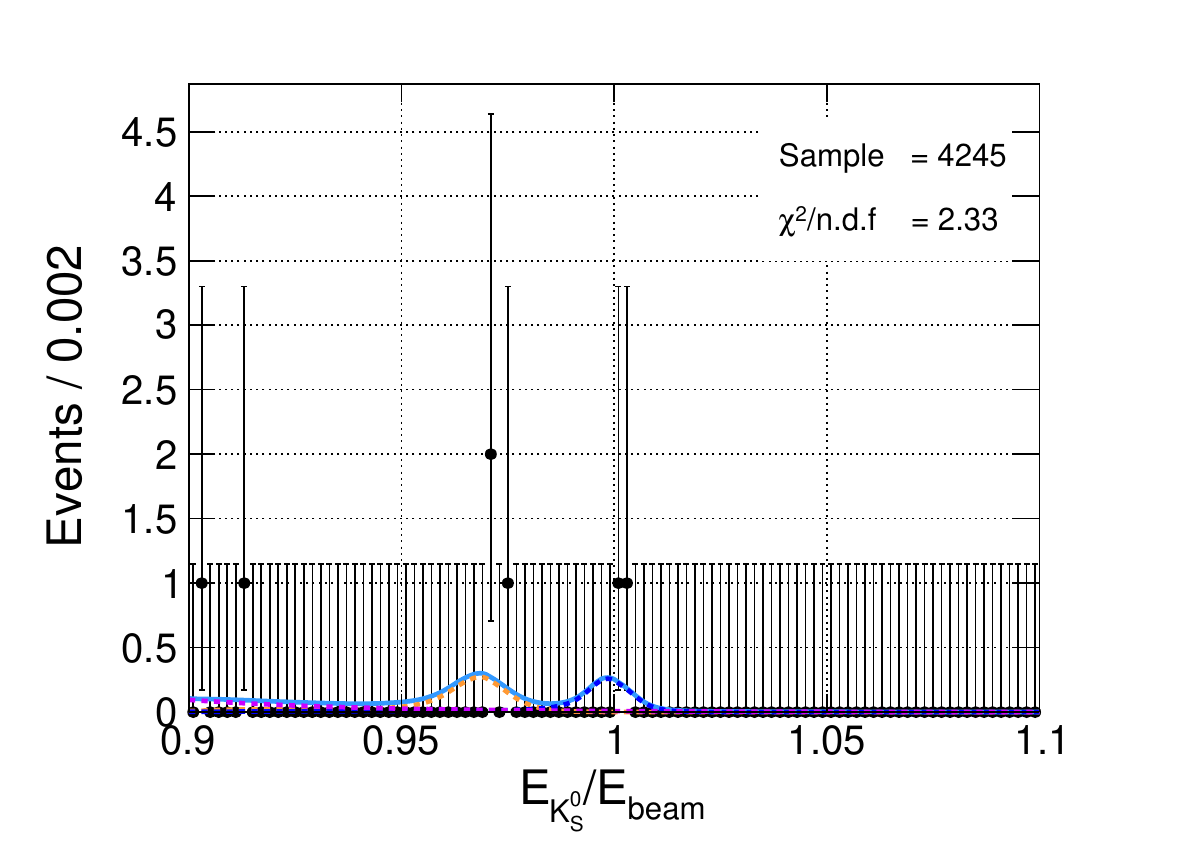}\\
  \includegraphics[width=0.3\textwidth]{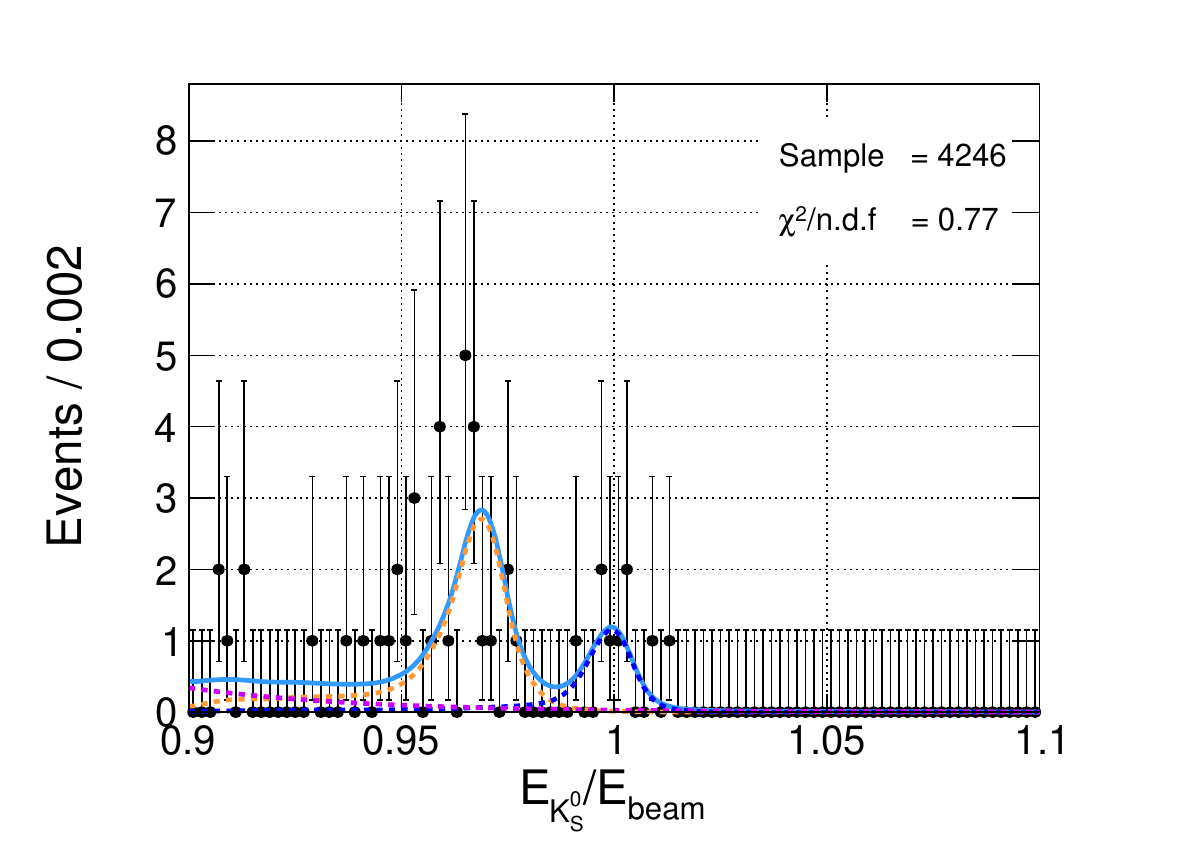}
  \includegraphics[width=0.3\textwidth]{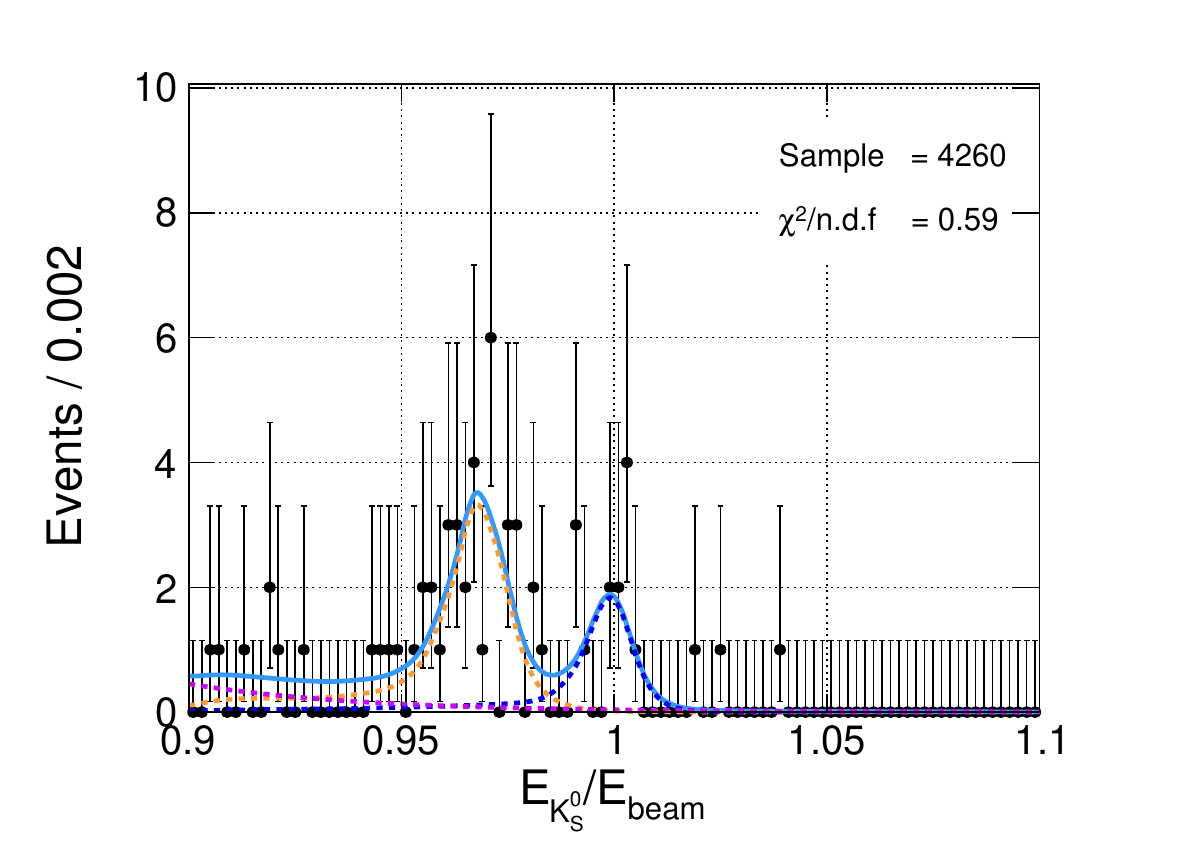}
  \includegraphics[width=0.3\textwidth]{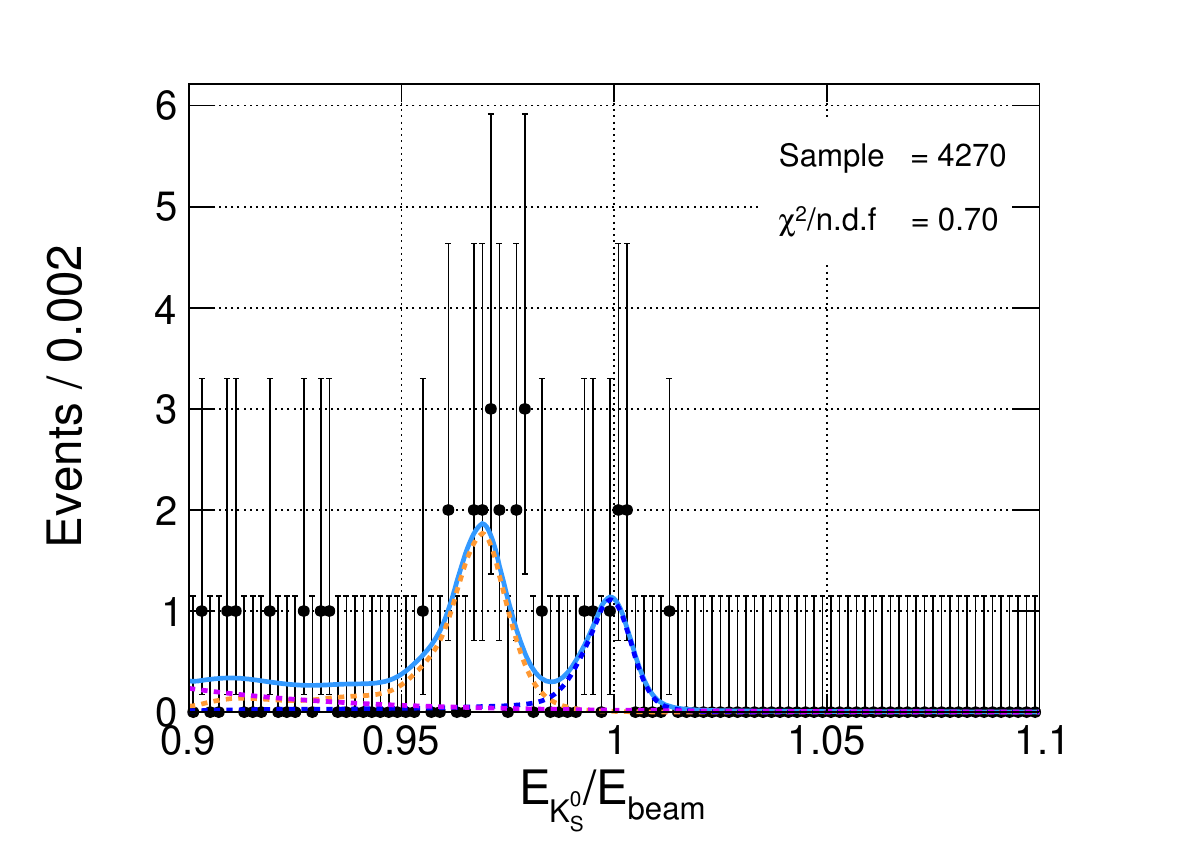}\\
  \includegraphics[width=0.3\textwidth]{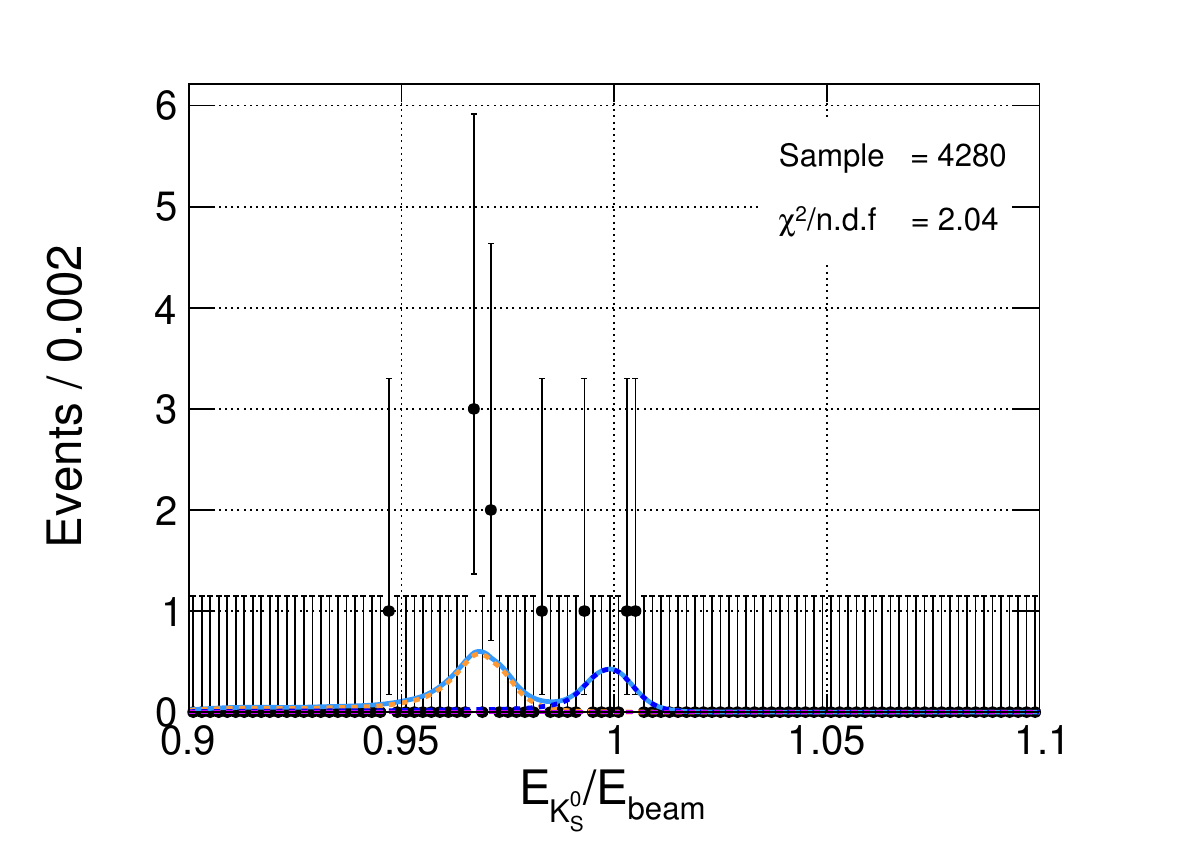}
  \includegraphics[width=0.3\textwidth]{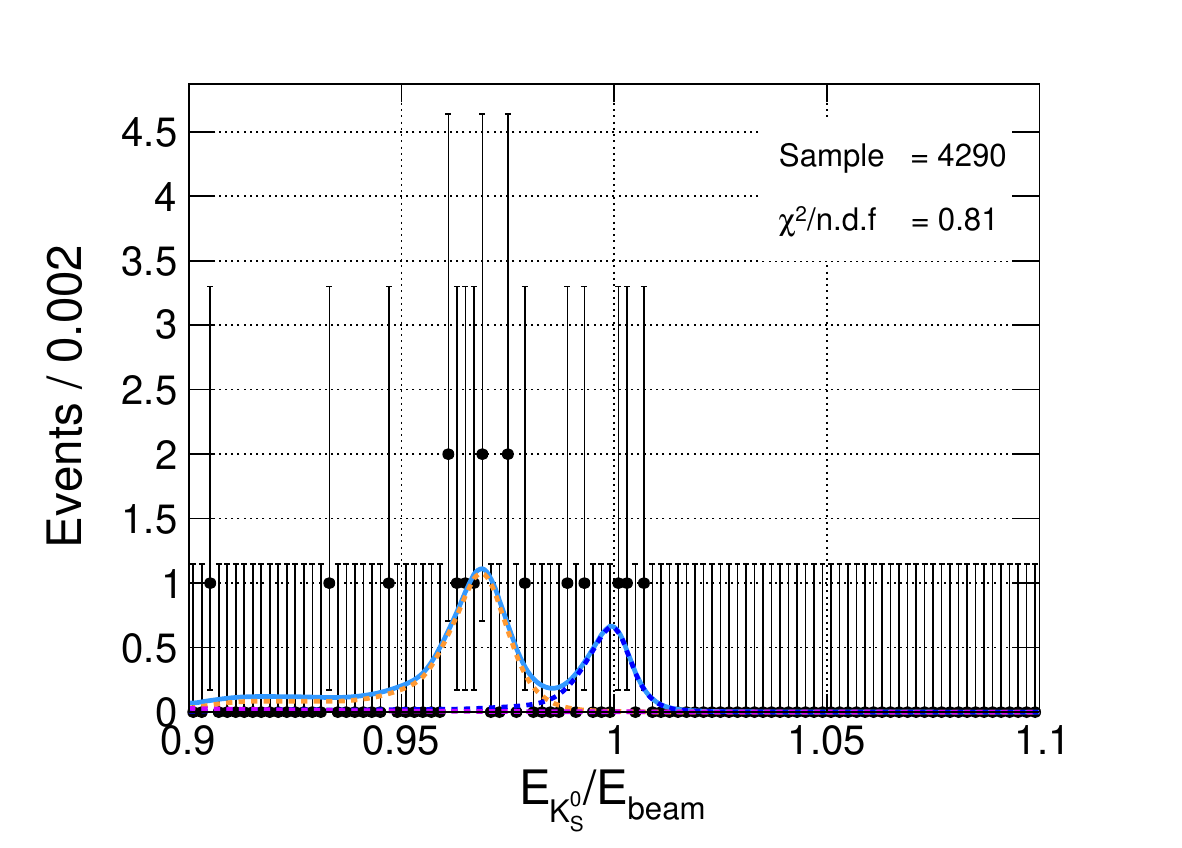}
  \includegraphics[width=0.3\textwidth]{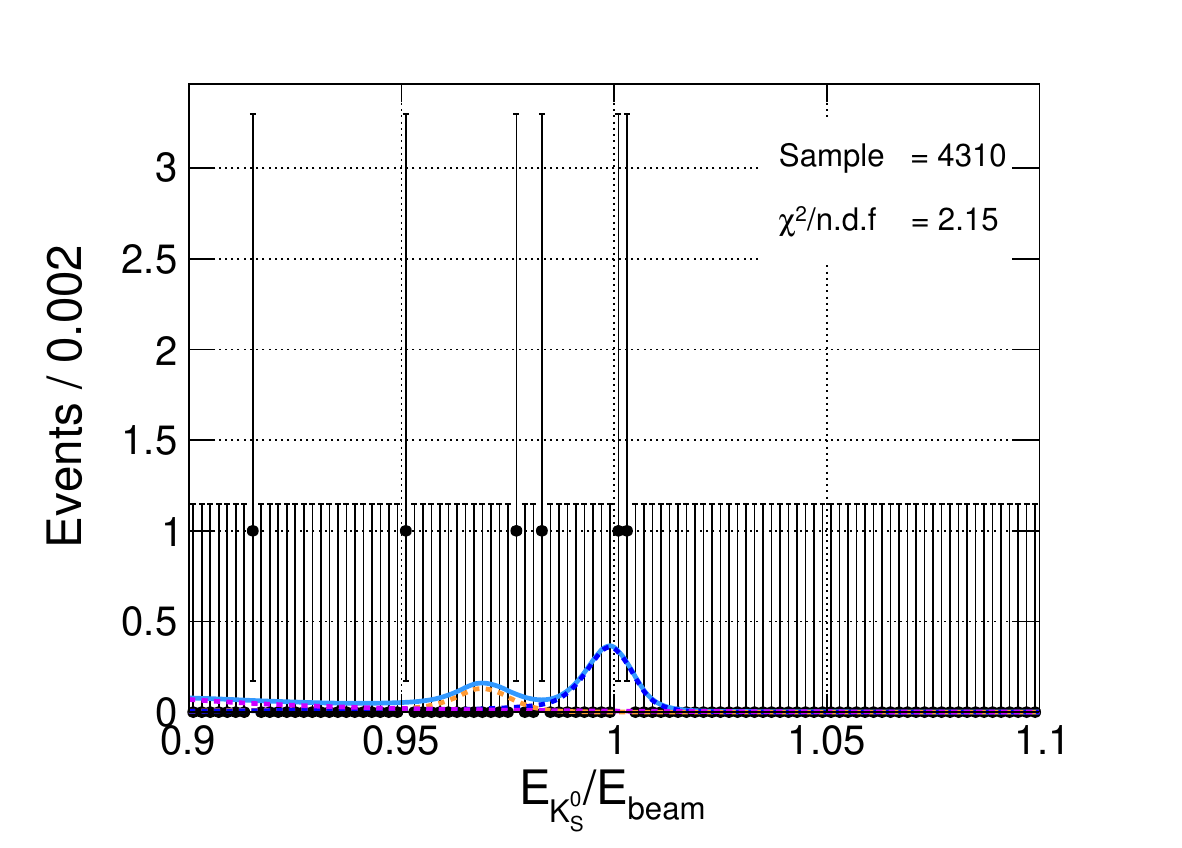}\\
  \includegraphics[width=0.3\textwidth]{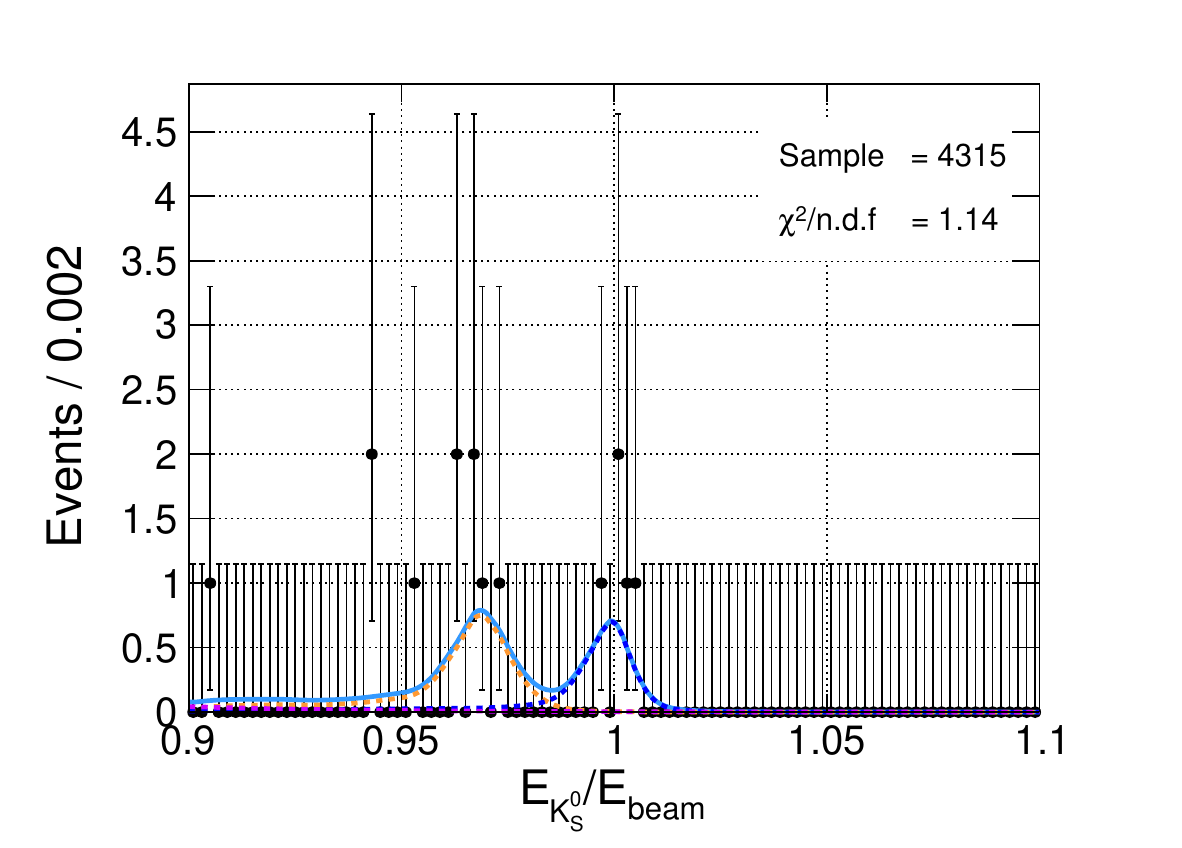}
  \includegraphics[width=0.3\textwidth]{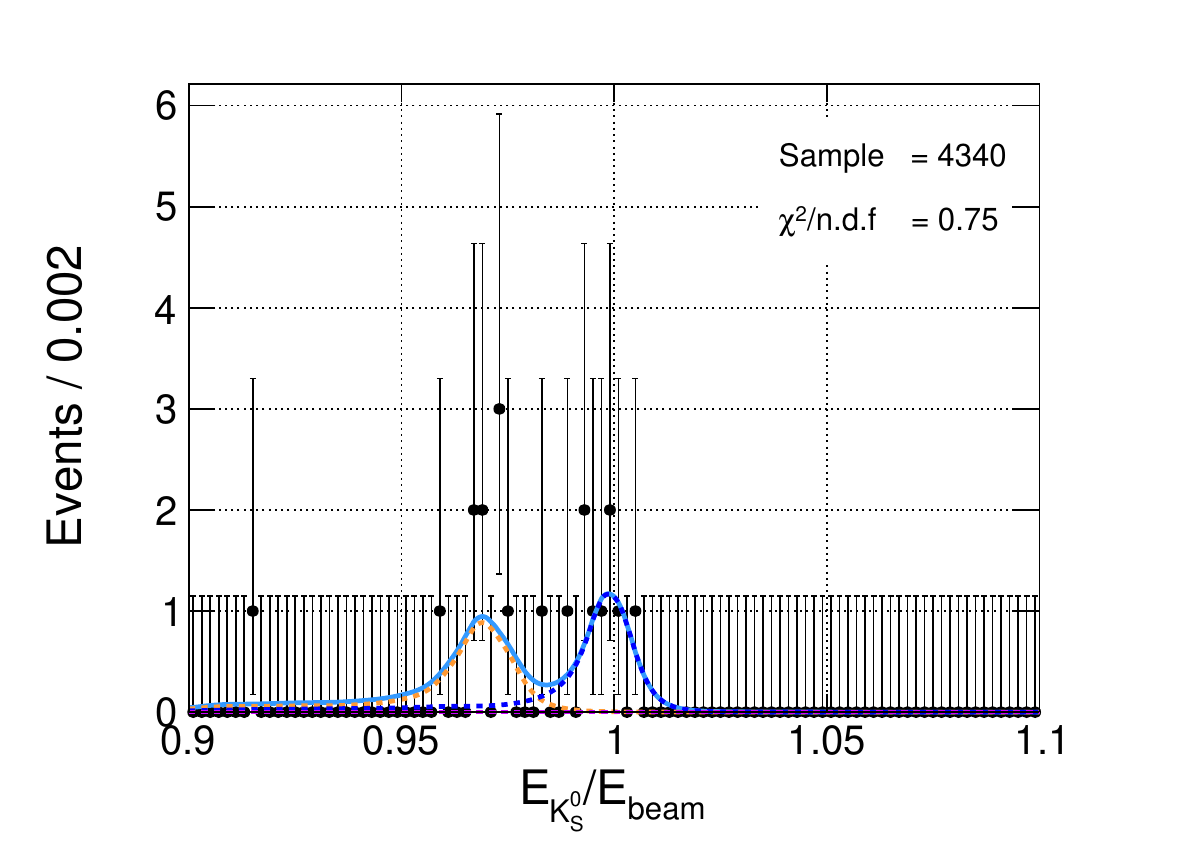}
  \includegraphics[width=0.3\textwidth]{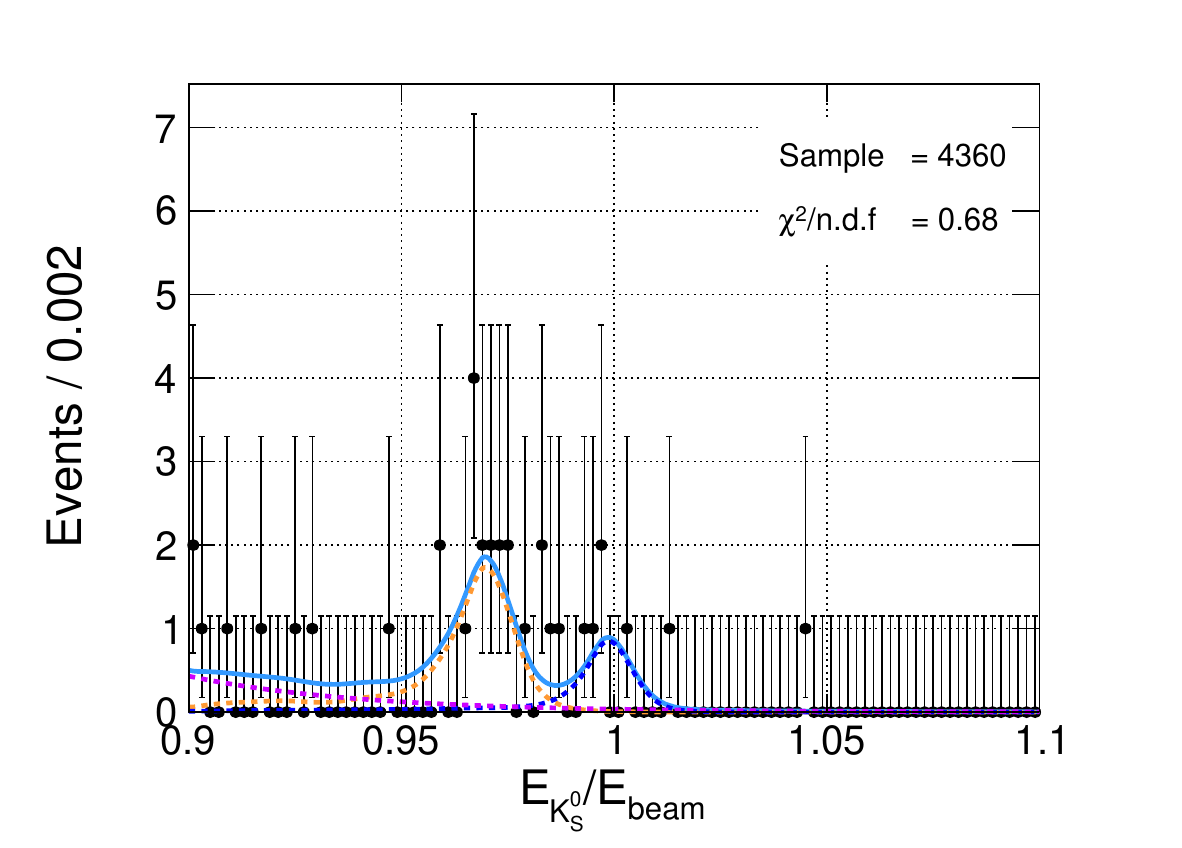}\\
  \caption{Distributions of $X\equiv E_{K_S^0}/E_{\rm beam}$. Dots with error bars are data. Blue solid, blue dashed, red dashed, orange dashed, and violet dashed lines are the total fit results, the signal MC samples, the MC samples of $e^+e^-\rightarrow \gamma^{ISR}\psi(3686)$, the MC samples of $e^+e^-\rightarrow K^{\ast0}(892)\bar{K^0}$, and the exponential functions describing the remaining background, respectively. Each MC simulated shape is convolved with a Gaussian function.}\label{data2}
\end{figure}

\newpage
\begin{figure}[tp]
  \centering
  %\ContinuedFloat
  \includegraphics[width=0.3\textwidth]{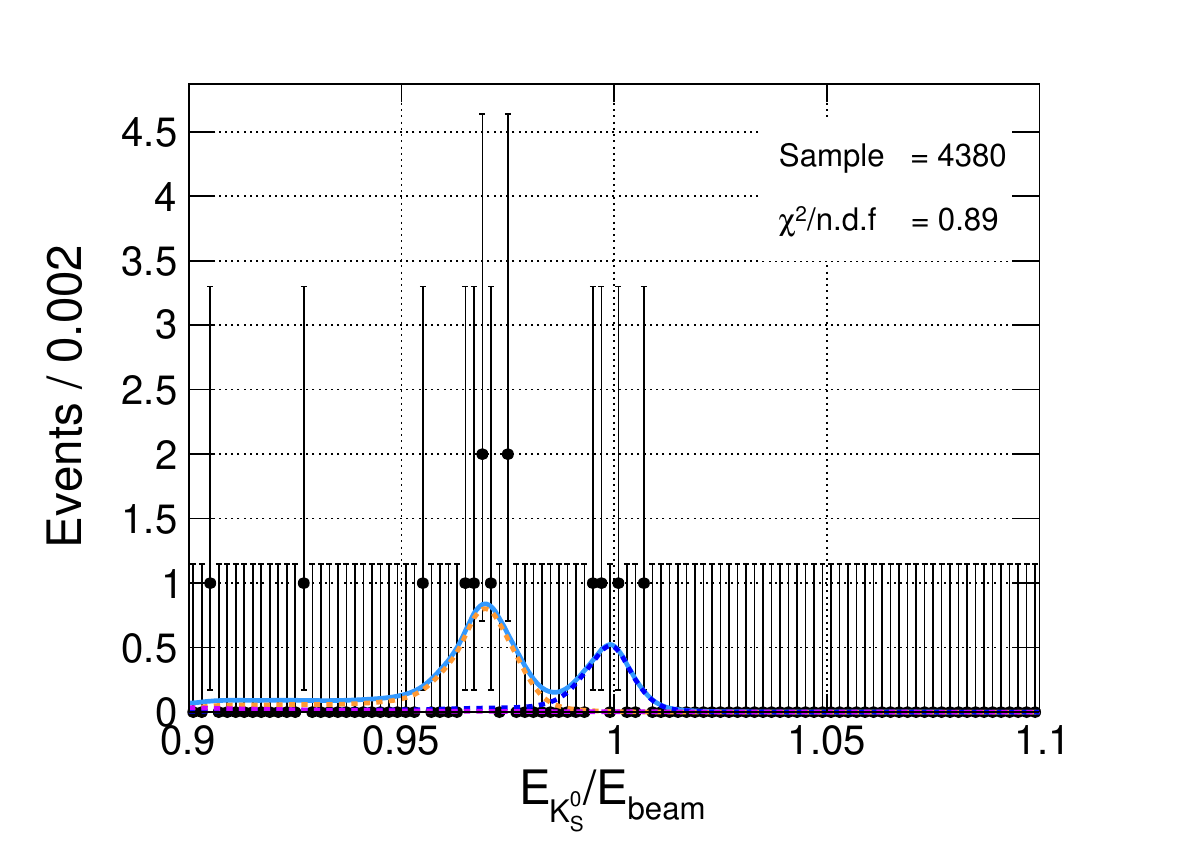}
  \includegraphics[width=0.3\textwidth]{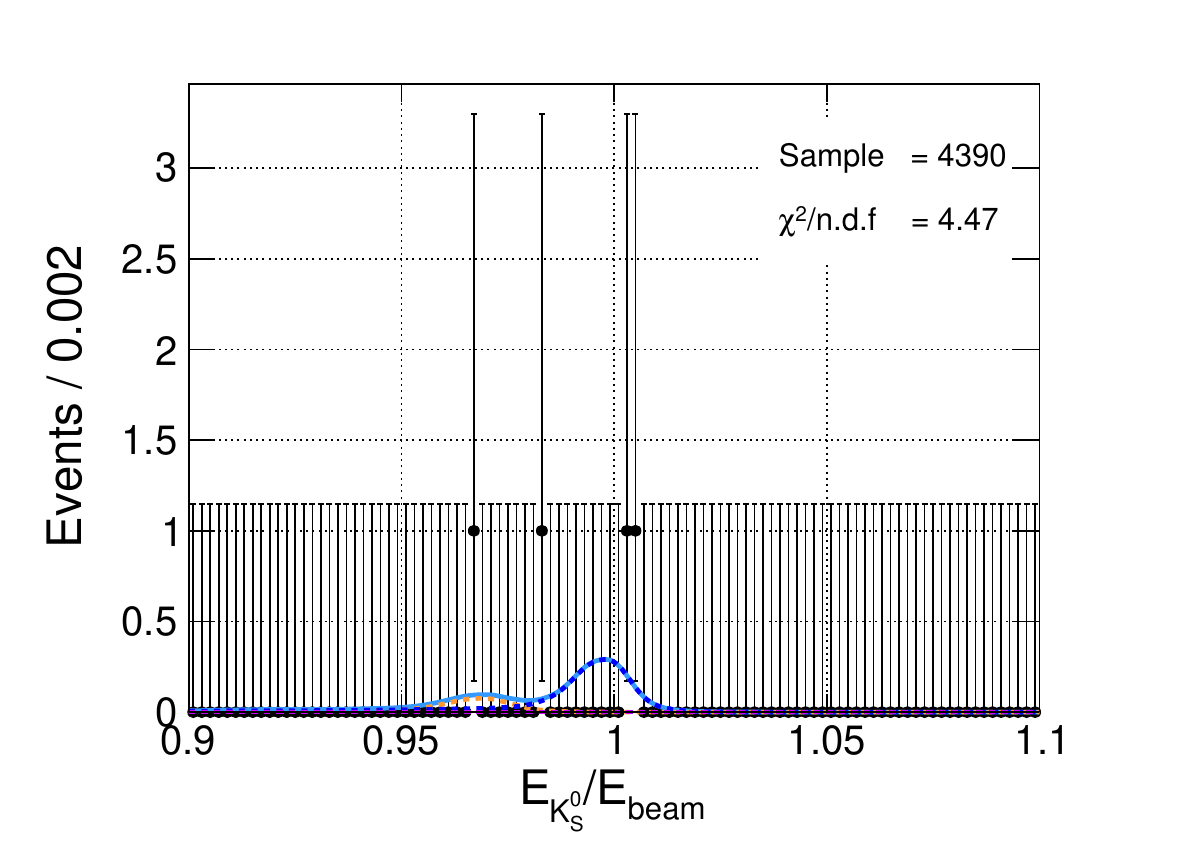}
  \includegraphics[width=0.3\textwidth]{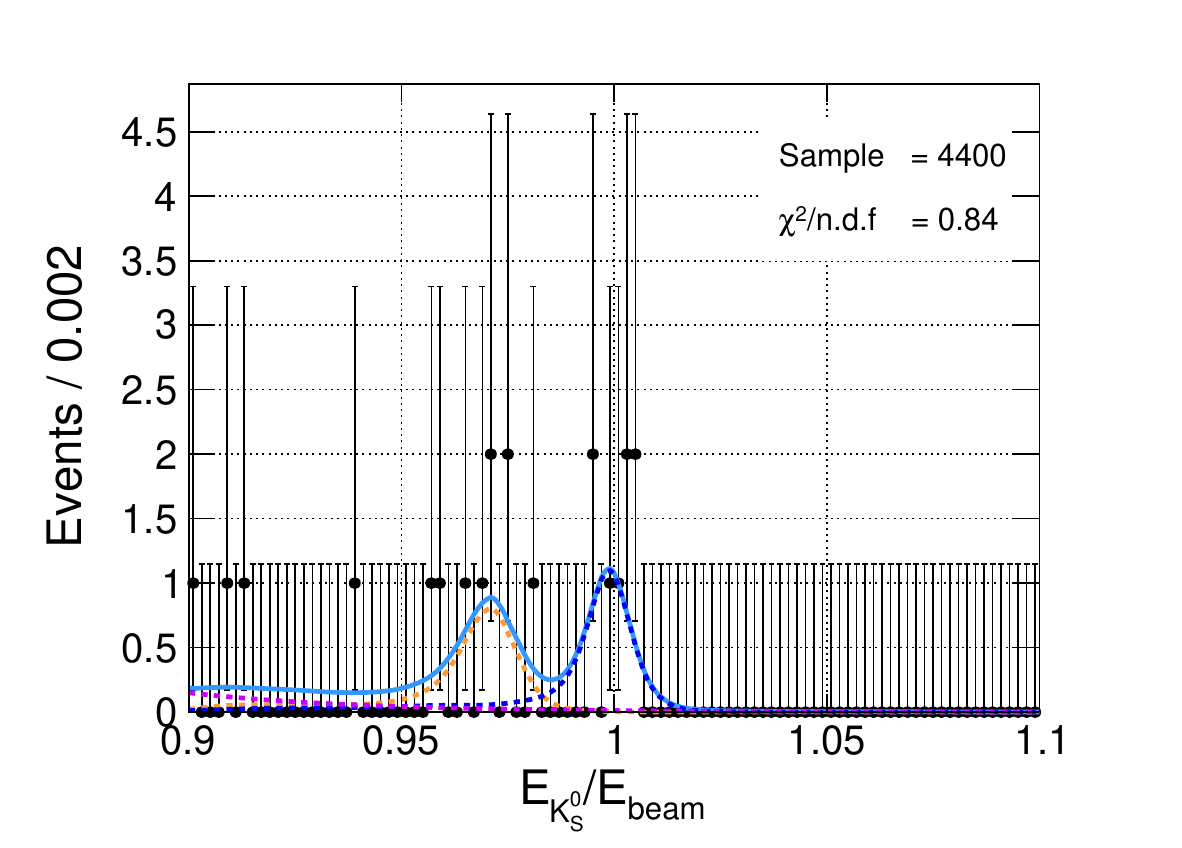}\\
  \includegraphics[width=0.3\textwidth]{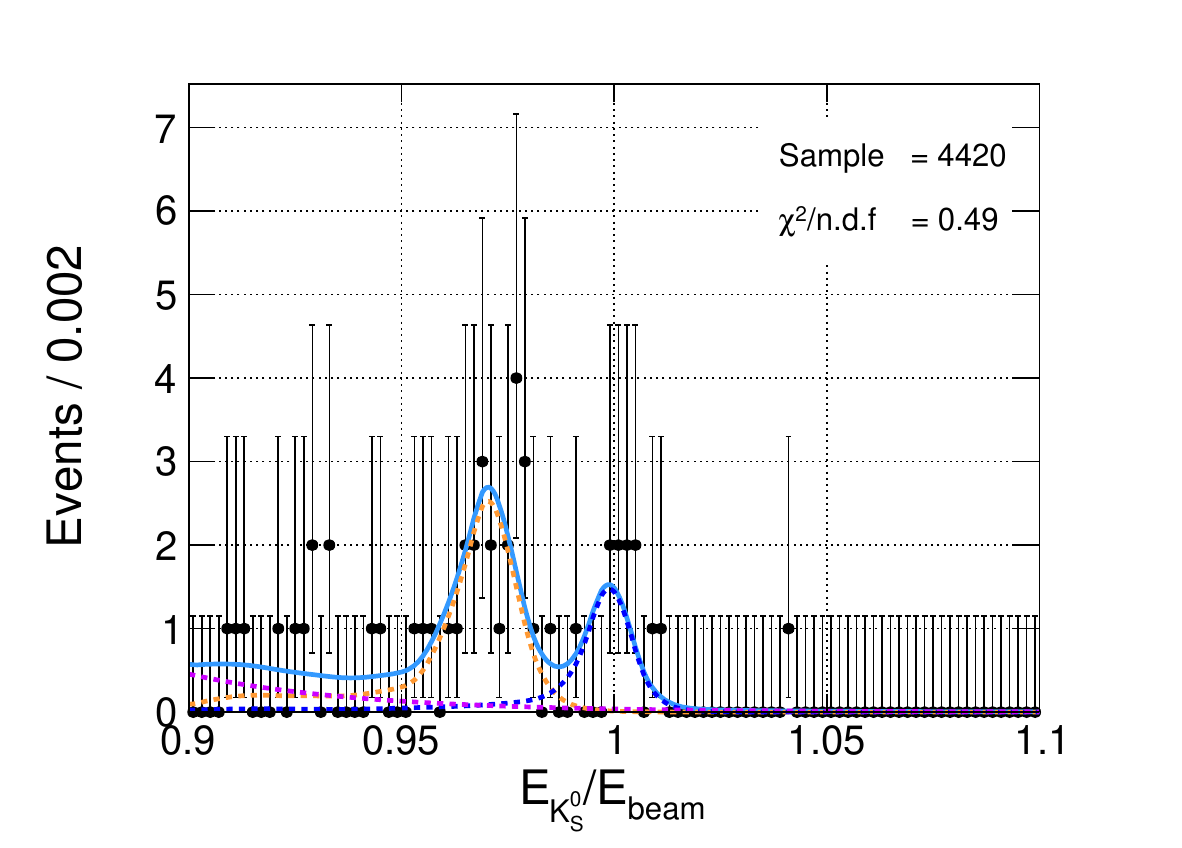}
  \includegraphics[width=0.3\textwidth]{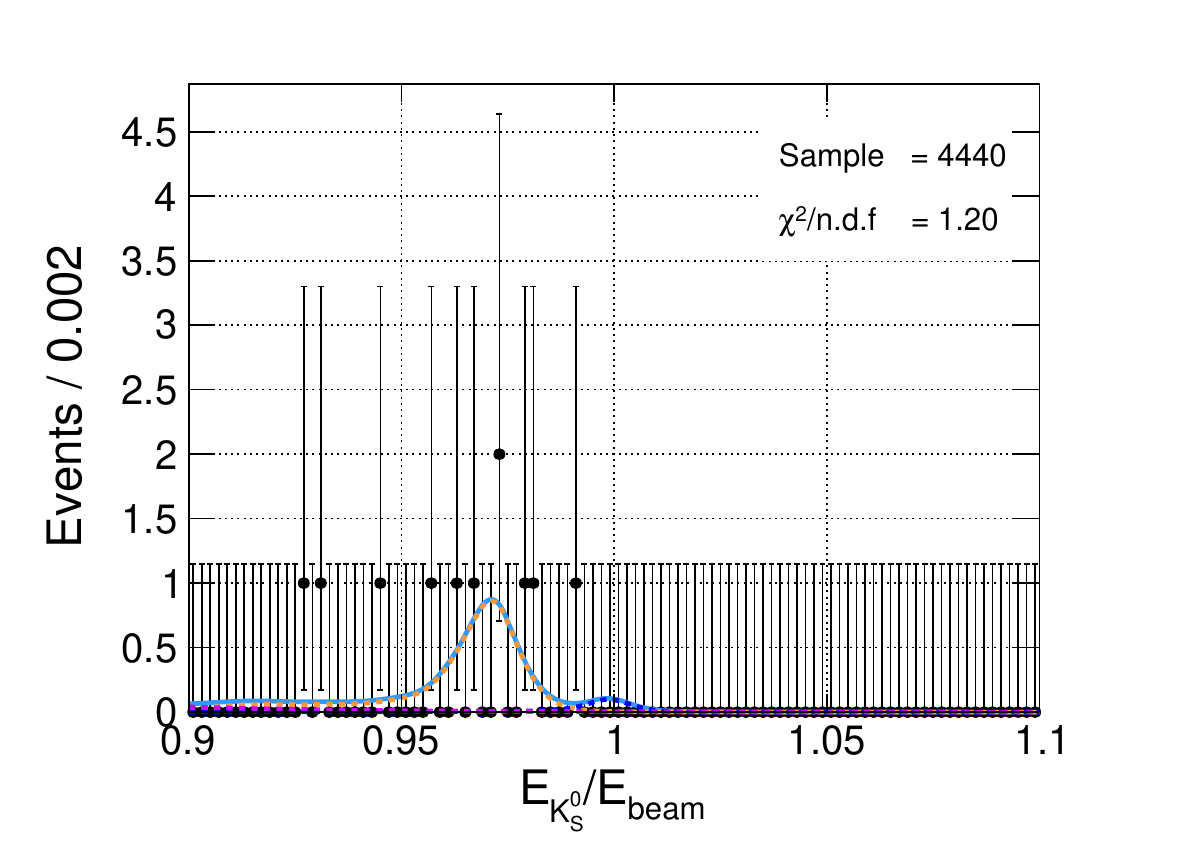}
  \includegraphics[width=0.3\textwidth]{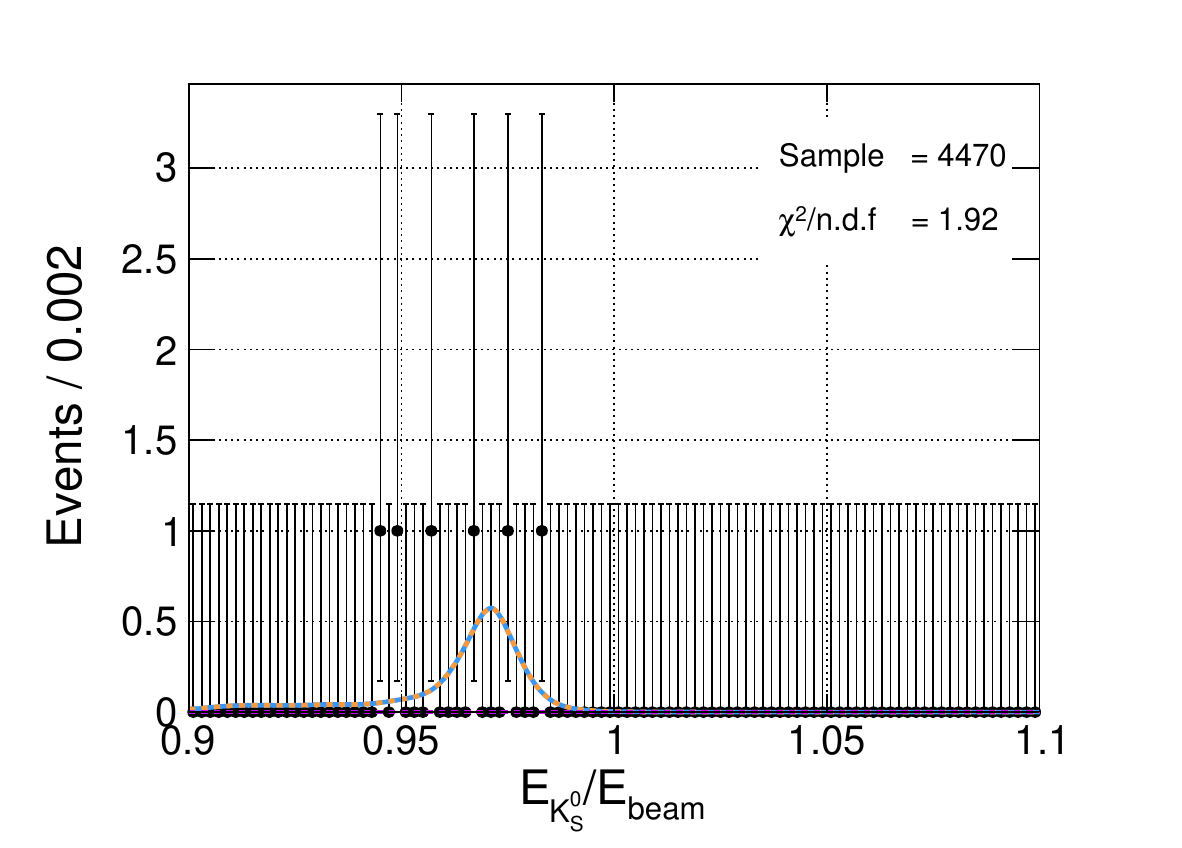}\\
  \includegraphics[width=0.3\textwidth]{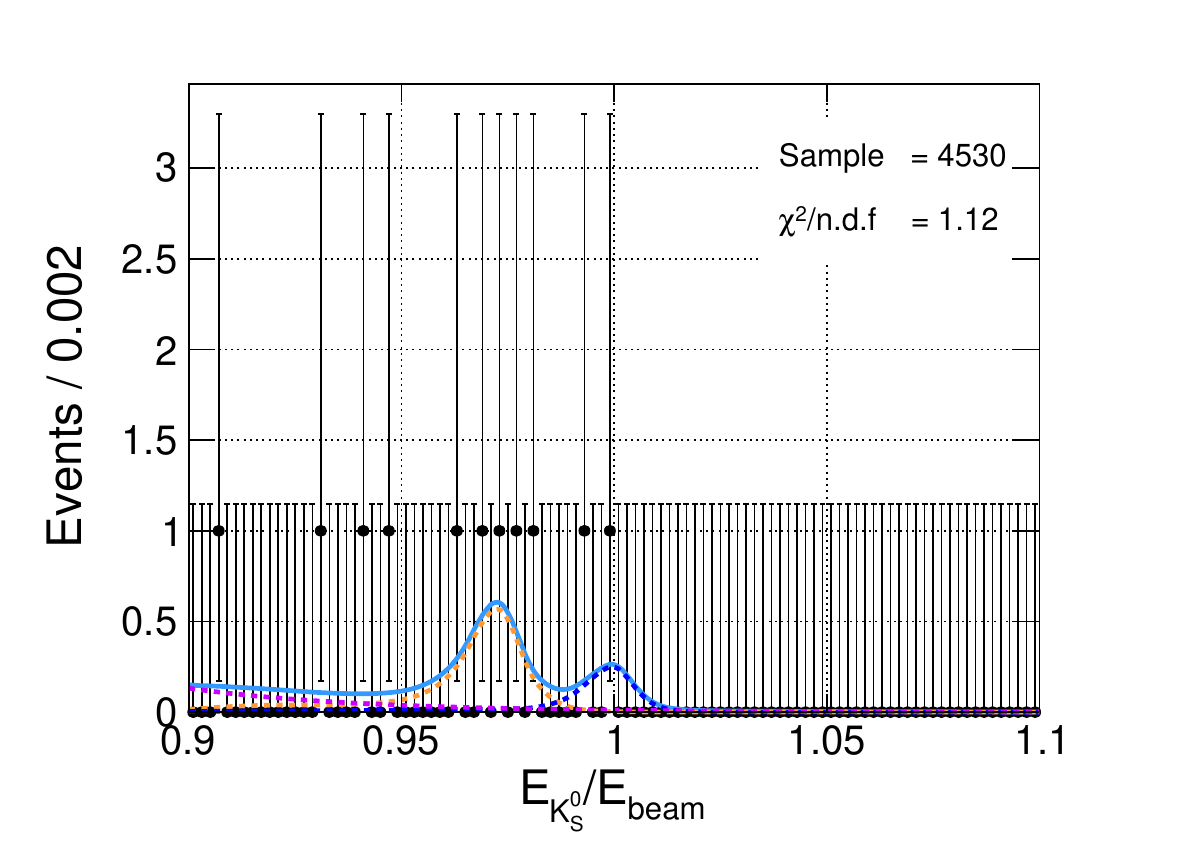}
  \includegraphics[width=0.3\textwidth]{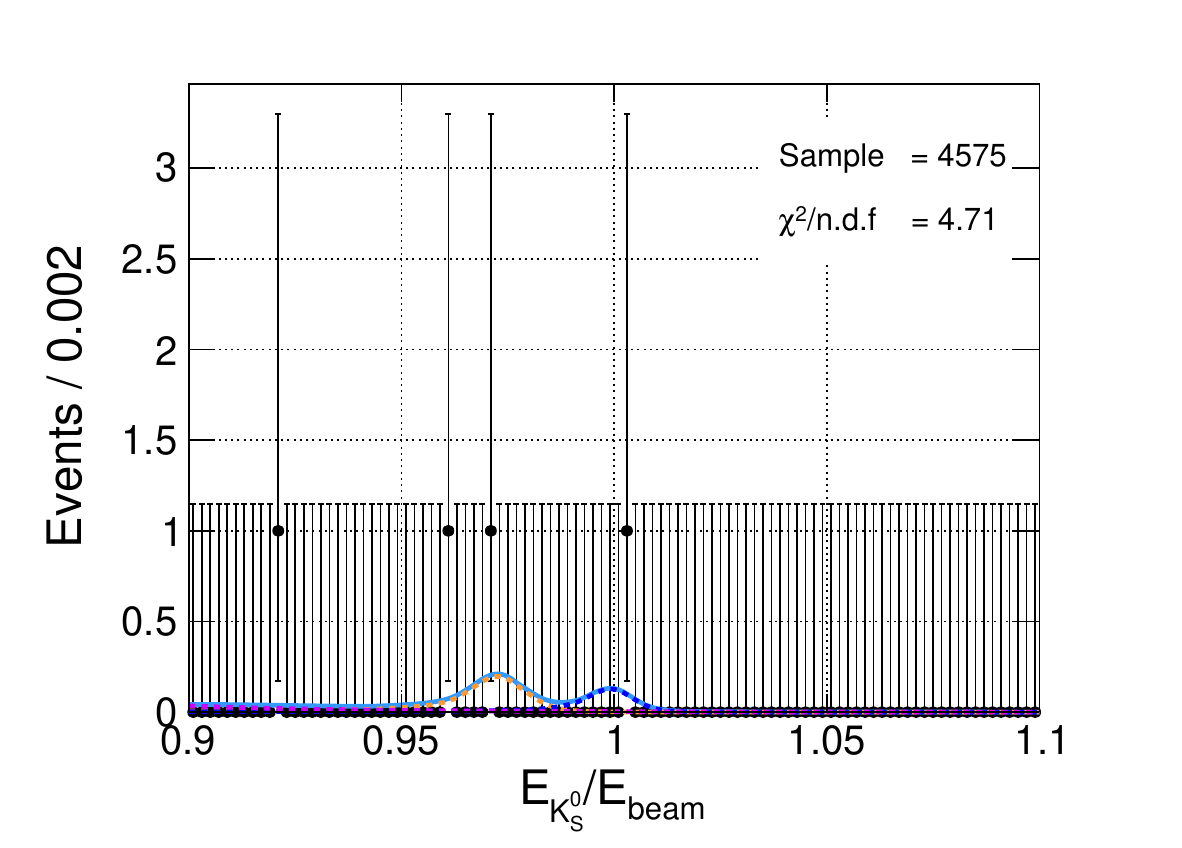}
  \includegraphics[width=0.3\textwidth]{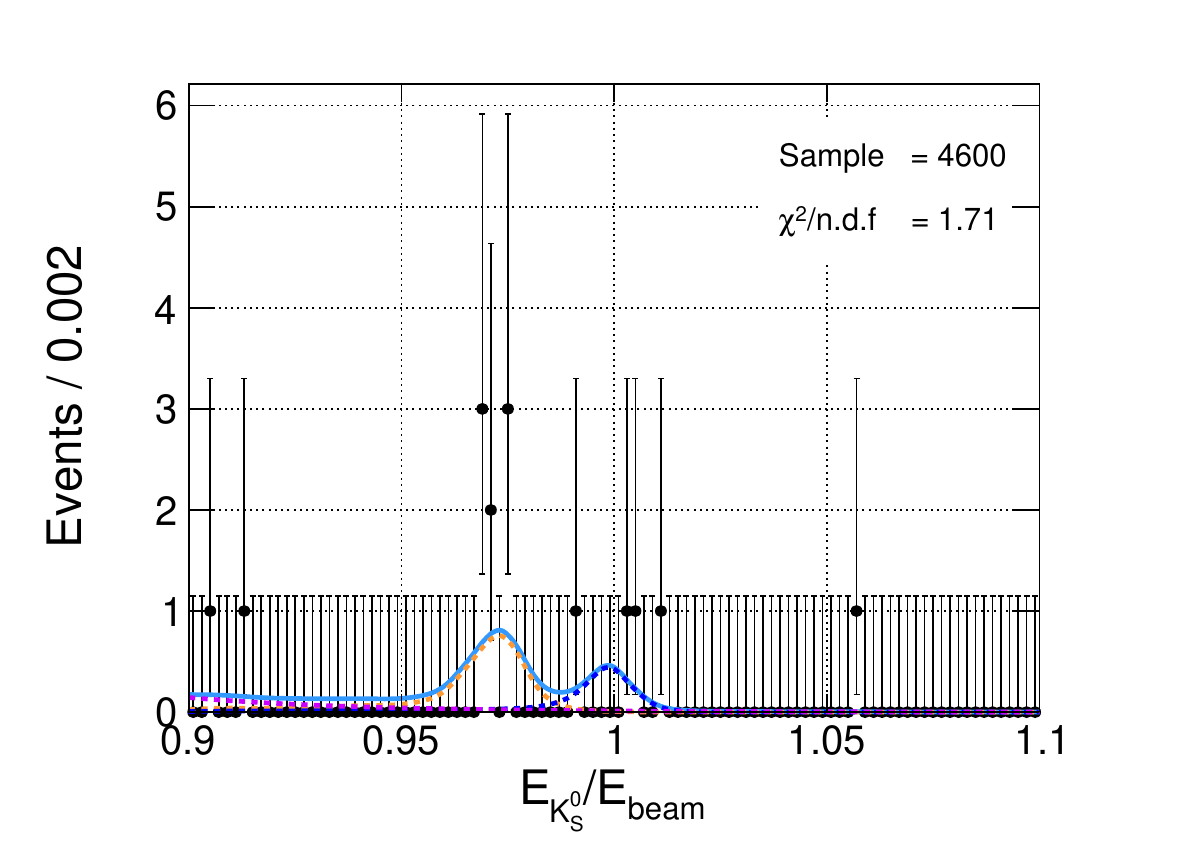}\\
  \includegraphics[width=0.3\textwidth]{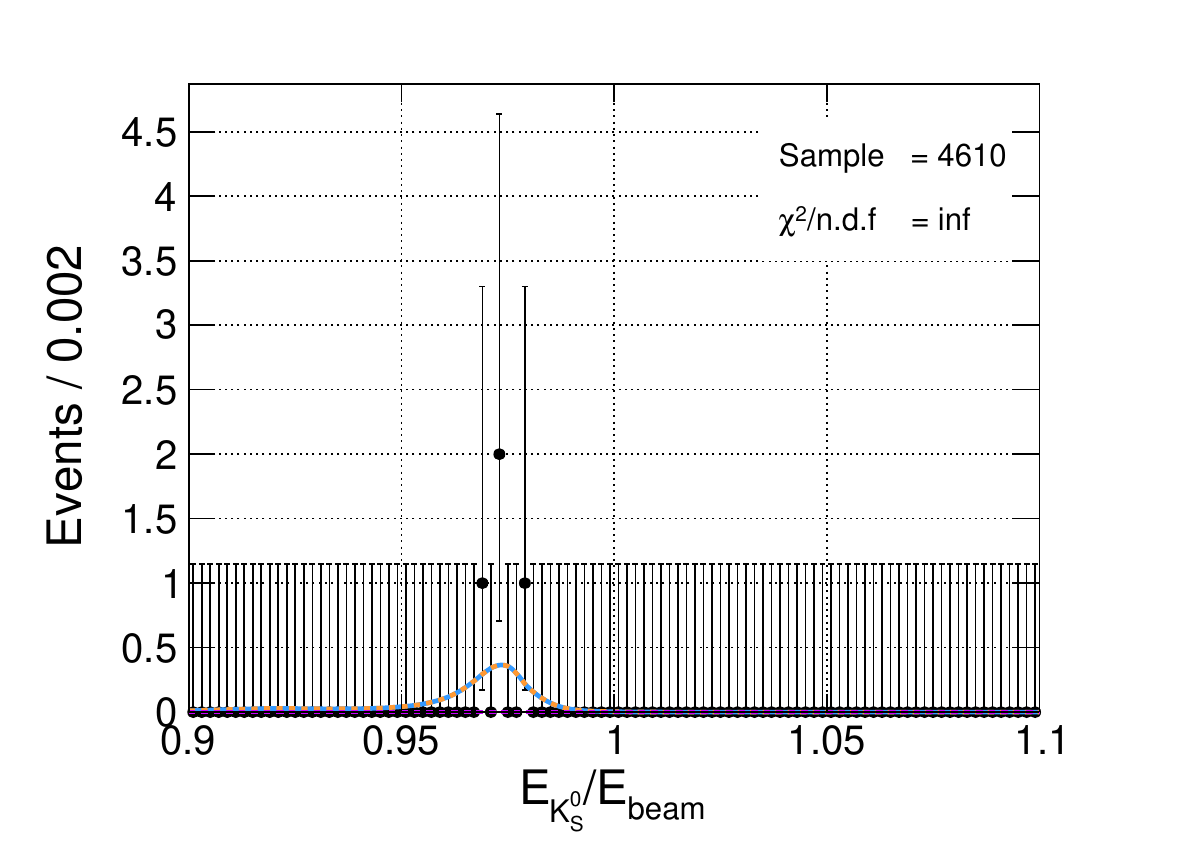}
  \includegraphics[width=0.3\textwidth]{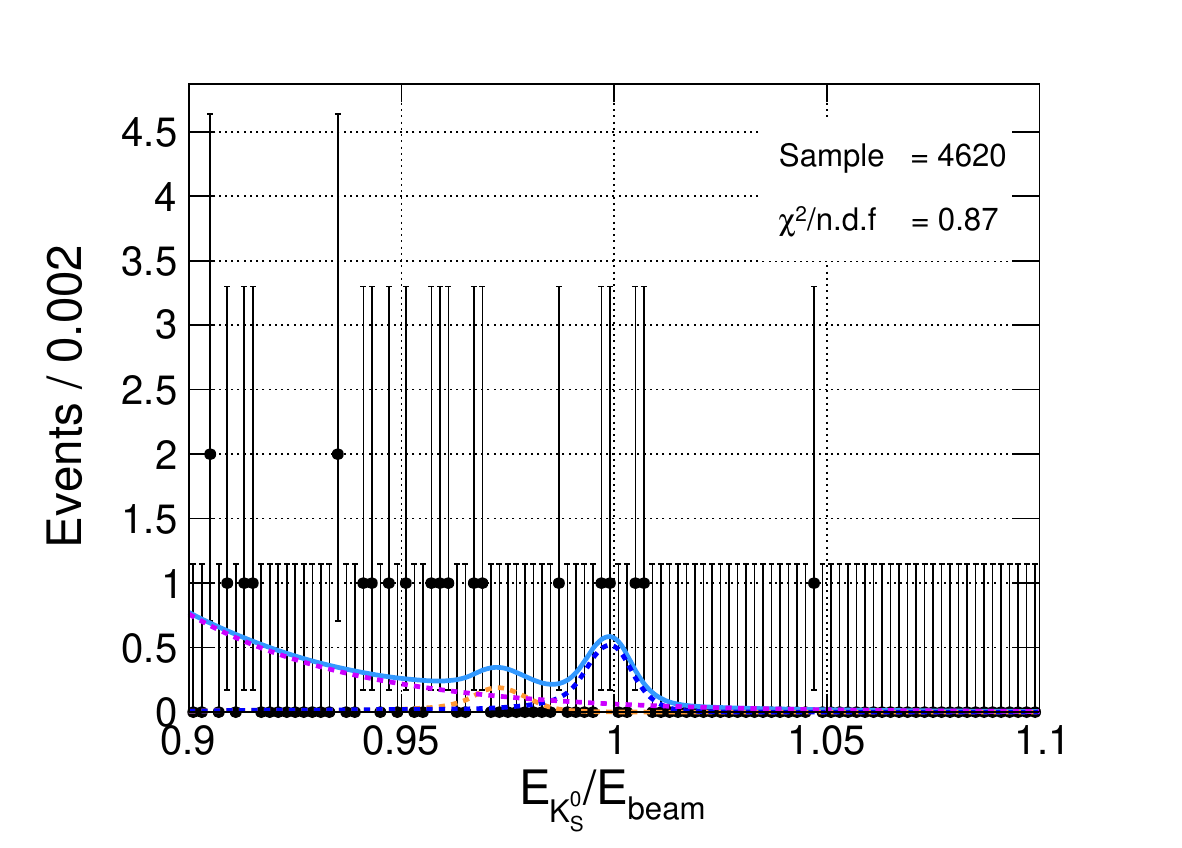}
  \includegraphics[width=0.3\textwidth]{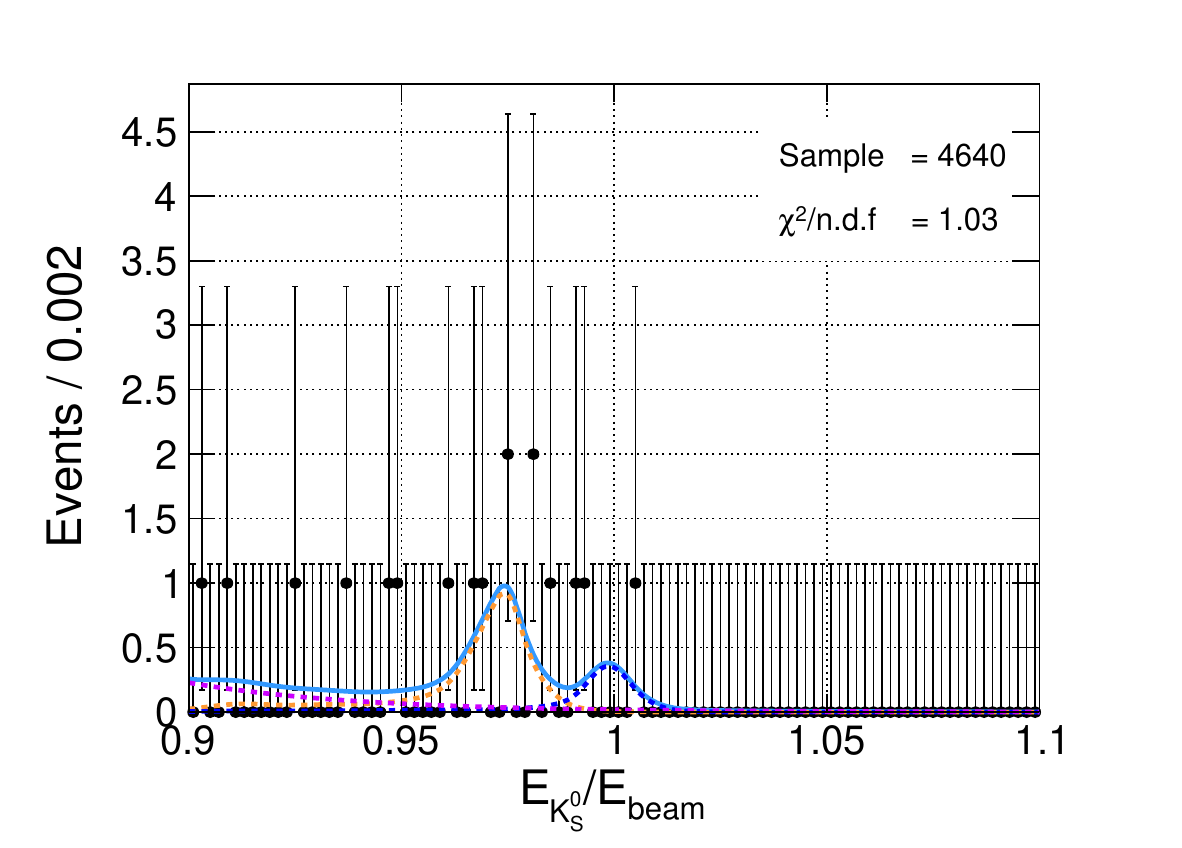}\\
  \includegraphics[width=0.3\textwidth]{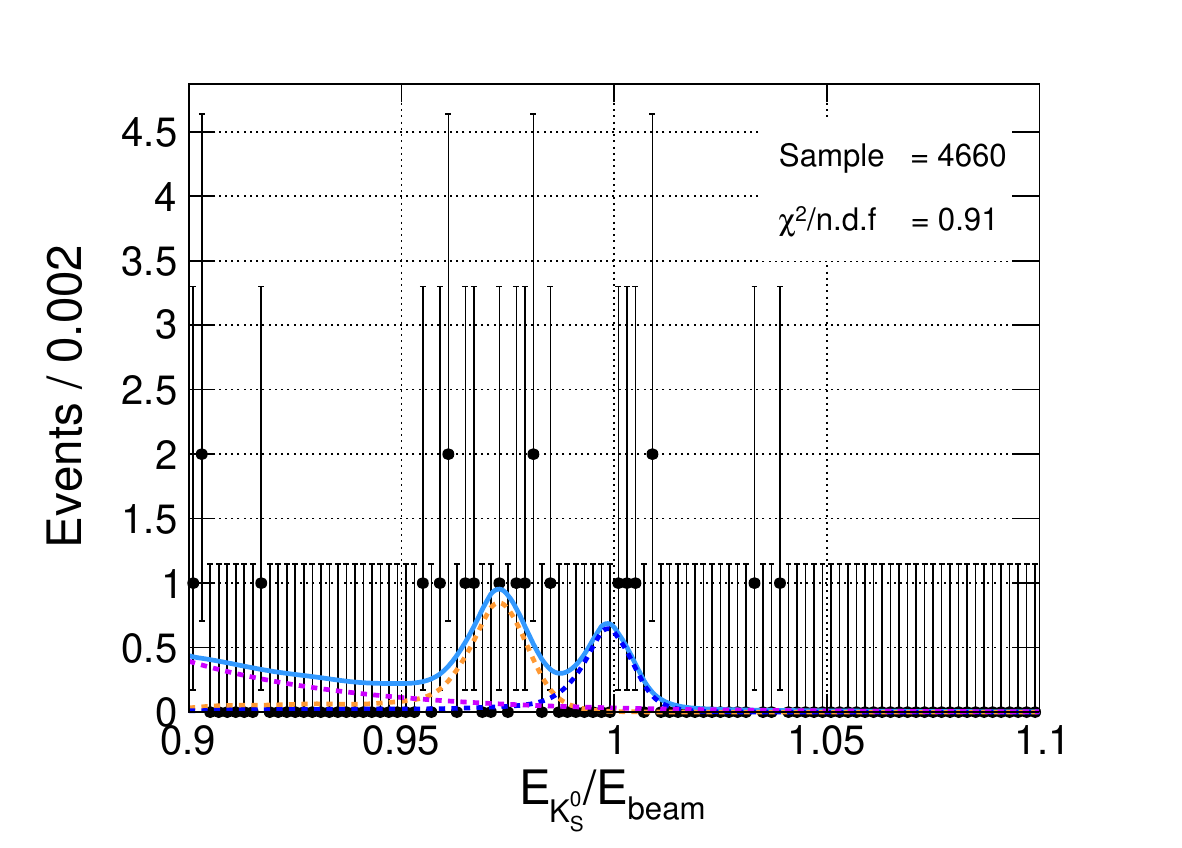}
  \includegraphics[width=0.3\textwidth]{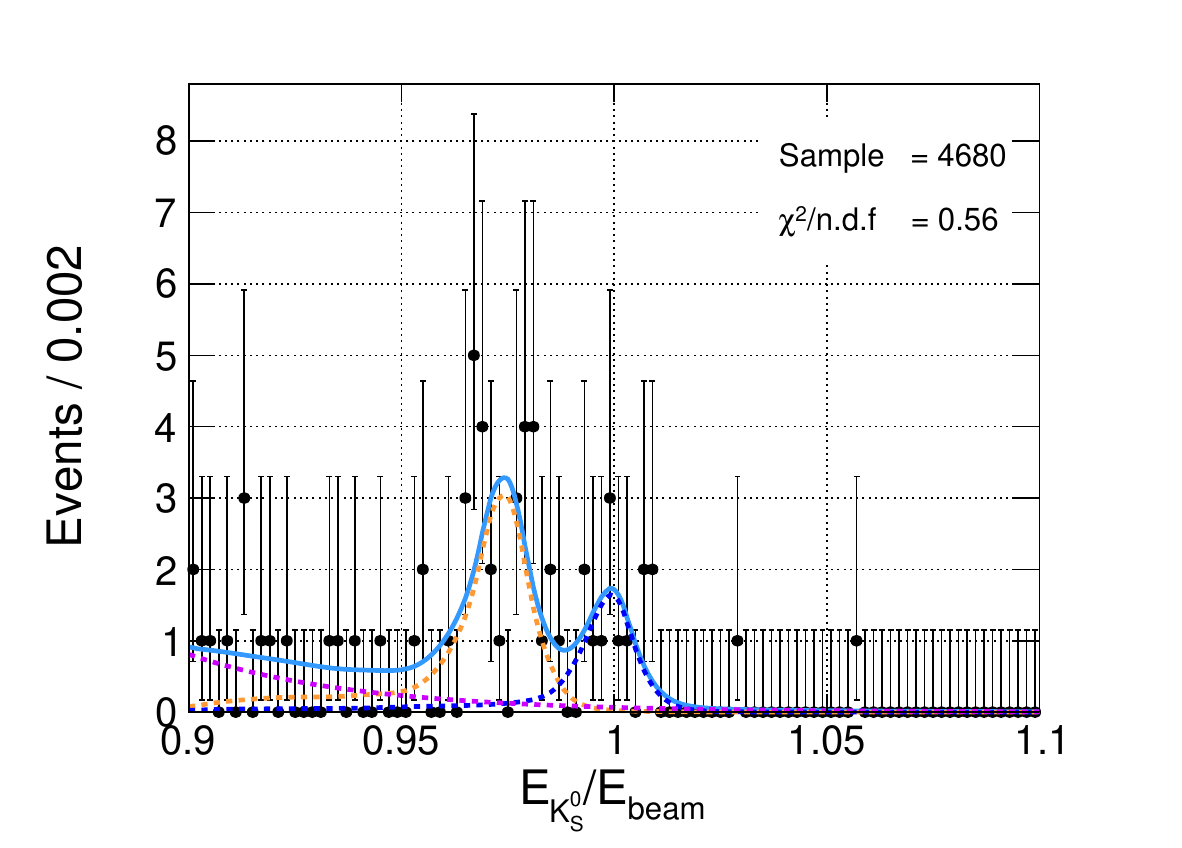}
  \includegraphics[width=0.3\textwidth]{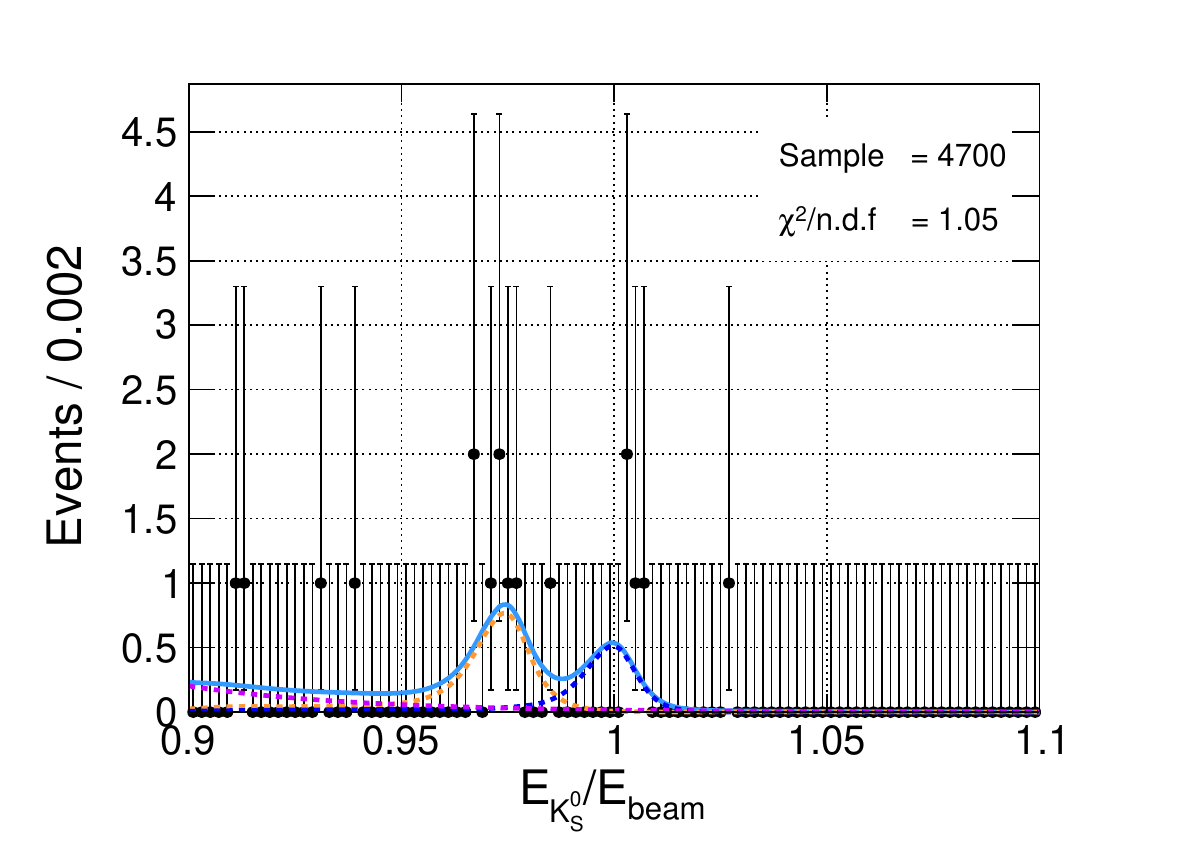}\\
  \caption{Distributions of $X\equiv E_{K_S^0}/E_{\rm beam}$. Dots with error bars are data. Blue solid, blue dashed, red dashed, orange dashed, and violet dashed lines are the total fit results, the signal MC samples, the MC samples of $e^+e^-\rightarrow \gamma^{ISR}\psi(3686)$, the MC samples of $e^+e^-\rightarrow K^{\ast0}(892)\bar{K^0}$, and the exponential functions describing the remaining background, respectively. Each MC simulated shape is convolved with a Gaussian function.}\label{data3}
\end{figure}

\newpage
\begin{figure}[tp]
  \centering
  %\ContinuedFloat
  \includegraphics[width=0.3\textwidth]{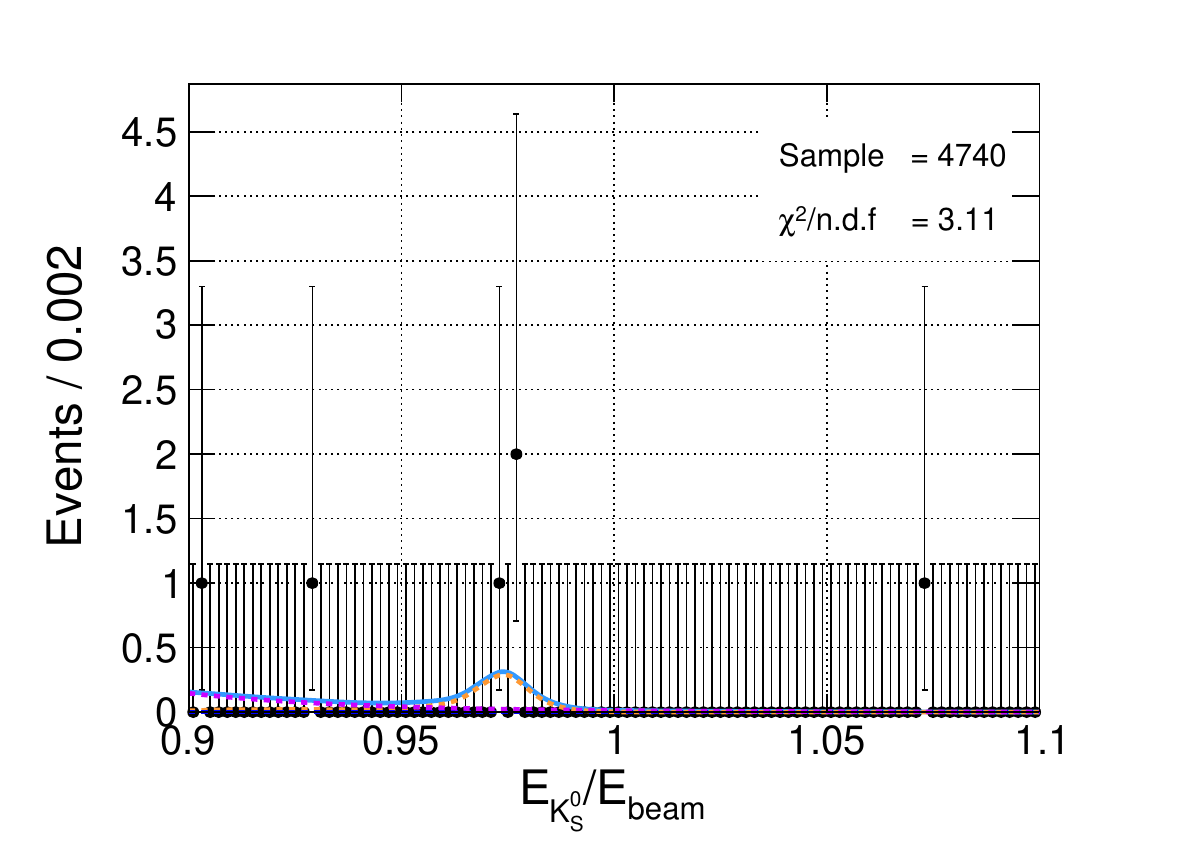}
  \includegraphics[width=0.3\textwidth]{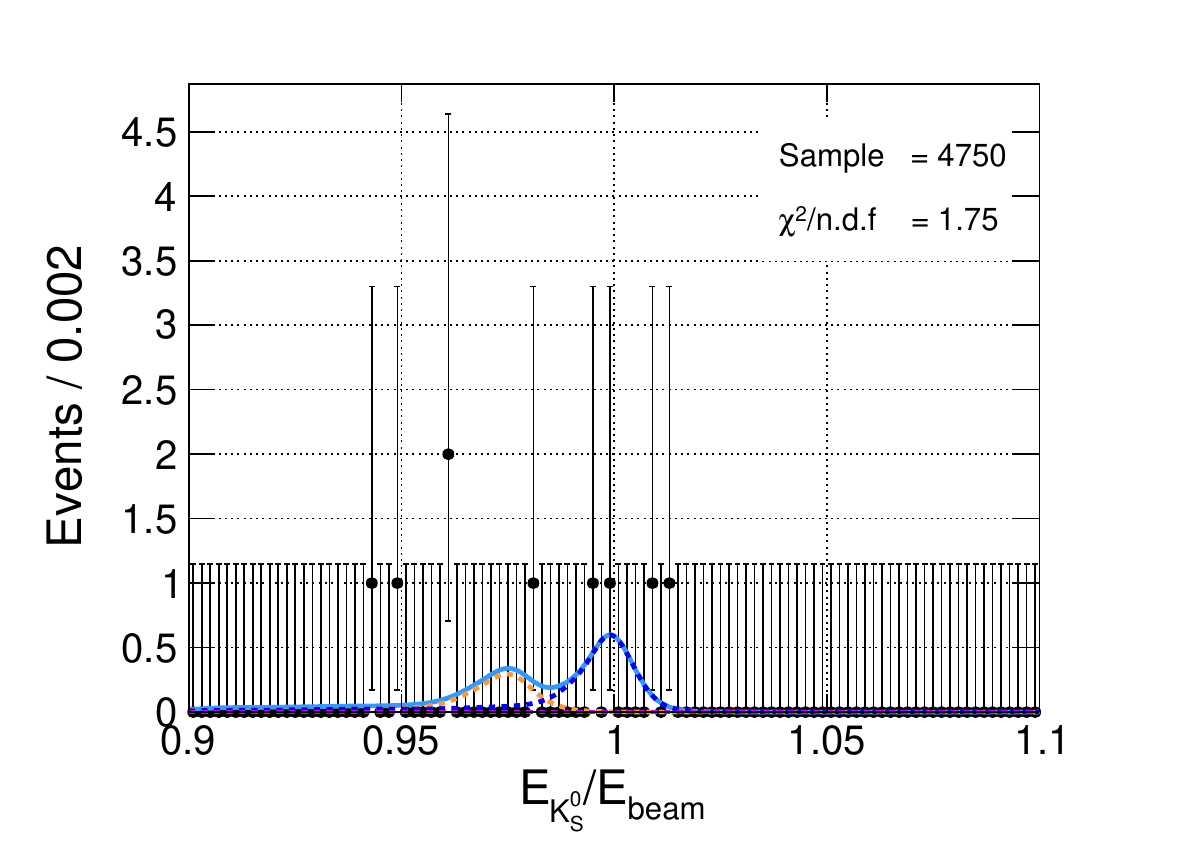}
  \includegraphics[width=0.3\textwidth]{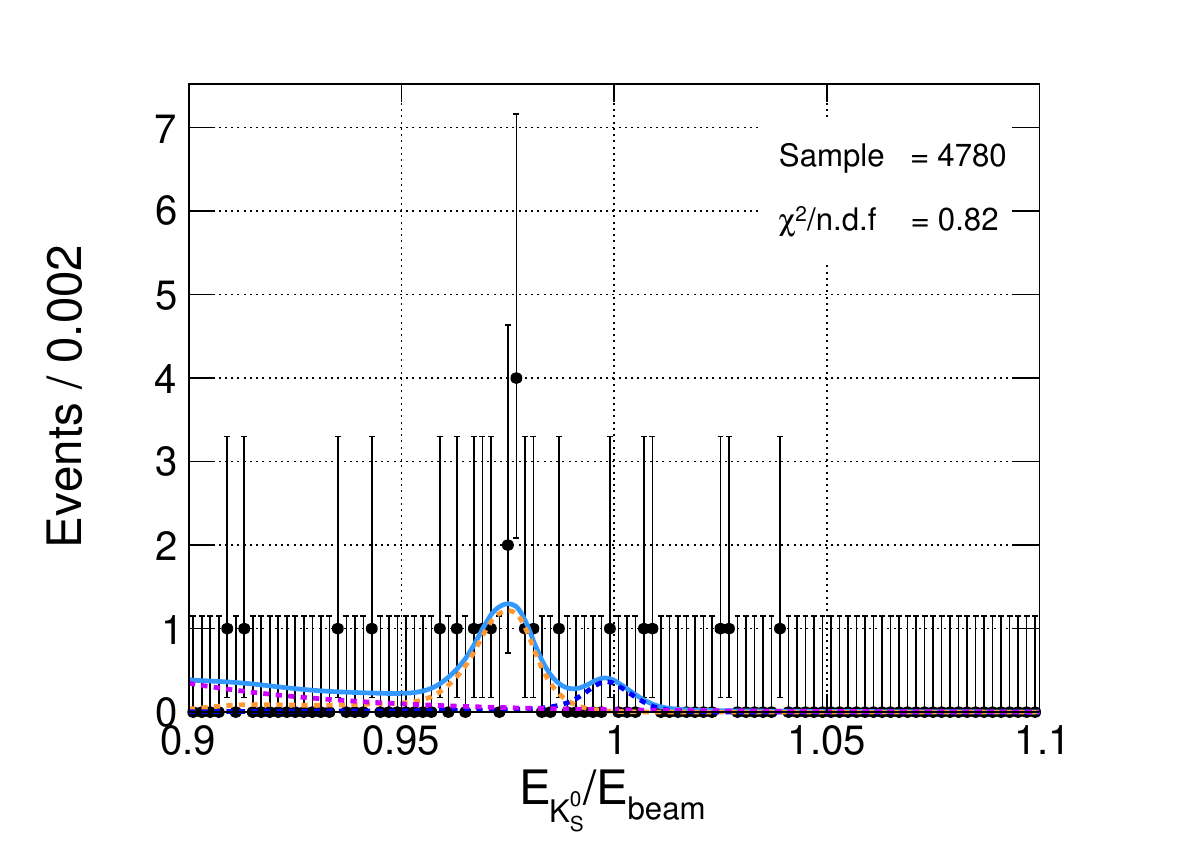}\\
  \includegraphics[width=0.3\textwidth]{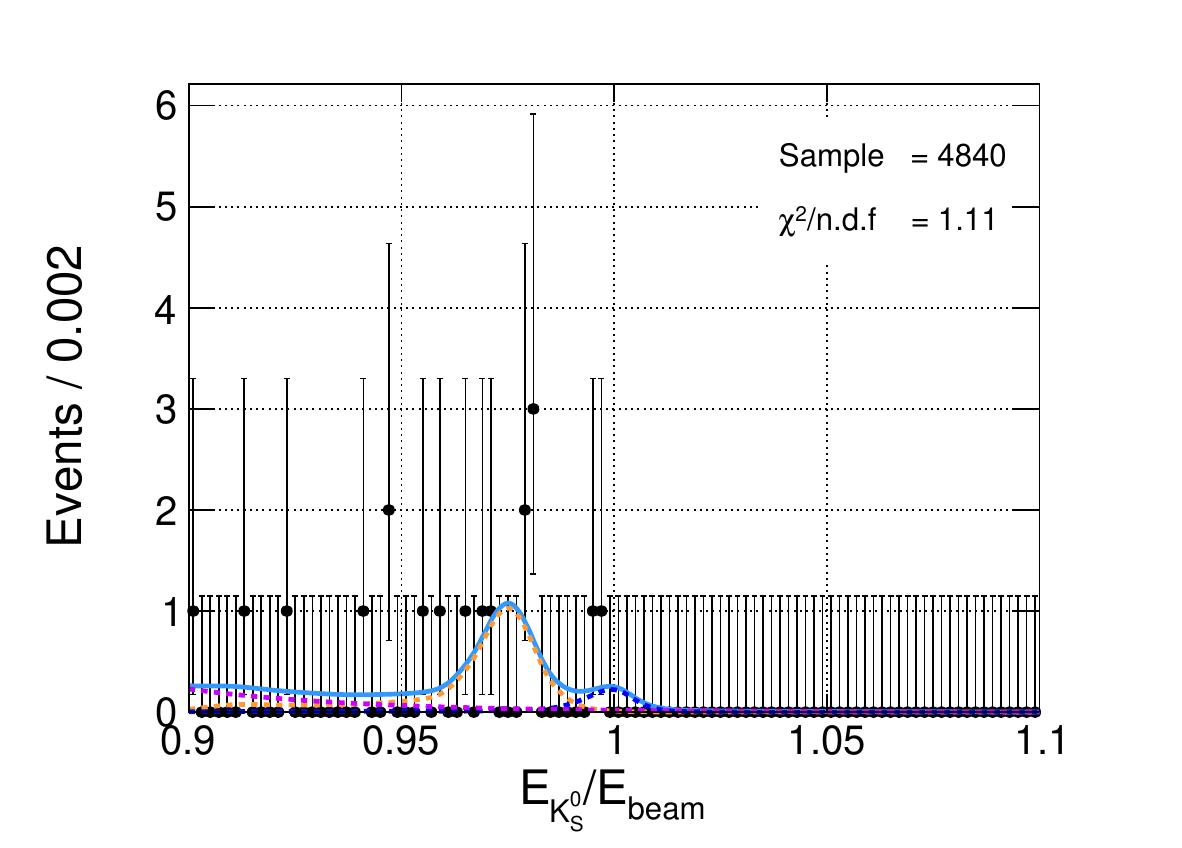}
  \includegraphics[width=0.3\textwidth]{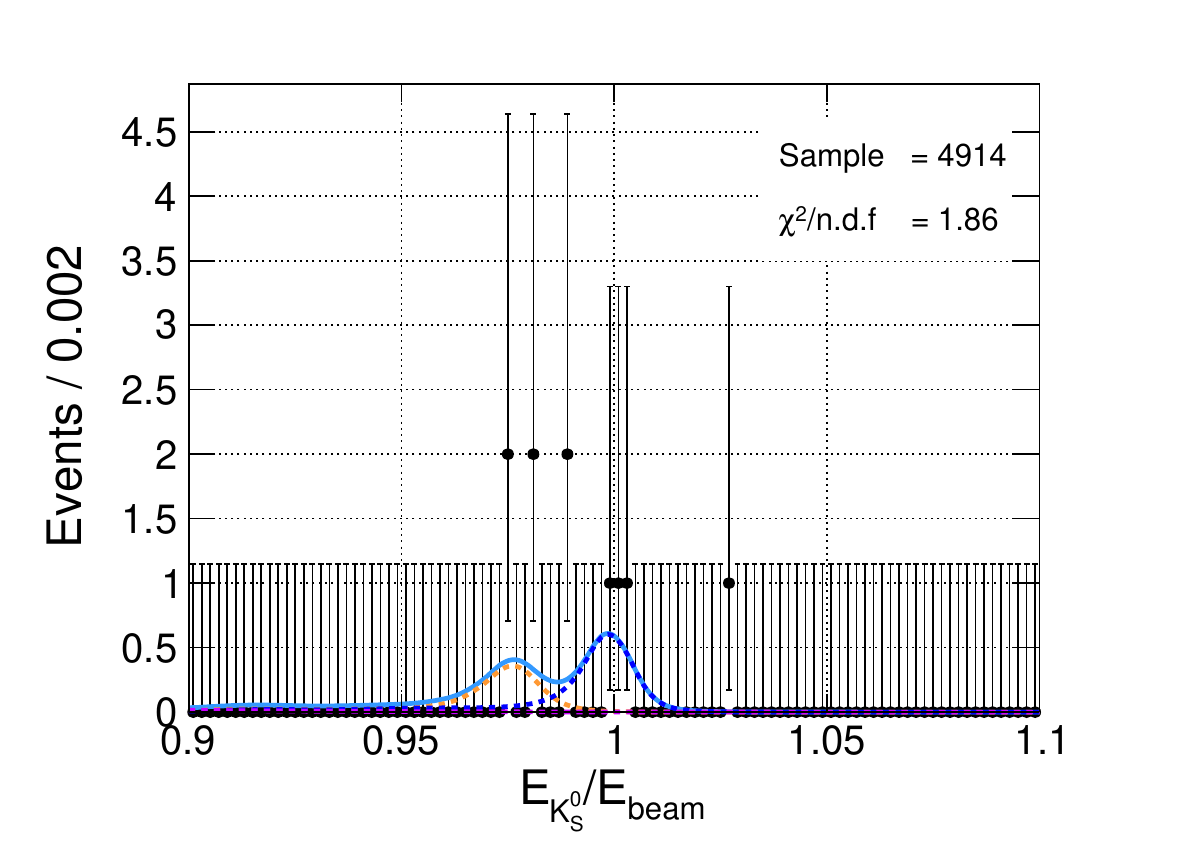}
  \includegraphics[width=0.3\textwidth]{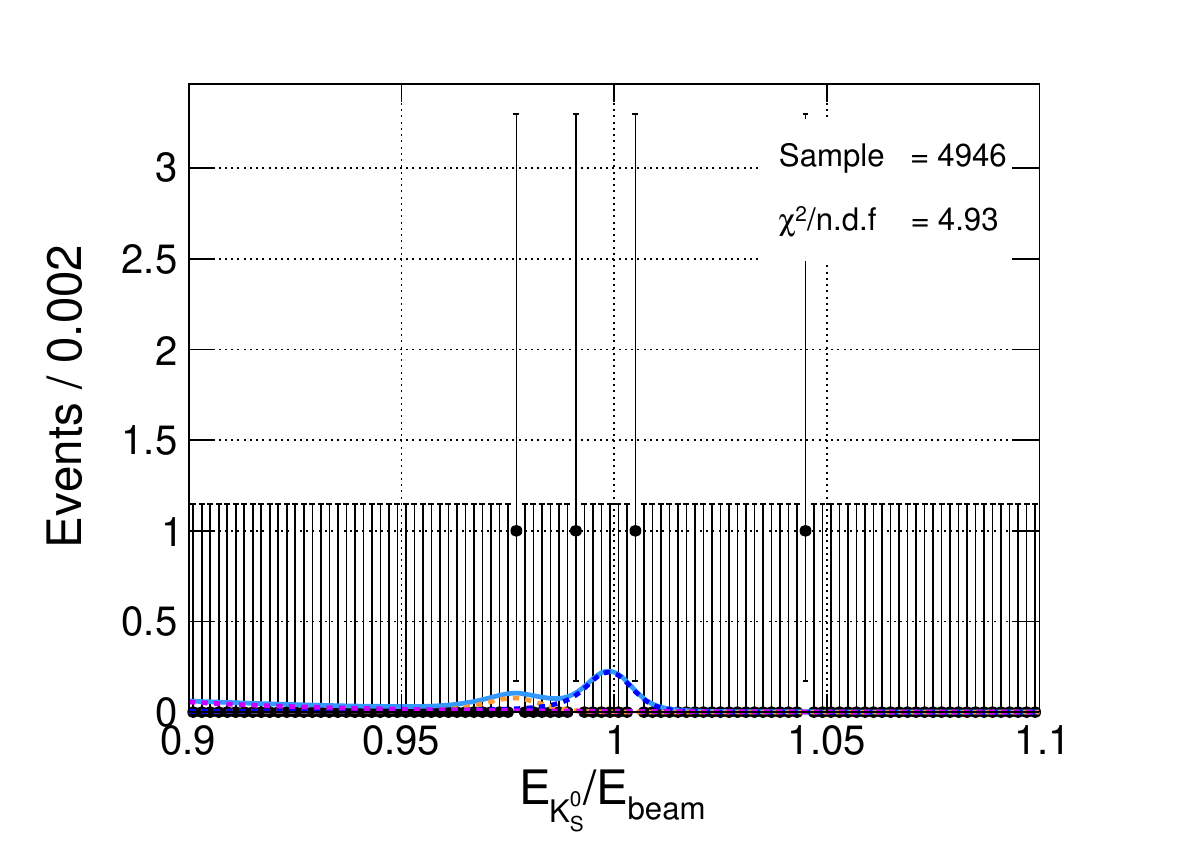}\\
  \caption{Distributions of $X\equiv E_{K_S^0}/E_{\rm beam}$. Dots with error bars are data. Blue solid, blue dashed, red dashed, orange dashed, and violet dashed lines are the total fit results, the signal MC samples, the MC samples of $e^+e^-\rightarrow \gamma^{ISR}\psi(3686)$, the MC samples of $e^+e^-\rightarrow K^{\ast0}(892)\bar{K^0}$, and the exponential functions describing the remaining background, respectively. Each MC simulated shape is convolved with a Gaussian function.}\label{data4}
\end{figure}

\clearpage

\nocite{*}
%\bibliography{draft}